\documentclass[fleqn,usenatbib]{mnras}

\usepackage{newtxtext,newtxmath}
\usepackage[T1]{fontenc}

\DeclareRobustCommand{\VAN}[3]{#2}
\let\VANthebibliography\thebibliography
\def\thebibliography{\DeclareRobustCommand{\VAN}[3]{##3}\VANthebibliography}

\usepackage{graphicx}	% Including figure files
\usepackage{amsmath}	% Advanced maths commands
\usepackage{comment}    % to comment whole paragraphs/sections
\usepackage{enumitem}   % to edit listings
\usepackage{pifont}     % for nice check marks
\usepackage[normalem]{ulem}  % to strike out text

\usepackage[flushleft]{threeparttable}  % annotation below table
\usepackage{caption}
\usepackage{booktabs}  % rules in tables
\usepackage{multirow}  % for multiple row cells in tables

\usepackage{xspace}     % for spacing at the end of defined variables (newcommand)

\newcommand*{\hcnone}{\ensuremath{\text{HCN(1--0)}}\xspace} % HCN(1-0)
\newcommand*{\hcopone}{\ensuremath{\text{HCO}^+\text{(1--0)}}\xspace} % HCO+(1-0)
\newcommand*{\cstwo}{\ensuremath{\text{CS(2--1)}}\xspace} % CS(2-1)
\newcommand*{\coone}{\ensuremath{\text{CO(1--0)}}\xspace} % CO(1-0)
\newcommand*{\cotwo}{\ensuremath{\text{CO(2--1)}}\xspace} % CO(2-1)
\newcommand*{\htwo}{\ensuremath{\text{H}_2}\xspace} % molecular hydrogen

\newcommand*{\intCO}{\ensuremath{W_{\text{CO(2--1)}}}\xspace}  % integrated CO(2-1) intensity
\newcommand*{\intHCN}{\ensuremath{W_{\text{HCN}}}\xspace}  % integrated HCN intensity
\newcommand*{\sigmol}{\ensuremath{\Sigma_{\text{mol}}}\xspace}  % molecular gas surface density
\newcommand*{\vdis}{\ensuremath{\sigma_{\text{mol}}}\xspace}  % velocity dispersion
\newcommand*{\avir}{\ensuremath{\alpha_{\text{vir}}}\xspace}  % virial parameter
\newcommand*{\Pturb}{\ensuremath{P_{\text{turb}}}\xspace}  % internal turbulent pressure
\newcommand*{\aCO}{\ensuremath{\alpha_{\text{CO}}}\xspace}  % CO-to-H2 conversion factor alpha_CO
\newcommand*{\aCOtwo}{\ensuremath{\alpha_{\text{CO(2--1)}}}\xspace}  % CO-to-H2 conversion factor alpha_CO
\newcommand*{\aHCN}{\ensuremath{\alpha_{\text{HCN}}}\xspace}  % HCN-to-dense gas conversion factor alpha_HCN
\newcommand*{\neff}{\ensuremath{n_{\text{eff}}}\xspace}  % effective critical density
\newcommand*{\seff}{\ensuremath{s_{\text{eff}}}\xspace}  % effective critical density in units of ln(density contrast)
\newcommand*{\nmean}{\ensuremath{n_0}\xspace}  % mean density
\newcommand*{\nSF}{\ensuremath{n_{\text{SF}}}\xspace}  % threshold density for grav. collapse and hence SF
\newcommand*{\sSF}{\ensuremath{s_{\text{SF}}}\xspace}  % same as above but in units of s
\newcommand*{\tffmean}{\ensuremath{t_{\text{ff},0}}\xspace}  % mean free-fall time
\newcommand*{\Mach}{\ensuremath{\mathcal{M}}\xspace}  % Mach number
\newcommand*{\sigsfr}{\ensuremath{\Sigma_{\text{SFR}}}\xspace}  % star formation rate surface density
\newcommand*{\mdense}{\ensuremath{M_{\text{dense}}}\xspace}  % dense gas mass
\newcommand*{\sigdense}{\ensuremath{\Sigma_{\text{dense}}}\xspace}  % dense gas surface density
\newcommand*{\fdense}{\ensuremath{f_{\text{dense}}}\xspace}  % dense gas fraction
\newcommand*{\sfedense}{\ensuremath{\text{SFE}_{\text{dense}}}\xspace}  % dense gas star formation efficiency
\newcommand*{\sfemol}{\ensuremath{\text{SFE}_{\text{mol}}}\xspace}  % molecular gas star formation efficiency
\newcommand*{\Xavg}{\ensuremath{\langle X\rangle}\xspace}  % weighted averages of some above quantities
\newcommand*{\intCOavg}{\ensuremath{\langle W_{\text{CO(2-1)}}\rangle}\xspace}
\newcommand*{\sigmolavg}{\ensuremath{\langle\Sigma_{\text{mol}}\rangle}\xspace}
\newcommand*{\vdisavg}{\ensuremath{\langle\sigma_{\text{mol}}\rangle}\xspace}
\newcommand*{\aviravg}{\ensuremath{\langle\alpha_{\text{vir}}\rangle}\xspace}
\newcommand*{\Pturbavg}{\ensuremath{\langle P_{\text{turb}}\rangle}\xspace}

\usepackage{siunitx}
\usepackage{xparse}  % for new commands
\DeclareSIUnit \parsec {pc}  % add parsec
\DeclareSIUnit \micron {\micro\metre}  % add short form of micrometer
\DeclareSIUnit \year {yr}  % add year
\DeclareSIUnit \jansky {Jy}  % add Jansky
\DeclareSIUnit \Msun {M_{\odot}}  % add solar mass
\DeclareSIUnit \Lsun {L_{\odot}}  % add solar luminosity
\DeclareSIUnit \Kkms {\kelvin\km\per\second}  % add Kelvin kilometers per second
\DeclareSIUnit \kB {\textit{k}_B}  % add Boltzmann constant
\DeclareSIUnit \dex {dex}  % add dex
\NewDocumentCommand\angRange{O{} m m}{\SIrange[parse-numbers=false, #1]{\ang[parse-numbers=true]{#2}}{\ang[parse-numbers=true]{#3}}{}}  % angular range (analog to SIrange)
\newcommand*{\ra}[2][]{{% extra pair of braces to keep the \definition local!
    \def\SIUnitSymbolDegree{\textsuperscript{h}}%
    \def\SIUnitSymbolArcminute{\textsuperscript{m}}%
    \def\SIUnitSymbolArcsecond{\textsuperscript{s}}%
    \ang[#1]{#2}}%
}

\newcommand{\cmark}{\textcolor{green}{\ding{51}}}
\newcommand{\xmark}{\textcolor{red}{\ding{55}}}

\newcommand{\ubonn}{Argelander-Institut f\"ur Astronomie, Universit\"at Bonn, Auf dem H\"ugel 71, 53121 Bonn, Germany}
\newcommand{\osu}{Department of Astronomy, The Ohio State University, 4055 McPherson Laboratory, 140 West 18th Ave, Columbus, OH 43210, USA}
\newcommand{\oan}{Observatorio Astronómico Nacional (IGN), C/ Alfonso XII, 3, E-28014 Madrid, Spain}
\newcommand{\inaf}{INAF — Osservatorio Astrofisico di Arcetri, Largo E. Fermi 5, I-50125, Florence, Italy}
\newcommand{\mpe}{Max-Planck-Institut f\"ur Extraterrestrische Physik (MPE), Giessenbachstr. 1, D-85748 Garching, Germany}
\newcommand{\zah}{Universit\"at Heidelberg, Zentrum f\"ur Astronomie, Institut f\"ur theoretische Astrophysik, Albert-Ueberle-Stra{\ss}e 2, 69120, Heidelberg, Germany}
\newcommand{\cool}{Cosmic Origins Of Life (COOL) Research DAO, coolresearch.io}
\newcommand{\anu}{Research School of Astronomy and Astrophysics, Australian National University, Canberra, ACT 2611, Australia}
\newcommand{\astrothreed}{ARC Centre of Excellence for All Sky Astrophysics in 3 Dimensions (ASTRO 3D), Australia}
\newcommand{\mpia}{Max Planck Institute for Astronomy, K\"onigstuhl 17, D-69117 Heidelberg, Germany}
\newcommand{\hsc}{Centro de Desarrollos Tecnol\'ogicos, Observatorio de Yebes (IGN), 19141 Yebes, Guadalajara, Spain}
\newcommand{\zw}{Universit\"{a}t Heidelberg, Interdisziplin\"{a}res Zentrum f\"{u}r Wissenschaftliches Rechnen, Im Neuenheimer Feld 205, 69120 Heidelberg, Germany}
\newcommand{\gent}{Sterrenkundig Observatorium, Universiteit Gent, Krijgslaan 281 S9, B-9000 Gent, Belgium}
\newcommand{\iram}{Institut de Radioastronomie Millim\'etrique (IRAM), 300 Rue de la Piscine, F-38406 Saint Martin d’Hères, France}
\newcommand{\lerma}{LERMA, Observatoire de Paris, PSL Research University, CNRS, Sorbonne Universit\'es, 75014 Paris, France}
\newcommand{\ucsd}{Center for Astrophysics and Space Sciences, Department of Physics, University of California San Diego, 9500 Gilman Drive, La Jolla, CA 92093, USA}
\newcommand{\mmu}{Department of Physics and Astronomy, McMaster University, 1280 Main Street West, Hamilton, ON L8S 4M1, Canada}
\newcommand{\cita}{Canadian Institute for Theoretical Astrophysics (CITA), University of Toronto, 60 St George Street, Toronto, ON M5S 3H8, Canada}
\newcommand{\wyo}{Department of Physics \& Astronomy, University of Wyoming, Laramie, WY, 82071, USA}
\newcommand{\alb}{Dept. of Physics, University of Alberta, Edmonton, Alberta, Canada T6G 2E1}
\newcommand{\ox}{Sub-department of Astrophysics, Department of Physics, University of Oxford, Keble Road, Oxford OX1 3RH, UK}
\newcommand{\ESO}{European Southern Observatory, Karl-Schwarzschild Stra{\ss}e 2, D-85748 Garching bei M\"{u}nchen, Germany}
\newcommand{\ljmu}{Astrophysics Research Institute, Liverpool John Moores University, 146 Brownlow Hill, Liverpool L3 5RF, UK}
\newcommand{\SAO}{Center for Astrophysics $\mid$ Harvard \& Smithsonian, 60 Garden St., 02138 Cambridge, MA, USA}
\title[HCN/CO and SFR/HCN vs. cloud-scale gas properties]{The ALMOND Survey: Molecular cloud properties and gas density tracers across 25 nearby spiral galaxies with ALMA}

\author[L.~Neumann et al.]{Lukas Neumann,$^{1}$\thanks{E-mail: lneumann@astro.uni-bonn.de}
Molly J.~Gallagher,$^{2}$
Frank Bigiel,$^{1}$
Adam K. Leroy,$^{2}$
Ashley T.~Barnes,$^{1,3}$
\newauthor
Antonio Usero,$^{4}$
Jakob S.~den Brok,$^{1,5}$
Francesco Belfiore,$^{6}$  % PHANGS team starts here
Ivana~Be\v{s}li\'c,$^{1}$
Yixian Cao,$^{7}$
\newauthor
M\'{e}lanie~Chevance,$^{8,9}$
Daniel~A.~Dale,$^{10}$
Cosima Eibensteiner,$^{1}$
Simon~C.~O.~Glover,$^{8}$
\newauthor
Kathryn Grasha,$^{11,12}$
Jonathan D.~Henshaw,$^{13, 14}$
Mar\'ia J. Jim\'enez-Donaire,$^{4,15}$
Ralf S.\ Klessen,$^{8,16}$
\newauthor
J.~M.~Diederik Kruijssen,$^{9}$
Daizhong Liu,$^{7}$
Sharon Meidt,$^{17}$
Jérôme~Pety,$^{18,19}$
Johannes Puschnig,$^{1}$
\newauthor
Miguel Querejeta,$^{4}$
Erik Rosolowsky,$^{20}$
Eva Schinnerer,$^{13}$
Andreas Schruba,$^{7}$
Mattia C.~Sormani,$^{8}$
\newauthor
Jiayi Sun,$^{2,21,22}$
Yu-Hsuan Teng,$^{23}$
and Thomas G.~Williams$^{11, 24}$
\\
$^{1}$\ubonn\\
$^{2}$\osu\\
$^{3}$\ESO\\
$^{4}$\oan\\
$^{5}$\SAO\\
$^{6}$\inaf\\
$^{7}$\mpe\\
$^{8}$\zah\\
$^{9}$\cool\\
$^{10}$\wyo\\
$^{11}$\anu\\
$^{12}$\astrothreed\\
$^{13}$\mpia\\
$^{14}$\ljmu\\
$^{15}$\hsc\\
$^{16}$\zw\\
$^{17}$\gent\\
$^{18}$\iram\\
$^{19}$\lerma\\
$^{20}$\alb\\\
$^{21}$\mmu\\
$^{22}$\cita\\
$^{23}$\ucsd\\
$^{24}$\ox\\
}

\date{Accepted XXX. Received YYY; in original form ZZZ}

\pubyear{2023}

\begin{document}
\label{firstpage}
\pagerange{\pageref{firstpage}--\pageref{lastpage}}
\maketitle

\begin{abstract}
We use new \hcnone data from the ALMOND (ACA Large-sample Mapping Of Nearby galaxies in Dense gas) survey to trace the kpc-scale molecular gas density structure and \cotwo data from PHANGS-ALMA to trace the bulk molecular gas across 25 nearby, star-forming galaxies. 
At \SI{2.1}{\kilo\parsec} scale, we measure the density-sensitive HCN/CO line ratio and the SFR/HCN ratio to trace the star formation efficiency in the denser molecular medium.
At \SI{150}{\parsec} scale, we measure structural and dynamical properties of the molecular gas via \cotwo line emission, which is linked to the lower resolution data using an intensity-weighted averaging method. 
We find positive correlations (negative) of HCN/CO (SFR/HCN) with the surface density, the velocity dispersion and the internal turbulent pressure of the molecular gas.
These observed correlations agree with expected trends from turbulent models of star formation, which consider a single free-fall time gravitational collapse.
Our results show that the kpc-scale HCN/CO line ratio is a powerful tool to trace the $\SI{150}{\parsec}$ scale average density distribution of the molecular clouds.
Lastly, we find systematic variations of the SFR/HCN ratio with cloud-scale molecular gas properties, which are incompatible with a universal star formation efficiency.  
Overall, these findings show that mean molecular gas density, molecular cloud properties and star formation are closely linked in a coherent way, and observations of density-sensitive molecular gas tracers are a useful tool to analyse these variations, linking molecular gas physics to stellar output across galaxy discs. 
\end{abstract}


%%%%%%%%%%%%%%%%%%%%%%%%%%%%%%%%%%%%%%%%%%%%%%%%%%

%%%%%%%%%%%%%%%%% BODY OF PAPER %%%%%%%%%%%%%%%%%%

%%%%%%%%%%%%%%%%%%%%%%%%%%%%%%%%%%%%%%
\section{Introduction}
%%%%%%%%%%%%%%%%%%%%%%%%%%%%%%%%%%%%%%

Star formation is at the heart of many astrophysical processes ranging from planet formation to the evolution of whole galaxies. Yet, the details of the star-forming process are far from being well understood. We know from observations inside the Milky Way (MW) and of other galaxies that the star formation rate (SFR) per unit area is tightly correlated to the gas surface density \citep[e.g.][]{Schmidt1959,Kennicutt1998,Bigiel2008,Schruba2011,Leroy2013b}. In more detail, observations of Milky Way star-forming regions show that stars form specifically within the densest parts of molecular clouds (MCs) and that the SFR of individual clouds correlates with the mass of dense gas\footnote{Here, the term "dense gas" refers to a density $n_{\text{H}_2}\gtrsim$ \SI{e4}{\per\cubic\centi\metre} and is primarily used to distinguish it from the lower-density molecular gas traced by low-J CO.} (\mdense) as traced by dust emission \citep[e.g.][]{Lada2003,Kainulainen2009,Andre2014} or emission of high excitation density lines \citep[e.g.][]{Wu2005,Wu2010,Stephens2016}. In a landmark paper, \citet{Gao2004} used HCN emission to trace \mdense from a large sample of external galaxies and found a linear relation between SFR and \mdense. Following up, \citet{Wu2005} studied HCN emission in local molecular clouds confirming the linear SFR-\mdense relation which, combining MC and integrated whole galaxy observations, spans \SI{10}{\dex}. These studies suggest that the star formation efficiency of dense gas ($\sfedense \equiv\text{SFR}/\mdense$) may be constant across this wide range of scales and environments.

However, the works by \citet{Usero2015}, \citet{Bigiel2016}, \citet{Gallagher2018a}, \citet{Jimenez-Donaire2019} and \citet{Bemis2019} on kpc-scale spectroscopic measurements find systematic variations of the HCN/CO line ratio and the SFR/HCN ratio with kpc-scale environmental properties, e.g. the molecular gas surface density or the stellar mass surface density. In addition, observations of the Milky Way's Central Molecular Zone (CMZ) show that the star formation efficiency of dense gas is much lower than is seen in the rest of the Galaxy \citep[see e.g.][]{Longmore2013,Barnes2017}. 
% In other words, the centre of the Milky Way is underproducing stars relative to that predicted by the relation found by \citet{Gao2004} by an order of magnitude or more \citep[see][]{Longmore2013,Barnes2017}. 
This apparent underproduction of stars follows naturally if the critical density of star formation is environmentally dependent, as predicted by turbulent star formation theories (e.g. \citealp{Kruijssen2014b}). One persistent question about these results is how HCN/CO or similar ratios (e.g. HCO$^+$/CO, CS/CO) trace density variations quantitatively in different environments when observed in other galaxies. In an attempt to address this, \citet{Gallagher2018b} took a novel step comparing the kpc-scale spectroscopic measurements with the $\sim$\SI{100}{\parsec}-scale molecular gas surface density in their five galaxies sample. They found systematic variations of the HCN/CO line ratio, a proxy for the fraction of dense molecular gas, as a function of the molecular gas surface density. This approach directly connects our two major methods of assessing density and gas properties in extragalactic systems: high resolution spectroscopic CO imaging and multi-species (HCN, HCO$^+$, CS) spectroscopy.

Combining multi-species spectroscopy with high resolution imaging has applications beyond only constraining density estimates. Turbulent theories of star formation predict that molecular cloud properties such as mean density, velocity dispersion or magnetic fields influence the density structure of the clouds, which regulates their ability to emit HCN \citep[e.g.][]{Krumholz2005,Padoan2011,Hennebelle2011,Federrath2012,Padoan2014}. Moreover, these same parameters also regulate the \sfedense of the clouds, thus providing a first order explanation of the observed correlations between the HCN/CO and SFR/HCN ratios and molecular cloud properties. 
%Following up on \citet{Gallagher2018b}, we study the connection between the coarse kpc-scale HCN spectroscopy and the molecular cloud properties as traced by $\sim$\SI{100}{\parsec}-scale CO observations. 

Until very recently, the exploration of such potential correlations was limited because high-resolution ($\sim$\SI{100}{\parsec}) CO imaging of the full molecular gas disc of galaxies has been almost as rare as kpc-scale and full-disc spectroscopy (see e.g. \citealp{Wong2002,Leroy2009} for kpc CO mapping, and e.g. \citealp{Usero2015,Jimenez-Donaire2019} for kpc HCN mapping). This situation was recently directly addressed in the Physics at High Angular resolution in Nearby GalaxieS project (PHANGS\footnote{http://phangs.org}), which uses the Atacama Large Millimeter/\linebreak[0]{}submillimeter Array (ALMA) to observe the molecular gas via the \cotwo line at $\sim$\angRange{;;1}{;;2} resolution in 90 nearby ($d<\SI{25}{\mega\parsec}$) galaxies \citep[PHANGS-ALMA;][]{Leroy2021b}. This survey allows access to the molecular gas distribution at $\sim$\SI{100}{\parsec} physical scales, which is close to the size of individual giant molecular clouds (GMCs). By combining PHANGS-ALMA with spectral mapping of dense gas tracers like \hcnone, we can explore the molecular cloud properties in the extragalactic regime and compare it to the kpc-scale dense gas spectroscopy. This technique bypasses the lack of extragalactic cloud-scale dense gas observations that are currently only available for a few galaxies (M51, \citealp{Querejeta2019} and NGC 3627, \citealp{Beslic2021}).

Tracing dense gas associated with star formation is challenging at extragalactic distances because tracers of dense gas that are currently popular in Galactic studies, e.g. $\text{N}_2\text{H}^+$ (see e.g. \citealp{Pety2017, Kauffmann2017, Barnes2020}), are too faint to be mapped at kpc scales across the discs of external galaxies with current instrumentation within reasonable time. Still, we can gain a lot of information about the dense gas by focusing on the brightest higher-critical density lines, i.e. \hcnone or \hcopone. The primary method to measure dense gas is based on the observation of various molecular emission lines with a range of effective critical densities (\neff; see e.g. \citealp{Leroy2017,Gallagher2018a}) To first order, the intensity of a line reflects the total gas mass above \neff, though see discussion in \citet{Shirley2015} and \citet{Mangum2015}. Therefore, the ratio of two lines with different critical densities reflects the ratio of gas masses above the two critical densities. For example, comparison between CO and HCN line emission yields an approximate gauge of the dense gas fraction (e.g. see \citealp{Usero2015, Bigiel2016} and reference therein), as the latter requires a significantly larger density for excitation.\footnote{$\neff(\hcnone)\approx$ \SIrange{2e4}{2e5}{\per\cubic\centi\metre}, $\neff(\cotwo)\approx\SI{1e3}{\per\cubic\centi\metre}$ \citep{Shirley2015,Mangum2015,Leroy2017,Onus2018}.}

Accordingly, in this paper we combine a large new HCN (along with HCO$^+$ and CS) data set with PHANGS-ALMA CO observations and use the \hcnone/\cotwo ratio to trace the fraction of dense gas. Because the targets were picked to overlap PHANGS-ALMA, we have cloud-scale gas properties, as well as IR- and UV-based SFR estimates across the whole sample. We explore the correlations of several cloud-scale structural and dynamical gas properties with both the HCN/CO ratio, a proxy for the dense gas fraction (\fdense), and the SFR/HCN ratio, a proxy for the dense gas star formation efficiency (\sfedense), across a sample of 25 galaxies. This builds on the study of \citet{Gallagher2018b}, who used a subset of these data (five galaxies) and considered only HCN/CO and cloud-scale molecular gas surface density (\sigmol), as well as on the works of \citet{Leroy2017b} and \citet{Utomo2018}, who compared CO-based cloud properties to the star formation efficiency in the bulk molecular medium traced by CO emission (\sfemol). We compare the kpc-scale HCN/CO and SFR/HCN to the cloud-scale molecular gas surface density (\sigmol), the velocity dispersion (\vdis), the virial parameter (\avir) and the internal turbulent pressure (\Pturb) as defined in Section~\ref{SEC:cloud_scale_props}. We measure \sigmol, \vdis, \avir and \Pturb using \cotwo data from the PHANGS-ALMA survey, and we measure HCN/CO and SFR/HCN using \hcnone data from new ALMA observations, called the ALMOND (ACA Large-sample Mapping Of Nearby galaxies in Dense Gas) survey. ALMOND uses the Morita Atacama Compact Array (ACA) to observe a sub-sample of 25 targets of the PHANGS-ALMA survey in dense molecular gas tracers like \hcnone, \hcopone or \cstwo. Our goal is to characterise the impact of these cloud-scale gas properties on the amount and star-forming ability of the dense gas and its connection with local environment.

This paper is organised as follows. First, we lay out the concept that motivates the studied correlations based on turbulent cloud models in Section~\ref{SEC:expectations}. Next, we describe our data products and methods in Section~\ref{SEC:data_and_methods}. In Section~\ref{SEC:results}, we present the main results where we compare the dense gas to cloud-scale molecular gas properties. We further analyse the findings in Section~\ref{SEC:analysis} where we separately look at the galaxies' centres. Finally, we summarise and discuss the results in Section~\ref{SEC:summary}. 
%In the online version, we provide supplementary figures.
%%%%%%%%%%%%%%%%%%%%%%%%%%%%%%%%%%%%%%
\section{Expectations}
\label{SEC:expectations}
%%%%%%%%%%%%%%%%%%%%%%%%%%%%%%%%%%%%%%

\subsection{Does HCN/CO trace dense gas fraction?}

The goal of this section is to set a qualitative, first order expectation of the relations between molecular cloud properties and the \intHCN/\intCO ratio (hereafter HCN/CO) as well as the \sigsfr/\intHCN integrated intensity ratio (hereafter SFR/HCN).
Using established models of star formation (e.g. \citealp{Krumholz2005}, see Section~\ref{SEC:models}), we model the probability distribution function (PDF) of the gas density of molecular clouds as a function of several cloud properties, i.e. the mean surface density \nmean, the Mach number \Mach and the virial parameter \avir (in Section~\ref{SEC:cloud_scale_props}, we explain our best empirical estimates of these molecular cloud properties). 
Then, based on the density PDF, we infer qualitative changes of HCN/CO and SFR/HCN as a function of the molecular cloud properties.
At the model level, we can infer the gas masses traced above certain density thresholds and thus estimate the dense gas fraction (\fdense) and formation efficiency (\sfedense). 
Therefore, to infer HCN/CO and SFR/HCN from the models we assume that \hcnone and \cotwo emission trace the gas mass above a certain effective critical density using a constant mass-to-light conversion factor. 
However, Galactic observations, albeit largely limited to selected local clouds or even sub-regions of these, \citep[e.g.][]{Pety2017,Kauffmann2017,Barnes2020,Evans2020} and simulations \citep[e.g.][]{Shirley2015,Mangum2015,Leroy2017,Onus2018,Jones2021} have clearly shown that reality is more complex. Rather than simply tracing gas above some fixed density threshold, HCN always traces a convolution of the density distribution and density-dependent emissivity, with additional complications offered by chemical abundance variations, variations in temperature, and possible excitation by collisions with electrons. Despite these concerns, the preponderance of evidence even in the studies above supports the use of the HCN/CO ratio as a tracer of the density distribution in a cloud, with higher HCN/CO reflecting denser gas.
%highlight that there is still a significant uncertainty in these density thresholds and conversion factors.

Given these uncertainties, in our analysis, we focus on the observational quantities, i.e. HCN/CO and SFR/HCN, rather than the less certain physical quantities, i.e. \fdense and \sfedense.
%Though, if the model expectations agree with the observational relations, we demonstrate indication for HCN/CO tracing \fdense and SFR/HCN tracing \sfedense.
In this section laying out basic theoretical expectations, we adopt the simpler picture that HCN emission has a step-function dependence on density and emits with a fixed mass-to-light ratio, or conversion factor, above that density threshold. 
The purpose is not to derive quantitative predictions about line emissivities but instead to discuss how currently popular models predict the directions of observed correlations between cloud-scale molecular gas properties and dense gas spectroscopy. 

We also note further alternative descriptions of the basic theoretical framework we adopt \citep[e.g.][]{Hennebelle2011, Federrath2012} and refer the reader to those works for more quantitative discussion of turbulent cloud models.
%As an additional caveat, we also note that several studies \citep[e.g.][]{Hennebelle2011, Federrath2012} have published refinements or alternatives to the basic theoretical framework that we adopt. Given that the purpose of this discussion is to develop intuition rather than offer a quantitative prediction, we consider it beyond the scope of this work to incorporate these more complex subsequent models into the discussion. We do refer the reader to those works for more quantitative discussion of turbulent cloud models.

\begin{figure*}
    \includegraphics{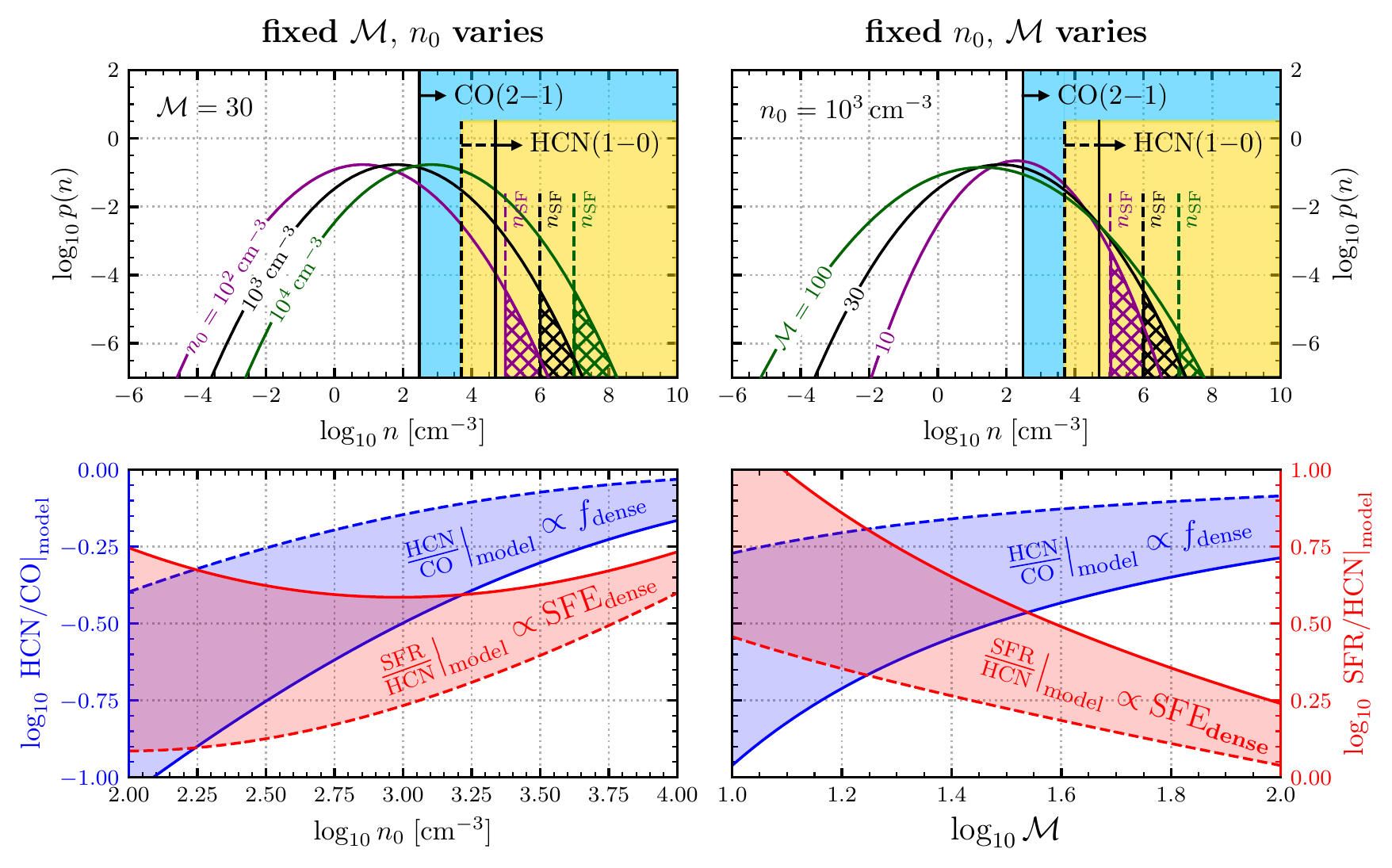}
    \caption{\textit{Top:} Volume-weighted probability distribution functions (PDFs) of the molecular cloud gas density, $n$, for varying mean density (\nmean, left panel) and varying Mach number (\Mach, right panel). The light blue shaded area indicates the density regime traced by \cotwo, i.e. all gas above $\neff(\text{CO})=\SI{3e2}{\per\cubic\centi\metre}$ \citep{Leroy2017}. Analogously, the yellow shaded area is the density regime traced by \hcnone, where we adopted two effective critical densities such that in one case (solid line) HCN traces all gas above $\neff(\text{HCN})=\SI{5e4}{\per\cubic\centi\metre}$ \citep{Leroy2017} and in the other case (dashed line) HCN traces gas above $\neff(\text{HCN})=\SI{5e3}{\per\cubic\centi\metre}$ \citep{Onus2018}. The dashed lines labelled with \nSF show the threshold density above which gas in clouds collapses to form stars. Thus, the hatched areas are a measure of the SFR per free-fall time. \textit{Bottom:} HCN/CO as a proxy for \fdense and SFR/HCN as a proxy for \sfedense estimated from the PDFs as a function of the mean density (left panel) and the Mach number (right panel) in accordance with the top panel plots. We compute HCN/CO as the ratio of the integrated mass-weighted PDFs within the assumed density regimes (Equation \ref{EQU:model_HCN_CO}). Similarly, SFR/HCN is obtained by integrating the mass-weighted PDF above \nSF accounting for the free-fall time at mean density and dividing with the area of the PDF traced by HCN (Equation \ref{EQU:model_SFR_HCN}). The solid line and dashed lines are in accordance with the density thresholds in the top panels.}
    \label{FIG:PDF_expectations}
\end{figure*}

\begin{figure*}
    \includegraphics{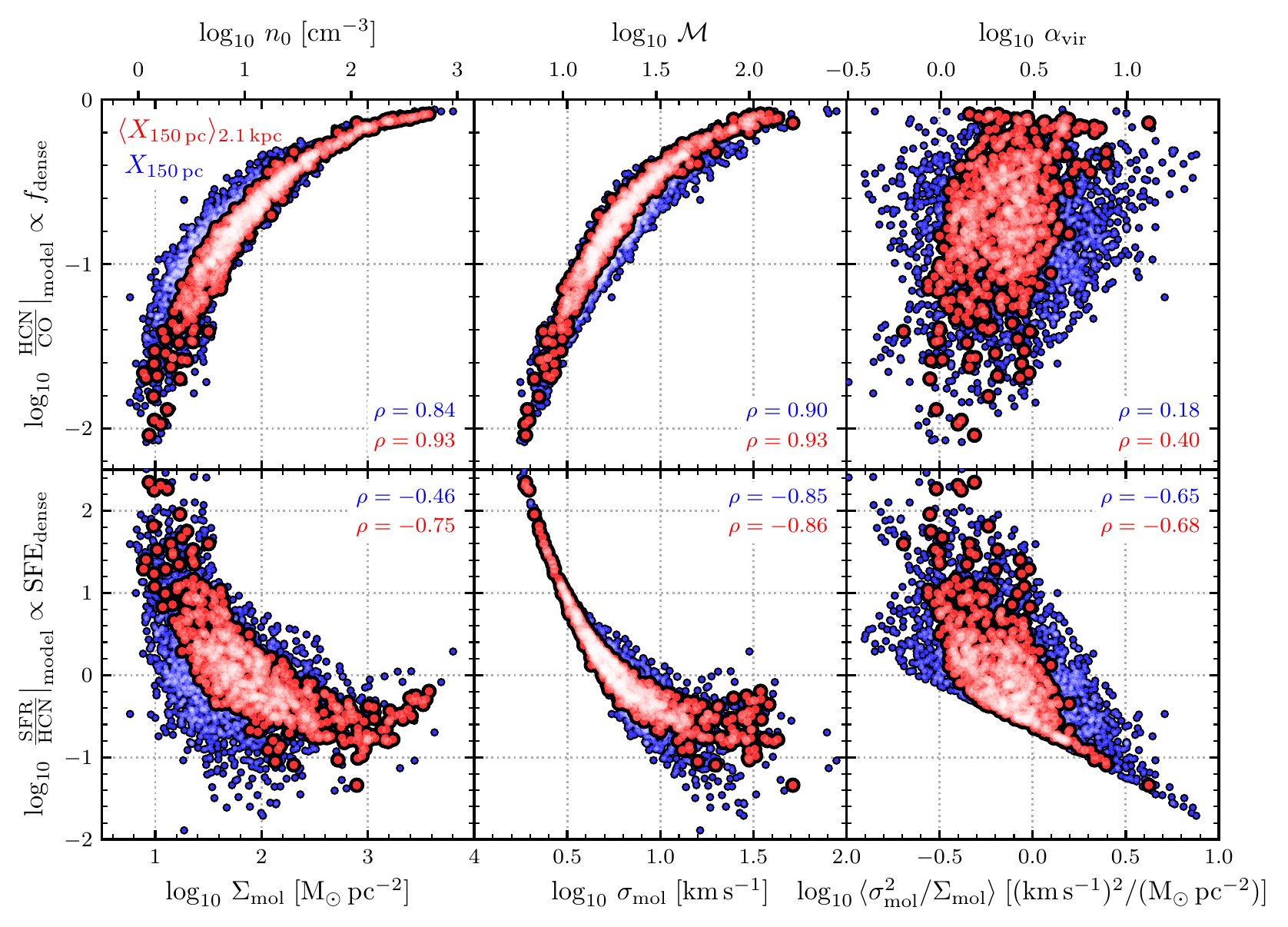}
    \caption{Model predictions of HCN/CO and SFR/HCN against the molecular cloud properties \sigmol, \vdis, \avir (similar to \citealp{Sun2018}). HCN/CO and SFR/HCN are computed as in Equations \ref{EQU:model_HCN_CO} and \ref{EQU:model_SFR_HCN} based on a log-normal PDF and assuming $\neff(\text{HCN})=\SI{5e3}{\per\cubic\centi\metre}$. The PDF parameters (\nmean, \Mach, \avir) are inferred from the observed \SI{150}{\parsec} molecular gas measurements (\sigmol, \vdis, $\avir\propto\vdis^2/\sigmol$). The red data points depict the intensity-weighted averages of the \SI{150}{\parsec} measurements (blue) at an averaging scale of \SI{2.1}{\kilo\parsec} (see Section \ref{SEC:weighted_avg} for more details). We show the Pearson correlation coefficient $\rho$ for both the original data and the weighted averages.}
    \label{FIG:PDF_expectations_data}
\end{figure*}

%%%%%%%%%%%%%%%%%%%%%%%%%%%%%%%%%%%%%%
\subsection{Turbulent Cloud Models}
\label{SEC:models}
In turbulent models of star formation \citep[e.g.][]{Padoan2002, Krumholz2005, Hennebelle2011, Padoan2011, Federrath2012, Federrath2013, Padoan2014} the probability distribution function (PDF; $p(n)$) of the molecular gas number density, $n$, is to first order described by a log-normal function, which can be written as
\begin{align}
    p(s) \text{d}s = \dfrac{1}{\sqrt{2\pi\sigma_s^2}} \exp\left[-\dfrac{(s-s_0)^2}{2\sigma_s^2}\right] \text{d}s \;,
\end{align}
where $s=\ln(n/\nmean)$ is the logarithmic number density in units of the mean number density, \nmean, and $s_0=-\sigma_s^2/2$ is the centre of the PDF. Note, that gravitational collapse and star formation will introduce a power-law tail at high densities \citep[see, e.g.,][]{girichidis2014,Burkhart2018}. This is particularly noticeable in the high-density gas of individual molecular clouds \citep[e.g.,][]{Kainulainen2009, schneider2015}. However, we expect the contribution of the power-law tail to the overall mass budget of the multi-phase ISM to be negligible at the larger scales of $\sim$\SI{150}{\parsec} and above (e.g. in entire gravitationally bound GMCs; e.g. \citep{klessen2016}).

For isothermal turbulent flows, the width of the log-normal PDF is quantified by the rms Mach number $\Mach\equiv\sigma_\text{3D}/c_s$ ($\sigma_\text{3D}$ is the three-dimensional velocity dispersion and $c_s$ is the sound speed of the molecular gas), the turbulence driving parameter, $b$, and the gas to magnetic pressure ratio, $\beta$ \citep[see e.g.][]{Padoan2011, Molina_2012}:
\begin{align}
    \sigma_s^2 = \ln\left(1+b^2\Mach^2\dfrac{\beta}{\beta +1}\right) \;.
\end{align}
The parameter $b$ depends on the ratio of compressive vs.\ solenoidal modes and on the dimensionality of the flow. For isotropic turbulence in isothermal gas with a natural mix of both modes contributing equally, simple theoretical considerations lead to $b = 3/4$ in two and $b = 2/3$ in three dimensions \citep{federrath2008}. Numerical simulations indicate somewhat smaller values \citep[][]{federrath2010}, however, with considerable scatter. We follow \citet{Padoan2002}, neglect magnetic fields ($\beta\rightarrow\infty$) and adopt $b \sim 0.5$ such that the width of the PDF becomes 
\begin{align}
    \sigma_s^2 = \ln\left(1+0.25\,\Mach^2\right)\,.
\end{align}
The above formalism implies a link between the distribution of mass above any given density and the mean properties of molecular clouds, i.e. for varying mean density (\nmean) or velocity dispersion (\vdis) as is illustrated in Figure \ref{FIG:PDF_expectations}. Here, we adopt the prescription from \citet{Krumholz2005} (hereafter KM theory) to compute the density threshold \nSF above which gas is considered to collapse and form stars within a free-fall time:
\begin{align}
    \dfrac{\nSF}{\nmean} = 0.82\,\avir\,\Mach^2 \;.
    \label{EQU:PDF_nSF}
\end{align}
Assuming a fixed virial parameter $\avir\approx1.3$ \citep{Krumholz2005}, the above equation reads: $\nSF/\nmean\approx 1.07 \Mach^2$. Thus, for fixed virial parameter, the physical interpretation drawn from Equation \eqref{EQU:PDF_nSF} is that stars form in local overdensities of the molecular clouds determined by the density contrast $\nSF/\nmean$ which shifts to higher overdensities if the turbulence (\Mach) of the molecular gas increases. Variations of the virial parameter are small ($\sim\SI{0.7}{\dex}$; \citealp{Sun2020b}) compared to variations of the mean density ($\sim\SI{3.4}{\dex}$) or the Mach number ($\sim\SI{1.7}{\dex}$) of molecular clouds which justifies assuming a fixed \avir to first order. However, variations of \avir are still evident and might also manifest in the spectroscopic observations, e.g. by affecting \nSF. In this simplified model, \avir does not affect the PDF and thus HCN/CO is unaffected by changes in \avir. On the contrary, based on Equation \ref{EQU:PDF_nSF}, \nSF increases for increasing \avir which would result in a negative correlation between SFR/HCN and \avir. In practice, in this study, we infer the virial parameter from observations by assuming a fixed cloud scale, such that $\avir\propto\vdis^2/\sigmol$ (see Section~\ref{SEC:virial_parameter}). In this case, \avir is correlated with \vdis (tracing \Mach) and \sigmol (tracing \nmean) making the effect of \avir on HCN/CO and SFR/HCN more complex. Still, we can estimate how $\vdis^2/\sigmol$ tracing \avir affects HCN/CO and SFR/HCN taking into account the distribution and thus the correlation of molecular cloud properties based on observations (see Sections \ref{SEC:models_HCN_CO_corr} and \ref{SEC:models_SFR_HCN_corr} and Figure \ref{FIG:PDF_expectations_data}).

%%%%%%%%%%%%%%%%%%%%%%%%%%%%%%%%%%%%%%
\subsection{Line Emissivity}
\label{SEC:line_emissivity}
In an ideal case, we can detect molecular lines, such as \hcnone or \cotwo, if a substantial fraction of the gas is at densities close to or above the so-called "critical density" for emission. 
Considering the simplest case of only collisional (de)excitation (e.g. within dense molecular clouds), this critical density can be defined as the density at which the collisional de-excitation rate and spontaneous de-excitation are equal, and hence above this density line emission is enhanced.
In general, the critical density of a certain line depends on the optical depth ($\tau$) of the line and the kinetic temperature ($T$) of the gas \citep[e.g.][]{Tielens2010,Draine2011,Mangum2015,Shirley2015,klessen2016}. 
% At typical cold, dense ISM conditions, i.e. $T\approx\SI{25}{\kelvin}$ and $\tau=1$ for \hcnone, $\tau=10$ for \cotwo, the effective critical densities are $\neff\approx$ \SIrange{2e4}{2e5}{\per\cubic\centi\metre} for \hcnone and $\neff=\SI{1e3}{\per\cubic\centi\metre}$ for \cotwo \citep{Shirley2015,Leroy2017,Onus2018}. 
% This means, \hcnone traces gas roughly one or two orders of magnitude denser than is traced by \cotwo. 
The concept of a critical density, above which all the line emission is associated with the gas mass above that critical density, is, however, somewhat limited in lower density gas, as sub-thermal excitation effects (e.g. \citealp{Pety2017}) and additional excitation mechanisms can be significant (e.g. see \citealp{Goldsmith2018}).  
Nonetheless, to first order, we consider all gas above a rescpective critical density to be traced by the respective molecular line emission. 
We select the density threshold based on the emissivity-density curves derived by \citet{Leroy2017} (their Figure 2). 
We define the threshold where their normalised emissivity ($\epsilon$) exceeds \SI{50}{\percent}, i.e. at $\neff(\text{HCN})=\SI{5e4}{\per\cubic\centi\metre}$ for \hcnone and $\neff(\text{CO})=\SI{3e2}{\per\cubic\centi\metre}$ for \cotwo as illustrated in Figure \ref{FIG:PDF_expectations} (left panels). 
The value of $\neff$ for HCN has, however, been the subject of some debate in the recent literature (e.g. \citealp{Kauffmann2017, Barnes2020}).
For example, numerical simulations from \citet{Onus2018} and \citet{Jones2021} find that \hcnone emission traces gas at densities of $\neff(\text{HCN})=$ \SIrange{e3}{e4}{\per\cubic\centi\metre}, which is around an order of magnitude lower than reported by \citet{Leroy2017}. Note however, that \citet{Leroy2017} uses a different definition of the effective critical density and that both results may be consistent which each other.
Nevertheless, to account for some variation of \neff, we adopt a second, lower critical density of $\neff(\text{HCN})=\SI{5e3}{\per\cubic\centi\metre}$ (dashed line in Figure \ref{FIG:PDF_expectations}).
\footnote{Note that we adopt a single critical density for CO emission, which could suffer from similar effects. Albeit given its already low critical density, which sits close to the density where molecular gas forms ($\sim\SI{e2}{\per\cubic\centi\metre}$), this effect should be less pronounced than with HCN.}
We then use these density regimes to infer the gas mass traced by \hcnone or \cotwo emission via integration of the mass-weighted PDF:
\begin{align}
    I_\text{line} \propto \int_{\seff(\text{line})}^{\infty} \dfrac{n}{\nmean}\, p(s)\, \text{d}s \,,
\end{align}
where $\seff(\text{line})$ is the effective critical line density in units of $s=\ln(n/\nmean)$ corresponding to $\neff(\text{line})$. Note that this formalism does not consider radiative transfer modelling and therefore only gives reasonable HCN/CO estimates in terms of comparative analysis.

%%%%%%%%%%%%%%%%%%%%%%%%%%%%%%%%%%%%%%
\subsection{HCN/CO correlations}
\label{SEC:models_HCN_CO_corr}

Turbulent cloud models predict the density distribution and star formation rate as a function of the molecular cloud properties. 
In the following, we adopt the description introduced in Section~\ref{SEC:models} and infer simplified line emissivities (Section~\ref{SEC:line_emissivity}).
In Figure~\ref{FIG:PDF_expectations}, we show how variations of molecular cloud properties affect the molecular gas density distribution, i.e. the PDF, and consequently the HCN/CO ratio.

At first, we keep the virial parameter fixed at $\avir=1.3$ and vary the mean density for fixed Mach number and vice-versa. In Figure \ref{FIG:PDF_expectations} (top left panel) we show how the cloud's PDF changes as a function of the mean density (\nmean), keeping the Mach number fixed at $\Mach=30$ which corresponds to $\vdis\approx\SI{5}{\kilo\metre\per\second}$ assuming a sound speed of $c_s=\SI{0.3}{\kilo\metre\per\second}$ (at $T\sim\SI{20}{\kelvin}$; \citealp{Krumholz2005}). We adopt typical molecular cloud densities, varying \nmean from \SIrange{e2}{e4}{\per\cubic\centi\metre} which results in a shift of the PDF to higher densities without changing the width of the PDF. We estimate the expected HCN/CO line ratio based on a simplified emissivity model and critical densities of \hcnone and \cotwo discussed above (Section~\ref{SEC:line_emissivity}) by integrating the PDF over the density ranges of the respective lines:
\begin{align}
    \left.\dfrac{\text{HCN}}{\text{CO}}\right\vert_\text{model} = \frac{\int_{\seff(\text{HCN})}^{\infty} \frac{n}{\nmean}\, p(s)\,\text{d}s}{\int_{\seff(\text{CO})}^{\infty} \frac{n}{\nmean}\, p(s)\,\text{d}s}\,,
    \label{EQU:model_HCN_CO}
\end{align}
where the respective HCN and CO effective critical densities are $\neff(\text{HCN})=\SI{5e4}{\per\cubic\centi\metre}$ \citep{Leroy2017} or $\neff(\text{HCN})=\SI{5e3}{\per\cubic\centi\metre}$ \citep{Onus2018} and $\neff(\text{CO})=\SI{2e3}{\per\cubic\centi\metre}$. This procedure computes the mass of gas which is traced by the different molecular lines, which serves as a first order estimate of the expected line intensities assuming a constant mass-to-light ratio. Note that equation \eqref{EQU:model_HCN_CO} does not account for the different \hcnone and \cotwo mass-to-light conversion factors. Thus we only claim to predict changes in HCN/CO. Moreover, we assume a fixed effective critical density of \hcnone and that the emissivity of the lines below \neff is zero. However, in reality, \neff can vary and the emissivity below \neff is not zero. Therefore, if the dense gas fraction is low, a significant fraction of the HCN emission could come from lower density gas. Thus, our toy model will predict a steeper correlation at low \nmean and low \Mach.

We find that the HCN/CO line ratio positively correlates with the mean density of the molecular cloud (see top right panel of Figure \ref{FIG:PDF_expectations}). The physical explanation is that at low mean densities $\nmean\sim\SI{e2}{\per\cubic\centi\metre}$ the \cotwo line is easily excited while only a small fraction of the cloud's gas is at densities high enough to produce \hcnone emission producing a low HCN/CO line ratio. Increasing \nmean leads to an increasing fraction of gas at the (effective) critical density of \hcnone thus increasing the \hcnone luminosity while the \cotwo luminosity is only marginally affected by increasing \nmean due to its low critical density. Thus, increasing the mean density of the cloud results in a higher HCN/CO line ratio. If we assume the \cotwo intensity to be a robust tracer of the surface density of the molecular gas at cloud-scales and further assume that the geometry of the clouds is similar such that surface density traces mean density, we expect a positive correlation between the surface density of molecular clouds and the HCN/CO line ratio as a proxy of the dense gas fraction. The connection between cloud-scale \sigmol and HCN/CO has already been tested by \citet{Gallagher2018b}, who found a positive correlation, thus supporting the model expectation.

Similarly, we vary the Mach number (and consequently the velocity dispersion) of the molecular cloud adopting typical values of $\Mach=$ \numrange{10}{100} while keeping the mean density fixed at $\nmean=$ \SI{e3}{\per\cubic\centi\metre}. Comparing with \citet{Krumholz2007}, the range of Mach numbers describes normal ($\Mach\sim30$) over intermediate ($\Mach\sim50$) to starburst galaxies ($\Mach\sim80$). We find that increasing the turbulence of the molecular cloud widens the PDF without significantly shifting its peak ($s_0=-\sigma_s^2/2$; see bottom left panel of Figure \ref{FIG:PDF_expectations}). As a result, at low velocity dispersion the PDF is narrow and centred around a density of order \SI{e2}{\per\cubic\centi\metre} such that only a small fraction of the gas is at high densities. Therefore, the \hcnone intensity is low while the \cotwo intensity is high, hence we expect a small HCN/CO line ratio. Increasing the velocity dispersion leads to a widening of the PDF such that a larger fraction of the gas is at higher densities thus increasing the \hcnone luminosity much more than the \cotwo which is less affected by the width of the PDF. Thus, assuming that the velocity dispersion is traced by the \cotwo line width, we expect a positive correlation between the line width and the HCN/CO line ratio as shown in the bottom right panel of Figure~\ref{FIG:PDF_expectations}.

As mentioned above, in this simplified model prescription, the actual (theoretical) virial parameter does not affect the PDF thus leaving HCN/CO unchanged. However, the empirical virial parameter, if measured as $\avir\propto\vdis^2/\sigmol$, might be connected to changes in HCN/CO. Therefore, under the assumption that the virial parameter is proportional to $\vdis^2/\sigmol$ (see Section \ref{SEC:virial_parameter}), we can study changes of HCN/CO with the empirically inferred virial parameter. In Figure~\ref{FIG:PDF_expectations_data} (upper row), we show how the model HCN/CO varies with the empirically based molecular cloud properties (see Appendix~\ref{SEC:appendix:cloud_props} for the distribution of the measured cloud-scale gas properties). Each data point corresponds to an aperture in one of our target galaxies. Blue dots indicate measurements at \SI{150}{\parsec}, while red dots indicate averages over \SI{2.1}{\kilo\parsec} apertures using a mass-weighting scheme (see Section~\ref{SEC:weighted_avg}). We predict HCN/CO as described above adopting the following data-to-model parameter conversions. We convert the observationally inferred \sigmol into \nmean assuming spherical clouds with radius $R$, such that the depth of the cloud is given by the beam size, e.g. $2R=\SI{150}{\parsec}$, leading to $\nmean/[\SI{}{\per\cubic\centi\metre}]=3/(4R\mu m_\text{H})\,\sigmol=0.144\times\sigmol/[\SI{}{\Msun\per\square\parsec}]$, where $\mu=2.8$ is the mean particle weight per hydrogen molecule assuming all hydrogen is $\text{H}_2$ \citep{Kauffmann2008} and $m_\text{H}$ is the mass of the hydrogen atom. Assuming a sound speed of $c_s=\SI{0.3}{\kilo\metre\per\second}$ we obtain $\Mach=\sqrt{3}\,\vdis/c_s = 5.8\times\vdis/[\SI{}{\kilo\metre\per\second}]$, where we assume an isotropic velocity dispersion, hence the factor $\sqrt{3}$. In accordance with the model predictions above, we find HCN/CO to positively correlate with \nmean and \Mach. In addition we observe a weak positive correlation of HCN/CO with the virial parameter (Pearson correlation $\rho=0.14$ for the \SI{150}{\parsec} measurements and $\rho=0.40$ for the \SI{2.1}{\kilo\parsec} scale weighted averages). Physically, the virial parameter is a measure of the gravitational boundedness, where higher \avir means less bound. The derived (weak) positive correlation between HCN/CO and \avir implies that less bound clouds tend to have more dense gas per molecular gas which seems counterintuitive given that one might expect a higher dense gas fraction for more bound clouds. However, high HCN/CO is also connected to highly turbulent clouds as is shown above. Indeed, we observe a steeper correlation of HCN/CO with Mach number than with \nmean, therefore a positive correlation between HCN/CO and \avir is indeed not surprising.

%%%%%%%%%%%%%%%%%%%%%%%%%%%%%%%%%%%%%%
\subsection{SFR/HCN correlations}
\label{SEC:models_SFR_HCN_corr}

Similar to the HCN/CO correlations above, we can make predictions about the SFR-to-HCN ratio as a function of molecular cloud properties. We model the SFR using Equation \eqref{EQU:PDF_nSF} where all gas above the threshold density \nSF is considered to form stars and \nSF is completely determined by the mean density (\nmean) and the Mach number (\Mach), $\nSF\propto\nmean\Mach^2$ at fixed $\avir=1.3$. This allows us to compute \nSF for any given tuple (\nmean, \Mach) or equivalently (\sigmol, \vdis). We add \nSF as vertical dashed lines in Figure \ref{FIG:PDF_expectations} and consider the cloud's gas above this threshold (hatched area) as the star forming gas. Similar to HCN/CO and following \citet{Krumholz2005}, we estimate SFR/HCN by integrating the PDF over the relevant density ranges:
\begin{align}
    \left.\dfrac{\text{SFR}}{\text{HCN}}\right\vert_\text{model} = \frac{\int_{\sSF}^{\infty} \frac{n}{\sqrt{\nmean}}\, p(s)\,\text{d}s}{\int_{\seff(\text{HCN})}^{\infty} \frac{n}{\nmean}\, p(s)\,\text{d}s}\,,
    \label{EQU:model_SFR_HCN}
\end{align}
where $\neff(\text{HCN})$ is defined as in Section \ref{SEC:models_HCN_CO_corr}. Equation \eqref{EQU:model_SFR_HCN} accounts for the (inverse) dependence of the SFR on the mean free fall time $\tffmean=\sqrt{3\pi/(32G\rho_0)}\propto\rho_0^{-1/2}\propto\nmean^{-1/2}$ \citep[e.g.][]{Padoan2014}. Again, we are only interested in relative changes of SFR/HCN so that the units have no physical meaning. We note that the prescription adopted here assumes a single free-fall time, while other pictures \citep[e.g.][]{Federrath2012} adopt a multi-free-fall approach that include an additional density dependent factor (a ratio of free-fall times) inside the integral in the numerator of Equation \eqref{EQU:model_SFR_HCN}. Multi-free-fall models can predict that \sfedense increases with Mach number, i.e. the reverse of single free-fall models predictions and the reverse of the trends found here at low \nmean, \Mach (Figure~\ref{FIG:PDF_expectations_data}). Given the sense of observed \sfedense trends examined in this work and by others (\citealp{Querejeta2015,Leroy2017}, Utomo et al. in prep.), we proceed with the single free-fall class of models in the following.

We explore the effect of the molecular cloud properties on SFR/HCN within the same parameter space as of HCN/CO. We find that \Mach negatively correlates with SFR/HCN, as is shown in the bottom right panel of Figure \ref{FIG:PDF_expectations}. This can be understood in the following way. At low velocity dispersion, HCN is a good tracer of the density regime where the stars are expected to form and thus the SFR/HCN ratio is high. For increasing turbulence the HCN luminosity becomes a less ideal tracer of the local overdensities and traces more of the bulk molecular gas leading to a decreasing SFR/HCN. For changes of SFR/HCN with the mean density the model predicts a decreasing trend at low \nmean and an increasing trend at high \nmean and hence no clear correlation between SFR/HCN and \nmean. We can understand the different dependencies in the following way. At low $\nmean\ll\neff$ an increase in \nmean leads to HCN tracing more of the bulk molecular gas such that SFR/HCN decreases leading to a negative correlation between SFR/HCN and \nmean similar to \Mach. Though, if \nmean reaches densities comparable to the critical density of \hcnone the ratio between the gas masses above $\neff$(HCN) and \nSF is barely affected by changes in \nmean. However, the SFR depends on the mean free-fall time such that a higher gas mass is converted into stars within a shorter time ($\tffmean\propto\nmean^{-1/2}$) leading to an increase of SFR/HCN with increasing \nmean. As a result, we expect a negative correlation between SFR/HCN at $\nmean\ll\neff(\text{HCN}),\nSF$ and a positive correlation at $\nmean\sim\neff(\text{HCN}),\nSF$.

Analogously to Section~\ref{SEC:models_HCN_CO_corr}, we additionally infer SFR/HCN for every data based triplet (\nmean, \Mach, \avir) meaning for each aperture, \nmean and \Mach are traced via \sigmol and \vdis, respectively, and \avir is proportional to $\vdis^2/\sigmol$. The resulting relations (SFR/HCN against cloud properties) are shown in Figure~\ref{FIG:PDF_expectations_data} (lower panels). Remarkably, we find a clear negative correlation between SFR/HCN and \nmean in contrast to the less clear relation shown in Figure~\ref{FIG:PDF_expectations}, where \nmean is varied at fixed \Mach. There are two reasons that we do not observe the upturn of SFR/HCN at higher \nmean. First, the \nmean values inferred from \sigmol are $\sim$\SIrange{1}{2}{\dex} lower than the adopted values in Figure \ref{FIG:PDF_expectations} so that \nmean is mostly lower than \neff(HCN) or \nSF and the dependence on the free fall time is less important. Second, the strong negative correlation between SFR/HCN and \Mach in combination with the positive correlation of \nmean and \Mach can overcompensate the SFR/HCN upturn at higher \nmean thus leading to a negative correlation between SFR/HCN and \nmean.  

In the KM model description, \avir affects \nSF without affecting the PDF and thus the line emissivity. This would result in a negative correlation between SFR/HCN and \avir. However, we measure \avir via $\vdis^2/\sigmol$ assuming a fixed cloud size (see Section \ref{SEC:virial_parameter}). Thus, \avir is constrained by the observational \sigmol and \vdis values and we want to explore variation of the model's SFR/HCN with $\vdis^2/\sigmol$. Analogously to Section \ref{SEC:models_HCN_CO_corr}, we infer SFR/HCN for every observationally based triplet (\nmean, \Mach, \avir) based on the same model description as above but also accounting for variations in \avir. The resulting relations (SFR/HCN against cloud properties) are shown in Figure \ref{FIG:PDF_expectations_data} (lower panels). Consistent with the results above we find very strong negative correlations of SFR/HCN with \nmean and \Mach. Moreover, we observe a moderate negative correlation of SFR/HCN with the virial parameter (Pearson correlation $\rho=-0.43$ for the \SI{150}{\parsec} scale measurement and $\rho=-0.57$ for the weighted averages). The virial parameter quantifies the gravitational boundedness of the cloud. The derived anti-correlation between \avir and SFR/HCN supports the concept that less bound clouds tend to be less efficient in producing stars from the dense gas (lower SFR/HCN).

%%%%%%%%%%%%%%%%%%%%%%%%%%
\section{Observations}
\label{SEC:data_and_methods}
%%%%%%%%%%%%%%%%%%%%%%%%%%

\begin{table*}
\begin{threeparttable}[t]
\centering
\caption{Galaxy Sample}
\label{TAB:galaxy_sample}
\begin{tabular}{ccccccccccc}
\hline\hline
\multirow{2}{*}{Galaxy} & R.A. & Dec. & $d$ & $i$ & $M_\star$ & $M_{\mathrm{H}_2}$ & SFR & SFR/$M_\star$ & \multirow{2}{*}{Bar} & \multirow{2}{*}{AGN} \\ 
& (J2000) & (J2000) & (\SI{}{\mega\parsec}) & (\SI{}{\degree}) & (\SI{e9}{\Msun}) & (\SI{e9}{\Msun}) & (\SI{}{\Msun\per\year}) & (\SI{e-10}{\per\year}) & & \\
(1) & (2) & (3) & (4) & (5) & (6) & (7) & (8) & (9) & (10) & (11) \\
\hline
 NGC 0628 &   \ra{1;36;41.7} &   \ang{+15;47;1.1} &   9.8 &   8.9 &   21.94 &   2.70 &   1.75 &       0.80 &   N &   N \\
 NGC 1097 &   \ra{2;46;18.9} &  \ang{-30;16;28.8} &  13.6 &  48.6 &   57.48 &   5.52 &   4.74 &       0.83 &   Y &   Y \\
 NGC 1365 &   \ra{3;33;36.4} &  \ang{-36;08;25.5} &  19.6 &  55.4 &   97.77 &  18.07 &  16.90 &       1.73 &   Y &   Y \\
 NGC 1385 &   \ra{3;37;28.6} &   \ang{-24;30;4.2} &  17.2 &  44.0 &    9.53 &   1.68 &   2.09 &       2.19 &   N &   N \\
 NGC 1511 &   \ra{3;59;36.6} &   \ang{-67;38;2.1} &  15.3 &  72.7 &    8.09 &   1.47 &   2.27 &       2.80 &   N &   N \\
 NGC 1546 &   \ra{4;14;36.3} &  \ang{-56;03;39.2} &  17.7 &  70.3 &   22.39 &   1.94 &   0.83 &       0.37 &   N &   N \\
 NGC 1566 &    \ra{4;20;0.4} &  \ang{-54;56;16.8} &  17.7 &  29.5 &   60.85 &   5.05 &   4.54 &       0.75 &   Y &   Y \\
 NGC 1672 &   \ra{4;45;42.5} &  \ang{-59;14;50.1} &  19.4 &  42.6 &   53.61 &   7.24 &   7.60 &       1.42 &   Y &   Y \\
 NGC 1792 &   \ra{5;05;14.3} &  \ang{-37;58;50.0} &  16.2 &  65.1 &   40.96 &   6.64 &   3.70 &       0.90 &   N &   N \\
 NGC 2566 &   \ra{8;18;45.6} &  \ang{-25;29;58.3} &  23.4 &  48.5 &   51.21 &   7.17 &   8.72 &       1.70 &   Y &   N \\
 NGC 2903 &   \ra{9;32;10.1} &   \ang{+21;30;3.0} &  10.0 &  66.8 &   43.02 &   3.74 &   3.08 &       0.71 &   Y &   N \\
 NGC 2997 &   \ra{9;45;38.8} &  \ang{-31;11;27.9} &  14.1 &  33.0 &   54.06 &   6.79 &   4.37 &       0.81 &   N &   N \\
 NGC 3059 &    \ra{9;50;8.2} &  \ang{-73;55;19.9} &  20.2 &  29.4 &   23.87 &   2.43 &   2.38 &       1.00 &   Y &   N \\
 NGC 3521 &  \ra{11;05;48.6} &   \ang{-00;02;9.4} &  13.2 &  68.8 &  105.21 &   5.90 &   3.72 &       0.35 &   N &   N \\
 NGC 3621 &  \ra{11;18;16.3} &  \ang{-32;48;45.4} &   7.1 &  65.8 &   11.38 &   1.15 &   0.99 &       0.87 &   N &   Y \\
 NGC 4303 &  \ra{12;21;54.9} &  \ang{+04;28;25.5} &  17.0 &  23.5 &   33.39 &   8.12 &   5.33 &       1.60 &   Y &   Y \\
 NGC 4321 &  \ra{12;22;54.9} &  \ang{+15;49;20.3} &  15.2 &  38.5 &   55.61 &   7.77 &   3.56 &       0.64 &   Y &   N \\
 NGC 4535 &  \ra{12;34;20.3} &  \ang{+08;11;52.7} &  15.8 &  44.7 &   33.96 &   3.99 &   2.16 &       0.64 &   Y &   N \\
 NGC 4536 &  \ra{12;34;27.1} &  \ang{+02;11;17.7} &  16.2 &  66.0 &   25.07 &   2.62 &   3.45 &       1.37 &   Y &   N \\
 NGC 4569 &  \ra{12;36;49.8} &  \ang{+13;09;46.4} &  15.8 &  70.0 &   64.04 &   4.55 &   1.32 &       0.21 &   Y &   Y \\
 NGC 4826 &  \ra{12;56;43.6} &  \ang{+21;40;59.1} &   4.4 &  59.1 &   17.40 &   0.41 &   0.20 &       0.12 &   N &   Y \\
 NGC 5248 &  \ra{13;37;32.0} &   \ang{+08;53;6.7} &  14.9 &  47.4 &   25.49 &   4.54 &   2.29 &       0.90 &   Y &   N \\
 NGC 5643 &  \ra{14;32;40.8} &  \ang{-44;10;28.6} &  12.7 &  29.9 &   21.69 &   2.66 &   2.59 &       1.20 &   Y &   Y \\
 NGC 6300 &  \ra{17;16;59.5} &  \ang{-62;49;14.0} &  11.6 &  49.6 &   29.45 &   1.90 &   1.89 &       0.64 &   Y &   Y \\
 NGC 7496 &  \ra{23;09;47.3} &  \ang{-43;25;40.3} &  18.7 &  35.9 &    9.92 &   1.81 &   2.26 &       2.28 &   Y &   Y \\
\hline\hline
\end{tabular}
\begin{tablenotes}
\small
\item \textbf{Notes.} (2) Right ascension, (3) declination, (4) distance \citep{Anand2021}, (5) inclination angle \citep{Lang2020}, (6) global stellar mass, (7) global H$_2$ mass and (8) global star formation rate. Integrated galaxy properties (6-8) are taken from \citet{Leroy2021b}. Columns (10) and (11) specify if a galaxy is barred (Y) or unbarred (N) \citep{Querejeta2021b} and if it contains an AGN (Y) or not (N) \citep{Veron2010}.
\end{tablenotes}
\end{threeparttable}
\end{table*}

In this study we link the kpc-scale dense gas spectroscopy with the cloud-scale molecular gas properties across 25 nearby galaxies. 
To enable this we present a new ALMA survey of high critical density molecular lines, which we call ALMOND (``ACA Large-sample Mapping of Nearby galaxies in Dense gas''). ALMOND aimed to detect emission from high critical density lines, \hcnone, \hcopone, \cstwo , from targets of the PHANGS--ALMA survey. 
Following standard practice for extragalactic work, \citep[e.g.][]{Gao2004, Usero2015, Bigiel2016, Gallagher2018a, Querejeta2019} ALMOND initially focuses on \hcnone (hereafter HCN), \hcopone , and \cstwo\ as our primary tracer of dense molecular gas.
%We include all PHANGS galaxies that are mapped in dense gas by the the recent ACA survey (Bigiel et al. in prep.). These galaxies have also been mapped in bulk molecular gas by PHANGS--ALMA \citep{Leroy2021b}. This forms an unmatched large sample (25 galaxies) of kpc-scale dense gas and cloud-scale bulk molecular gas maps. 
We designed ALMOND with the goal of detecting these high critical density tracers, and as a result began by targeting the more massive and actively star-forming PHANGS--ALMA targets.
All targets are nearby ($d<\SI{25}{\mega\parsec}$), relatively massive ($\SI{e10}{\Msun}\lesssim M_\star\lesssim\SI{e11}{\Msun}$) gas-rich ($\SI{e9}{\Msun}\lesssim M_{\mathrm{H}_2}\lesssim\SI{e10}{\Msun}$), star-forming ($\SI{1}{\Msun\per\year}\lesssim \text{SFR}\lesssim\SI{10}{\Msun\per\year}$) galaxies, selected based on the PHANGS--ALMA CO~(2-1) maps and mid-IR emission so that we expected the ACA to be able to achieve significant detections of the high critical density rotational lines near $\nu \approx 85{-}100$~GHz, \hcnone, \hcopone, \cstwo , at least in the galaxy centres and across spiral arms. 
At these nearby distances, even the moderate angular resolution of the ACA allows us to resolve key environmental features (centre, bar, spiral arms) in both the bulk and dense molecular gas. 
Our diverse sample covers a variety of morphology, including 16 barred (9 unbarred) galaxies and 11 galaxies containing (14 without) an active galactic nucleus (AGN).
Table \ref{TAB:galaxy_sample} lists the galaxy sample along with their physical properties. 
We summarise the used data products in Table \ref{TAB:observations}.

%%%%%%%%%%%%%%%%%%%%%%%%%%%%%%%
\subsection{New \hcnone\ observations}
\label{SEC:dense_gas_observations}

%In this work, we present new ALMA observations referred to as the ALMOND (ACA Large-sample Mapping of Nearby galaxies in Dense Gas) survey. This survey mapped 25 nearby spiral galaxies in dense molecular gas tracers using the Morita Atacama Compact Array (ACA).}
%\textcolor{red}{\sout{The HCN data is taken from the ALMOND (ACA Large-sample Mapping of Nearby galaxies in Dense Gas; Bigiel et al. in prep.) 
ALMOND observed 25 nearby galaxies in dense molecular gas tracers using the Morita Atacama Compact Array (ACA) as part of the ALMA facility.
The ACA consists of four 12-m dishes which operate in single dish ("total power", TP) mode and an array of 14 7-m telescopes. 
The spectral setup is similar to the one described in \citet{Gallagher2018a}, 
%who observed \hcnone, \hcopone, and \cstwo
%, $^{13}$CO(1-–0), and C$^{18}$O(1-–0) 
% in four galaxies using ALMA’s main array of 12-m antennas.
%at an angular resolution of \ang{;;8}.}
and covers the brightest high critical density lines, \hcnone, \hcopone and \cstwo as well as a a suite of fainter lines. At these frequencies, the ACA has a native resolution of \angRange{;;17}{;;22} which, for our targets, relates to physical scales of $\sim$ \SIrange{1}{2}{\kilo\parsec}. 
In total, ALMOND currently includes 7-m+TP observations of 25 targets (projects 2017.1.00230.S, 2018.1.01171.S, 2019.2.00134.S), which we combine in this analysis with additional 7-m+TP observations of NGC\,2903 (project 2021.1.00740.S) and NGC\,4321 (project 2017.1.00815.S).
The data consist of a homogeneous set of ACA observations of a large sample of 23 galaxies with exceptionally deep observations of NGC\,2903 and NGC\,4321, for a total of 25 galaxies, which we believe to be the largest or one of the largest-ever mapping surveys targeting these high critical density lines.
The data reduction was carried out using the PHANGS--ALMA pipeline \citep[for more details see][]{Leroy2021a}, which uses the the standard ALMA data reduction package, \texttt{CASA} \citep{casa_package}.

The resulting PPV (position-position-velocity) cubes have typical spectral resolution of \SI{10}{\kilo\metre\per\second} and typical noise per channel of \SI{1}{\milli\kelvin} for the deeper observations (NGC\,2903 and NGC\,4321) and $\sim$ \SI{3}{\milli\kelvin} for the other 23 galaxies. 
The good sensitivity of the ACA allows us to detect \hcnone, \hcopone and \cstwo emission in the centres of all targets and in individual locations across the molecular spiral arms in some of the ALMOND galaxies. 
Across all galaxies, we observe in total 4566 independent sightlines, whereof 242 sightlines show significant HCN emission, i.e. integrated intensities with $\text{S/N}\geq3$.

\begin{figure*}
    \centering
    \includegraphics{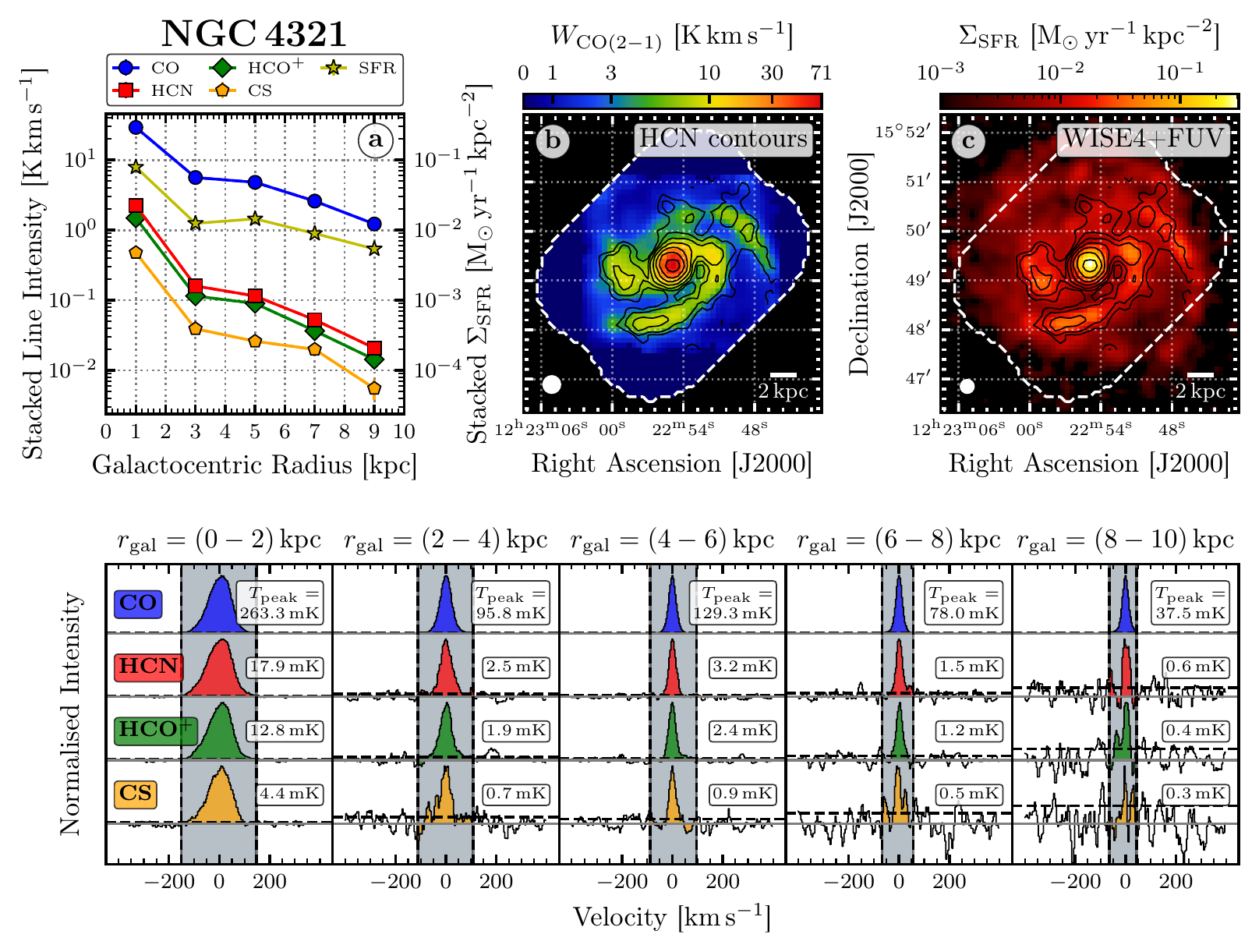}
    \caption{Spectral stacking across NGC\,4321. \textit{a)} Integrated intensities of radially stacked spectra in bins of $r_{\rm gal}=\SI{2}{\kilo\parsec}$. Shown are \cotwo from PHANGS--ALMA \citep{Leroy2021b}, \hcnone, \hcopone, \cstwo from ALMOND (this work) and SFR surface density from z0MGS \citep{Leroy2019}. Solid points indicate significant data ($\text{S/N}>3$). \textit{b)} \cotwo moment-0 map, computed as described in Section~\ref{SEC:mom0_maps}, overlaid with \hcnone contours in S/N levels of $2^n\;\text{for}\; n\in\{0,1,2,...,7\}$, both at a common spatial resolution of \ang{;;19.7}. \textit{c)} SFR map at \ang{;;15} resolution, computed as described in Section~\ref{SEC:star_formation_rate} from a linear combination of the WISE4 IR and GALEX FUV data. \textit{Bottom:} Stacked spectra, obtained as described in Appendix~\ref{SEC:appendix:stacking} corresponding to the integrated intensities shown in a). The grey shaded area indicates the velocity-integration mask. The spectra are normalised by their peak intensity for each bin and each line individually. The respective peak intensities (measured inside the integration mask) are shown in the box next to each spectra. The horizontal dotted line indicates the rms, i.e. the standard deviation of the spectrum outside the integration mask. 
    %We made the above plot for all 25 galaxies available online.
    We show analogues plots for the other 24 galaxies in Appendix~\ref{SEC:appendix:supplements}.
    }
    \label{FIG:ngc4321_stacking}
\end{figure*}

Beyond the individual detections, the survey covers a large area and we know the likely location and velocity of the faint \hcnone\ emission. This allows us to achieve widespread detections of these faint lines via stacking, e.g., constructing sensitive radial profiles. 
In Appendix~\ref{SEC:appendix:stacking}, we show that via spectral stacking HCN can be detected in the central \SI{2}{\kilo\parsec} in all galaxies, out to \SI{4}{\kilo\parsec} and \SI{6}{\kilo\parsec} in 21 and 9 of the 25 galaxies, respectively. 
In Figure~\ref{FIG:ngc4321_stacking}, we illustrate this radial stacking spectra procedure and show the integrated intensities for NGC\,4321. 
These are our deepest observations, and so are not typical, but they nicely illustrate the nature of the ALMOND data and the stacking procedures. 
For more details on the stacking method see Appendix~\ref{SEC:appendix:stacking}. 
%The complete atlas of maps and stacked spectra are made available online.
The complete atlas of maps and stacked spectra are presented in Appendix~\ref{SEC:appendix:supplements}.

%%%%%%%%%%%%%%%%%%%%%%%%%%%%%%%
\subsubsection{\cotwo\,-- Bulk Molecular Gas}
\label{SEC:molecular_gas_observations}

We trace the bulk molecular gas via the \cotwo emission line as observed by the PHANGS--ALMA survey \citep{Leroy2021b}. ALMA produced \cotwo line maps with \angRange{;;1}{;;2} resolution corresponding to physical scales of \SIrange{25}{180}{\parsec}, \SI{2.5}{\kilo\metre\per\second} velocity resolution and \SIrange{0.2}{0.3}{\kelvin} noise per channel. It combines interferometric and single-dish data from the 12-m array and the ACA consisting of the 7-m array and four 12-m dishes observing in total power mode. Thus, it should recover information on all physical scales. In Section~\ref{SEC:cloud_scale_props}, we infer various dynamical properties of the molecular gas following a series of studies \citep{Sun2018,Sun2020a,Sun2020b} which extensively analysed the molecular gas in PHANGS--ALMA.

%%%%%%%%%%%%%%%%%%%%%%%%%%%%%%%%%%%%%%
\subsection{UV + IR -- Star Formation Rate}
\label{SEC:star_formation_rate}

We use star formation rate maps from the "z = 0 Multiwavelength Galaxy Synthesis" study (z0MGS; \citealp{Leroy2019}) adopting a combination of \SI{22}{\micron} (WISE4) and GALEX-FUV \SI{154}{\nano\metre} emission. \citet{Leroy2019} present an atlas of IR and UV images of $\sim$ 15,750 local ($d\lesssim\SI{50}{\mega\parsec}$) galaxies at a matched resolution of \ang{;;7.5} and \ang{;;15}. \citet{Leroy2019} find a linear combination of WISE4 and FUV to be their most robust tracer of the SFR:
\begin{align}
    \left(\dfrac{\sigsfr}{\SI{}{\Msun\per\year\per\square\kilo\parsec}}\right) \approx (T_\text{WISE4} + T_\text{FUV}) \cdot \cos{i} \; ,
    \label{EQU:SFR}
\end{align}
where
\begin{align}
    T_\text{WISE4} = \num{3.24e-3} \left(\dfrac{\log_{10}C_\text{WISE4}}{-42.7}\right)  \left(\dfrac{I_\text{WISE4}}{\SI{}{\mega\jansky\per\steradian}}\right) \; ,
\end{align}
and
\begin{align}
    T_\text{FUV} = \num{1.04e-1} \left(\dfrac{\log_{10}C_\text{FUV}}{-43.42}\right)  \left(\dfrac{I_\text{FUV}}{\SI{}{\mega\jansky\per\steradian}}\right) \; .
\end{align}
We refer to \citet{Kennicutt2012} for a comparative discussion of SFR tracers. In Equation \eqref{EQU:SFR}, $i$ is the galaxy's inclination as listed in Table \ref{TAB:galaxy_sample} and the $\cos{i}$ term corrects for the projection effect due to the galaxy's inclination. For galaxies without GALEX coverage \cite{Leroy2019} also prescribe formulas using only WISE4. Table \ref{TAB:observations} lists the available SFR tracers for our sample. The coefficients $\log_{10}C_\text{WISE4}$ depend on the galaxy and were benchmarked to \citet{Salim2016} and \citet{Salim2018} (see \citet{Leroy2019} for details). We downloaded the SFR maps for our galaxy sample at a resolution of \ang{;;15} from the public z0MGS repository\footnote{irsa.ipac.caltech.edu/data/WISE/z0MGS}. These maps are then convolved to the spatial resolution of the ACA maps (\SI{2.1}{\kilo\parsec} $\sim$ \ang{;;20}).

\begin{table*}
%\resizebox{\textwidth}{!}{
\begin{threeparttable}[t]
\centering
\caption{Data/Observations}
\label{TAB:observations}
\begin{tabular}{ccccccccc}
\hline\hline
\multirow{2}{*}{Galaxy} & \multicolumn{3}{c}{CO observations} && \multicolumn{3}{c}{HCN observations} & \multirow{2}{*}{SFR tracers} \\ \cline{2-4}\cline{6-8}
 & Survey & Res. & Res. && Survey & Res. & Res. & \\
 & & (\SI{}{\arcsecond}) & (\SI{}{\parsec}) && & (\SI{}{\arcsecond}) & (\SI{}{\kilo\parsec}) & \\
(1) & (2) & (3) & (4) && (5) & (6) & (7) & (8) \\
\hline
 NGC 0628 &  PHANGS--ALMA &      1.12 &        53 &       &  ALMOND &       18.6 &        0.89 &  WISE4, FUV  \\
 NGC 1097 &  PHANGS--ALMA &      1.70 &       112 &       &  ALMOND &       19.4 &        1.28 &  WISE4, FUV  \\
 NGC 1365 &  PHANGS--ALMA &      1.38 &       131 &       &  ALMOND &       20.6 &        1.96 &  WISE4, FUV  \\
 NGC 1385 &  PHANGS--ALMA &      1.27 &       106 &       &  ALMOND &       19.9 &        1.67 &  WISE4, FUV  \\
 NGC 1511 &  PHANGS--ALMA &      1.45 &       107 &       &  ALMOND &       17.6 &        1.30 &  WISE4, FUV  \\
 NGC 1546 &  PHANGS--ALMA &      1.28 &       110 &       &  ALMOND &       19.0 &        1.63 &  WISE4, FUV  \\
 NGC 1566 &  PHANGS--ALMA &      1.25 &       108 &       &  ALMOND &       19.8 &        1.69 &  WISE4, FUV  \\
 NGC 1672 &  PHANGS--ALMA &      1.93 &       182 &       &  ALMOND &       17.7 &        1.67 &  WISE4, FUV  \\
 NGC 1792 &  PHANGS--ALMA &      1.92 &       151 &       &  ALMOND &       18.8 &        1.47 &  WISE4, FUV  \\
 NGC 2566 &  PHANGS--ALMA &      1.28 &       145 &       &  ALMOND &       18.6 &        2.11 &       WISE4  \\
 NGC 2903 &  PHANGS--ALMA &      1.45 &        71 &       &  ALMOND &       18.4 &        0.89 &  WISE4, FUV  \\
 NGC 2997 &  PHANGS--ALMA &      1.77 &       121 &       &  ALMOND &       20.4 &        1.39 &  WISE4, FUV  \\
 NGC 3059 &  PHANGS--ALMA &      1.22 &       120 &       &  ALMOND &       16.8 &        1.64 &       WISE4  \\
 NGC 3521 &  PHANGS--ALMA &      1.33 &        85 &       &  ALMOND &       21.2 &        1.36 &       WISE4  \\
 NGC 3621 &  PHANGS--ALMA &      1.82 &        62 &       &  ALMOND &       18.9 &        0.65 &  WISE4, FUV  \\
 NGC 4303 &  PHANGS--ALMA &      1.81 &       149 &       &  ALMOND &       20.3 &        1.67 &  WISE4, FUV  \\
 NGC 4321 &  PHANGS--ALMA &      1.67 &       123 &       &  ALMOND &       19.7 &        1.45 &  WISE4, FUV  \\
 NGC 4535 &  PHANGS--ALMA &      1.56 &       119 &       &  ALMOND &       22.9 &        1.75 &  WISE4, FUV  \\
 NGC 4536 &  PHANGS--ALMA &      1.48 &       116 &       &  ALMOND &       21.6 &        1.70 &  WISE4, FUV  \\
 NGC 4569 &  PHANGS--ALMA &      1.69 &       129 &       &  ALMOND &       19.3 &        1.47 &  WISE4, FUV  \\
 NGC 4826 &  PHANGS--ALMA &      1.26 &        27 &       &  ALMOND &       18.8 &        0.40 &  WISE4, FUV  \\
 NGC 5248 &  PHANGS--ALMA &      1.29 &        93 &       &  ALMOND &       19.9 &        1.44 &  WISE4, FUV  \\
 NGC 5643 &  PHANGS--ALMA &      1.30 &        80 &       &  ALMOND &       18.1 &        1.11 &       WISE4  \\
 NGC 6300 &  PHANGS--ALMA &      1.08 &        60 &       &  ALMOND &       17.7 &        1.00 &       WISE4  \\
 NGC 7496 &  PHANGS--ALMA &      1.68 &       152 &       &  ALMOND &       17.9 &        1.63 &  WISE4, FUV  \\
\hline\hline
\end{tabular}
\begin{tablenotes}
\small
\item \textbf{Notes.} (2-4) \cotwo data from PHANGS--ALMA \citep{Leroy2021b} along with their native resolutions (full-width half-maximum) in arcseconds and parsecs, (5-7) analog for the \hcnone data taken from ALMOND (this work), (8) applied star formation rate tracers from WISE \citep{Wright2010} and GALEX \citep{Martin2005}. The data has been spatially homogenised. The CO observations from PHANGS--ALMA have been convolved to a physical resolution of \SI{150}{pc} and the HCN observations from ALMOND as well as the SFR maps have been convolved to \SI{2.1}{\kilo\parsec}.
\end{tablenotes}
\end{threeparttable}
%}
\end{table*}

%%%%%%%%%%%%%%%%%%%%%%%%%%%%%%%%%%%%%%
\section{Methods}
\label{SEC:methods}
%%%%%%%%%%%%%%%%%%%%%%%%%%%%%%%%%%%%%%

The aim of this work is to compare the kpc-scale dense gas and SFR observations with the cloud-scale molecular gas properties.
To do so we need to determine the integrated intensities of each line (Section~\ref{SEC:mom0_maps}).
We then estimate the cloud-scale properties from the \SI{150}{\parsec} scale \cotwo data (Section~\ref{SEC:cloud_scale_props}), and the dense gas quantities from the coarser \hcnone and SFR data at \SI{2.1}{\kilo\parsec} scale (Section~\ref{SEC:dense_gas_props}). 
Next, we explain the weighted averaging method, which is used to compare these two scales (Section~\ref{SEC:weighted_avg}), and the data binning that is used to improve signal-to-noise (Section~\ref{SEC:binning}).
Finally, we introduce the fitting scheme, which is used to constrain a first order relation between the kpc- and cloud-scale quantities.

%%%%%%%%%%%%%%%%%%%%%%%%%%%%%%%%%%%%%%
\subsection{Integrated Intensity Maps}
\label{SEC:mom0_maps}
%%%%%%%%%%%%%%%%%%%%%%%%%%%%%%%%%%%%%%

We produce integrated intensity maps from the original \cotwo, \hcnone (analogously with \hcopone, \cstwo) PPV cubes for all galaxies. At first, we convolve the data cubes to the target resolution using the respective cloud-scale resolution for the \cotwo data and the kpc-scale resolution for the \cotwo and \hcnone (\hcopone, \cstwo) cubes.  Then, we put the voxels on hexagonal grids, using one sample per beam (FWHM) for the kpc-scale maps, and two samples per beam (FWHM) for the cloud-scale maps. We use a higher sampling rate (satisfying the Nyquist–Shannon sampling theorem) for the cloud-scale maps in order to avoid losing information in computing the weighted averages (see Section \ref{SEC:weighted_avg}). After conducting the weighted averages, we resample to match the kpc-resolution maps which are sampled at the beam size to get statistically independent data points for further processing.

We use the \cotwo data to create position-position-velocity masks, where we apply customised scripts that have been utilised in previous large program studies (e.g. EMPIRE, \citealp{Jimenez-Donaire2019}) and is based on the methodology introduced by \citet{Rosolowski2006}. We first identify pixels with high signal-to-noise ratio (S/N; $\text{S/N}\ge 4$) in at least three adjacent velocity channels. In addition we build a low S/N mask requiring at least three adjacent velocity channels with  $\text{S/N}\ge 2$. Then we iteratively grow the identified high S/N regions to include adjoining regions with moderate S/N as defined by the low S/N mask. In doing so, we recover the more extended 2-sigma detection belonging to a 4-sigma core and thus recover regions of bright CO emission that one would also identify by eye. Finally, we collapse the masked data cubes along the velocity axis by summing the mask-selected channels (in K) multiplied by the channel width (in \SI{}{\kilo\metre\per\second}) to produce integrated intensity maps (in \SI{}{\kelvin\kilo\metre\per\second}).

We extract the HCN (analogously with HCO$^+$ and CS) emission via the the CO-based position-position-velocity masks and produce the integrated intensity maps as described above. \cotwo is easy to excite and the brightest line observed here, being detected with a much higher S/N compared to the faint dense gas tracers, e.g. HCN. As such, CO emission unveils the regions of molecular gas where we also expect to find emission of the dense molecular gas as traced by \hcnone (or \hcopone, \cstwo).

For each line of sight, we compute the statistical uncertainties in the integrated intensity $\sigma_I$ from the rms in the emission-free (not selected by the mask) channels via:
\begin{align}
    \left(\dfrac{\sigma_I}{\SI{}{\Kkms}}\right) = \left(\dfrac{\text{rms}}{\SI{}{\kelvin}}\right) \times \left(\dfrac{\Delta v_\text{channel}}{\SI{}{\kilo\metre\per\second}}\right) \times \sqrt{N}
\end{align}
where $\Delta v_\text{channel}$ is the channel width and $N$ is the number of mask-selected voxels along the line of sight.

%%%%%%%%%%%%%%%%%%%%%%%%%%%%%%%
\subsection{kpc-Scale Dense Gas Properties}
\label{SEC:dense_gas_props}
%%%%%%%%%%%%%%%%%%%%%%%%%%%%%%%

\subsubsection{Dense Gas Fraction}
\label{SEC:fdense}
In Sections \ref{SEC:results} and \ref{SEC:analysis}, we focus on the observed ratio $\intHCN/\intCO$, which we expect to be sensitive to density with some additional dependence on physical parameters like abundances, temperature, and opacities. In the discussion section we also comment on implications for the actual dense gas fraction (\fdense), which is a simple recasting of this ratio using common mass-to-light ratios for both lines. We compute \fdense as the ratio of the dense gas surface density (\sigdense) and the molecular gas surface density (\sigmol) which is traced by $\intHCN/\intCO$:
\begin{align}
    \fdense = \dfrac{\sigdense}{\sigmol} = \dfrac{\aHCN\intHCN}{\aCO R_\text{21}^{-1}\intCO} \approx 2.1\dfrac{\intHCN}{\intCO}  \, .
\end{align}
The kpc-scale integrated intensity maps are obtained as described in Section~\ref{SEC:mom0_maps}. \sigmol is measured via \intCO assuming a constant mass-to-light ratio $\aCO=4.3\, \text{M}_\odot \,\text{pc}^{-2} \,(\text{K}\,\text{km}\,\text{s}^{-1})^{-1}$ \citep{Bolatto2013} and a \cotwo-to-\coone line ratio of $R_{21}=0.64$ \citep{denBrok2021,Leroy2021_line_ratios}. For more details on \aCO and $R_{21}$, see Section \ref{SEC:molecular_gas_surface_density}. Similarly, \sigdense is obtained via \intHCN adopting a more uncertain $\alpha_\text{HCN}\approx14\, \text{M}_\odot \,\text{pc}^{-2} \,(\text{K}\,\text{km}\,\text{s}^{-1})^{-1}$ (uncertain by at least $\sim\SI{0.3}{\dex}$) tracing gas above $n_{\text{H}_2}\approx \SI{5e3}{\centi\metre\tothe{-3}}$ \citep{Onus2018}. For comparison, but not used in this work, previous studies assumed a lower value of $\alpha_\text{HCN}\approx10\, \text{M}_\odot \,\text{pc}^{-2} \,(\text{K}\,\text{km}\,\text{s}^{-1})^{-1}$ and that HCN traces gas above a higher density of \SI{3e4}{\centi\metre\tothe{-3}} (following \citealp{Gao2004}, also see \citealp{Jones2021}).

%%%%%%%%%%%%%%%%%%%%%%%%%%%%%%%
\subsubsection{Dense Gas Star Formation Efficiency}
\label{SEC:sfedense}
We compute the star formation efficiency of the dense gas via the ratio of star formation rate surface density and dense gas surface density:
\begin{align}
    \sfedense = \dfrac{\sigsfr}{\sigdense} = \aHCN^{-1}\,\dfrac{\sigsfr}{\intHCN} \, .
    \label{EQU:sfedense}
\end{align}
Note that here \sigmol, \sigsfr and \sigdense are not corrected for the galaxies' inclinations because we are only interested in the ratio of surface densities such that the deprojection term $\cos{i}$ cancels out. For typical units and by adopting $\alpha_\text{HCN}\approx14\, \text{M}_\odot \,\text{pc}^{-2} \,(\text{K}\,\text{km}\,\text{s}^{-1})^{-1}$ like in Section \ref{SEC:sfedense}, above Equation \eqref{EQU:sfedense} becomes:
\begin{align}
    \left(\dfrac{\sfedense}{\SI{}{\per\mega\year}}\right) = \num{7.1e-1} \left(\dfrac{\sigsfr}{\SI{}{\Msun\per\year\per\square\kilo\parsec}}\right) \left(\dfrac{\intHCN}{\SI{}{\Kkms}}\right)^{-1} \, .
\end{align}

\begin{figure*}
    \includegraphics[scale=0.95]{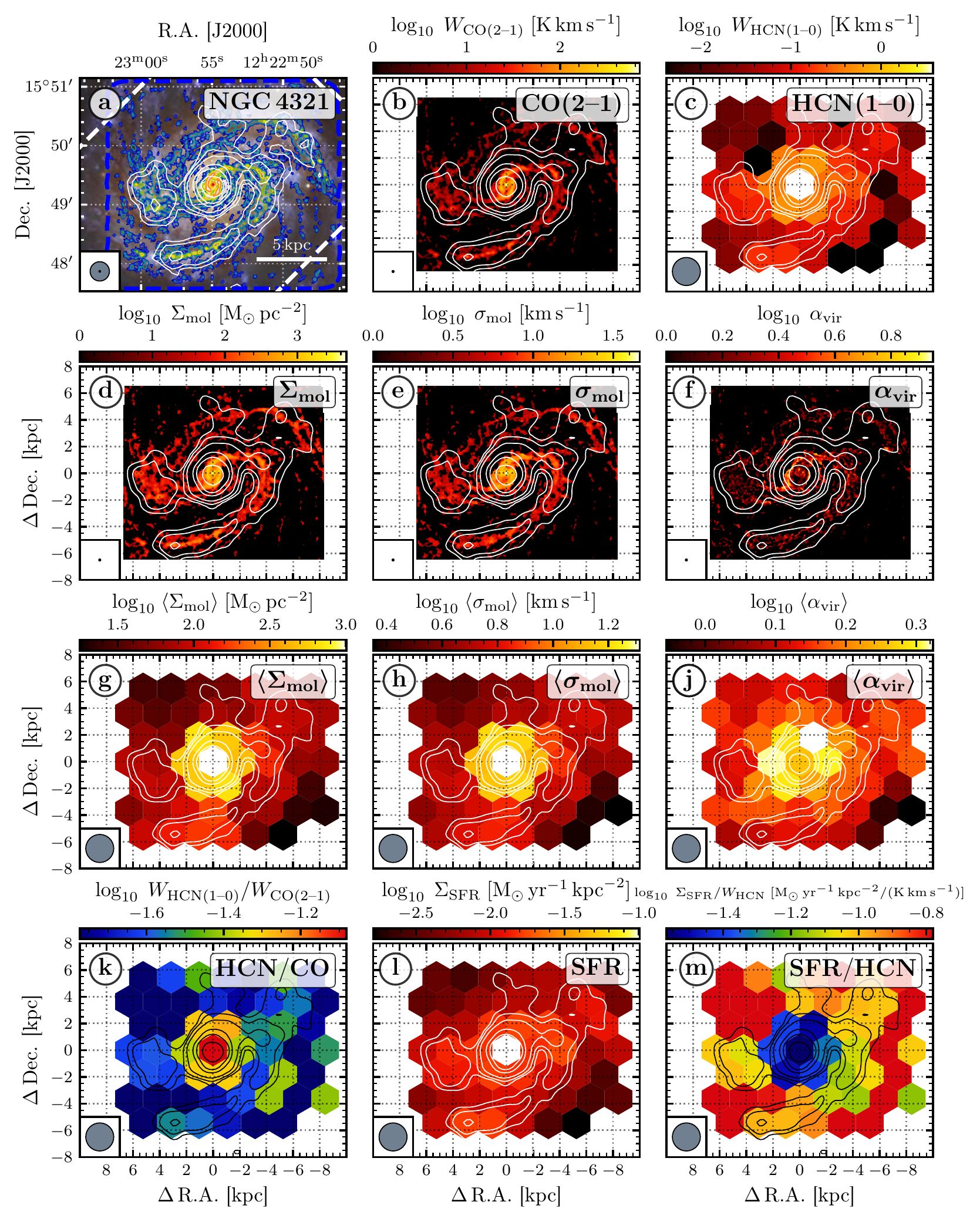}
    \caption{Data products compilation of NGC\,4321 (one of the deepest observations) at cloud- and kpc-scale resolutions \SI{150}{\parsec} and \SI{2.1}{\kilo\parsec} respectively: \textit{a)} ESO three colour image composed of \SI{648}{\nm} (red), \SI{544}{\nm} (green), and \SI{433}{\nm} (blue) wideband emission (Image credit: ESO/IDA/Danish 1.5 m/R. Gendler, J.-E. Ovaldsen, C. C. Thöne and C. Féron). Overlaid are colored \cotwo contours and white \hcnone contours, respectively in S/N levels of  ${3, 5, 10, 30, 50, 100, 300}$ (the same HCN contours are used throughout the other maps). The white and blue dashed contour indicates the ALMOND and PHANGS--ALMA FOV, respecpectively. \textit{b)} Integrated \cotwo intensity at \SI{150}{\parsec} resolution from PHANGS--ALMA and \textit{b)} integrated \hcnone intensity at \SI{2.1}{\kilo\parsec} resolution from ALMOND as obtained from the data cubes according to Section~\ref{SEC:mom0_maps}. \textit{d)-f)} Cloud-scale properties of the molecular gas (molecular gas surface density \sigmol, velocity dispersion \vdis and virial parameter \avir) computed from \cotwo as described in Section~\ref{SEC:cloud_scale_props}. \textit{g)-i)} \cotwo intensity weighted averages \sigmolavg, \vdisavg, \aviravg of the above cloud-scale properties based on the formalism described in Section~\ref{SEC:weighted_avg}. \textit{k)} HCN/CO tracing \fdense, \textit{l)} SFR surface density from FUV (GALEX), \textit{m)} IR (WISE) and SFR/HCN tracing \sfedense following Section~\ref{SEC:dense_gas_props}.}
    \label{FIG:ngc4321_maps}
\end{figure*}

%%%%%%%%%%%%%%%%%%%%%%%%%%%%%%%%%%%
\subsection{Cloud-Scale Molecular Gas Properties}
\label{SEC:cloud_scale_props}
%%%%%%%%%%%%%%%%%%%%%%%%%%%%%%%%%%%

We compute the four cloud-scale properties molecular gas surface density (\sigmol), velocity dispersion (\vdis), virial parameter (\avir) and internal turbulent pressure (\Pturb) using PHANGS--ALMA \cotwo data (see Section~\ref{SEC:molecular_gas_observations}) following \citet{Sun2018}. We measure the cloud-scale properties at beam sizes of \SI{150}{\parsec} using pixel-by-pixel values instead of identifying individual molecular clouds. Based on comparisons of the two approaches by \citet{Sun2020b} and \citet{Rosolowsky2021}, statistically we expect similar results for the molecular gas properties as measured at cloud-scale compared to cloud properties as obtained for individually identified clouds. 
In Appendix~\ref{SEC:appendix:res_configs} we also discuss sub-samples, where we have access to higher resolutions, i.e. \SI{75}{\parsec} for five galaxies and \SI{120}{\parsec} for twelve galaxies, respectively. 
We confirm that the results do not significantly depend on the resolution at which the cloud-scale properties are measured.

%%%%%%%%%%%%%%%%%%%%%%%%%%%%%%
\subsubsection{Molecular Gas Surface Density}
\label{SEC:molecular_gas_surface_density}
We trace \sigmol at \SI{150}{\parsec} resolution via \intCO using a constant mass-to-light ratio conversion factor:
\begin{align}
    \sigmol = \aCO \, R_\text{21}^{-1} \, \intCO \, .
\end{align}
We adopt a constant, Milky Way-like CO-to-\htwo conversion factor of $\aCO=4.3\, \text{M}_\odot \,\text{pc}^{-2} \,(\text{K}\,\text{km}\,\text{s}^{-1})^{-1}$ (uncertainty of $\pm 0.1\,\text{dex}$) as suggested by \citet{Bolatto2013} and a constant \cotwo-to-\coone line ratio of $R_{21}=0.64$ (uncertainty of $\pm 0.06\,\text{dex}$) as recently constrained by \citet{denBrok2021} and for a larger sample including many of these targets by \citet{Leroy2021_line_ratios}, which yields: 
\begin{align}
    \left(\dfrac{\sigmol}{\SI{}{\Msun\per\square\parsec}}\right) = \num{6.7e2} \left(\dfrac{\intCO}{\SI{e2}{\Kkms}}\right) \, .
    \label{EQU:Sigmol_units}
\end{align}
Note that some of the \aCO and $R_{21}$ uncertainty can be attributed to variations as a function of the galactocentric radius, where \aCO is found to be lower in the centres of galaxies \citep{Sandstrom2013}, while $R_{21}$ is higher towards galaxy centres \citep{denBrok2021}. To account for systematic variations of \aCO with metallicity $Z^{\prime}$\footnote{$Z^{\prime}$ is the metallicity normalised to the solar metallicity $[12 + \log_{10}(\text{O}/\text{H})=8.69]$ \citep{Allende-Prieto2001}}, recent studies (as in \citealp{Sun2020b}) adopt a metallicity-dependent $\aCO\propto Z^{\prime\,-1.6}$, which leads to lower \aCO in the central region of galaxies. However, metallicity variations can only partly explain the low \aCO in centres. \citet{Sandstrom2013} conclude that the physical conditions in the centres of galaxies (ISM pressure, gas temperature) are responsible for lowering \aCO by roughly a factor of two. Thus, by adopting a constant \aCO, we may overestimate \sigmol in the central regions of galaxies and underestimate \sigmol at larger galactocentric radii. We still adopt a constant \aCO in analogy to previous studies \citep[e.g.][]{Gallagher2018a,Gallagher2018b,Sun2018} and discuss in Section~\ref{SEC:centres_vs_discs} how lowering \aCO by a factor of two in the centres of galaxies affects the studied relations.

%%%%%%%%%%%%%%%%%%%%%%%%%%%%%%
\subsubsection{Velocity Dispersion}
\label{SEC:velocity_dispersion}
We characterise the line width using the "effective width" according to the prescription of \citet{Heyer2001}, calculated via:
\begin{align}
    \sigma_\text{measured} = \dfrac{\intCO}{\sqrt{2\pi}\,T_\text{peak}} \, ,
\end{align}
where $T_\text{peak}$ (in units of K) is obtained as the maximum intensity of the cubes' spectra for each line of sight. Then, for a Gaussian line profile with peak intensity $T_\text{peak}$ the effective width is equal to the rms velocity dispersion of the line (\vdis). In order to correct for the line broadening caused by the instrument (finite channel width, spectral response curve width) we subtract the contribution of the instrument's response following \citet{Rosolowski2006} and \citet{Sun2018}:
\begin{align}
    \vdis = \sqrt{\sigma_\text{measured}^2 - \sigma_\text{response}^2} \; .
\end{align}
Here, $\sigma_\text{response}$ is estimated from the channel width and the channel-to-channel correlation coefficient, following \citet{Leroy2016} and \citet{Sun2018}.

%%%%%%%%%%%%%%%%%%%%%%%%%%%%%%
\subsubsection{Virial Parameter}
\label{SEC:virial_parameter}
The virial parameter of GMCs is typically defined as $\avir\equiv 2K/U_g$, where $K$ is the kinetic energy and $U_g$ is its self-gravitational potential energy of the cloud such that \avir quantifies deviations from virial equilibrium. Virialised clouds have $\avir=1$, if surface pressure or magnetic support can be neglected. For unbound clouds \avir moves to higher values.

Following \citet{Bertoldi1992}, under the assumption of spherical clouds, the virial parameter can be expressed as:\footnote{Note that this approach neglects contributions from the magnetic energy density or the cosmic ray flux. Moreover, it ignores any surface terms \citep[see e.g.][]{McKee1992,Ballesteros-Paredes2006}}
\begin{align}
    \avir \equiv \dfrac{2K}{U_g} = \dfrac{5\vdis^2R}{fGM} \; ,
\label{EQU:virial_parameter}
\end{align}
where $M$, $R$ and \vdis are the cloud's mass, radius and velocity dispersion, $G$ is the gravitational constant and $f$ is a geometrical factor specifying the density profile of the cloud. We adopt $f=10/9$ which assumes a density profile of the form $\rho\propto r^{-1}$ (e.g. following \citealp{Rosolowski2006}). Given that the cloud-scale resolutions are at the scale of GMCs we take the beam size as the relevant size scale ($R=D_\text{beam}/2$), such that equation \eqref{EQU:virial_parameter} implies:
\begin{align}
    \avir = \dfrac{5}{2fG}\dfrac{\vdis^2 D_\text{beam}}{\sigmol A_\text{beam}} = \dfrac{10\ln{2}}{\pi fG}\dfrac{\vdis^2}{\sigmol D_\text{beam}}
\end{align}
Here, \sigmol is the molecular gas surface density, computed in Section \ref{SEC:molecular_gas_surface_density}, \vdis is the velocity dispersion (see Section \ref{SEC:velocity_dispersion}) and $D_\text{beam}$ is the FWHM of the beam, i.e. \SI{150}{\parsec}. Normalising by typical numbers, we obtain:
\begin{align}
    \avir = 3.1 \left(\dfrac{\sigmol}{\SI{e2}{\Msun\per\square\parsec}}\right)^{-1} \left(\dfrac{\vdis}{\SI{10}{\kilo\metre\per\second}}\right)^2 \left(\dfrac{D_\text{beam}}{\SI{150}{\parsec}}\right)^{-1}
\end{align}
Note, that above formalism is likely to produce uncertainties in \avir reaching factors of a few. 
However, following the approach of e.g. \citet{Sun2018,Sun2020b}, we are interested in measuring $\vdis^2/\sigmol$ for comparative analysis and consider it as a tracer of \avir, where the conversion factor is uncertain by a factor of a few. In other words, we measure \avir in units of $\vdis^2/\sigmol$ for fixed physical scale.

%%%%%%%%%%%%%%%%%%%%%%%%%%%%%%
\subsubsection{Internal Turbulent Pressure}
\label{SEC:Pturb}
We infer the internal turbulent pressure, \Pturb, from the \cotwo observations. Following \citet{Sun2018}, the internal pressure in molecular gas with line-of-sight depth $\sim 2R$ can be expressed as:
\begin{align}
    \Pturb \approx \rho_\text{mol}\vdis^2 \approx \dfrac{1}{2R}\sigmol\vdis^2 \;.
    \label{EQU:turbulent_pressure}
\end{align}
Similar to the virial parameter computation in Section \ref{SEC:virial_parameter}, we aim to measure the quantity $\sigmol\vdis^2$ in order to trace \Pturb at a scale of $R=D_\text{beam}/2$ with the purpose of comparative analysis. \Pturb is linked to $\sigmol\vdis^2$ via a proportionality factor:
\begin{align}
    \left(\dfrac{\Pturb}{\SI{}{\kB\kelvin\per\cubic\centi\metre}}\right) \approx\; & \num{3.3e5}\left(\dfrac{\sigmol}{\SI{e2}{\Msun\per\square\parsec}}\right) \nonumber\\ 
    & \times \left(\dfrac{\vdis}{\SI{10}{\kilo\metre\per\second}}\right)^2\left(\dfrac{D_\text{beam}}{\SI{150}{\parsec}}\right)^{-1}\;,
\end{align}
where \sigmol and \vdis are taken from Sections \ref{SEC:molecular_gas_surface_density} and \ref{SEC:velocity_dispersion}, respectively.

%%%%%%%%%%%%%%%%%%%%%%%%%%%%%%
\subsection{Weighted Averages}
\label{SEC:weighted_avg}
%%%%%%%%%%%%%%%%%%%%%%%%%%%%%%

In order to connect the cloud-scale - \sigmol, \vdis, \avir, \Pturb - measurements to the kpc-scale - \fdense, \sfedense - measurements, we calculate the intensity-weighted averages of \sigmol, \vdis, \avir, \Pturb inside each kpc-scale beam. These weighted averages - \sigmolavg, \vdisavg, \aviravg, \Pturbavg - measure the cloud-scale \sigmol, \vdis, \avir, \Pturb, respectively, from which the average CO photon emerges within the kpc-scale resolution beam. In practice we compute (following \citealp{Leroy2016}):
\begin{align}
    \langle X \rangle = \dfrac{(X\cdot \intCO)\ast\Omega}{\intCO\ast\Omega} \; .
\label{EQU:weighted_convolution}
\end{align}
Here, \intCO is the \cotwo integrated intensity and $X$ is the quantity to be averaged, both at cloud-scale resolution (in this work, \SI{150}{pc}). $X$ is weighted with \intCO (via multiplication) and convolved to the kpc-scale resolution (here, \SI{2.1}{\kilo\parsec}) indicated by the asterisk using a Gaussian kernel $\Omega$. Finally, the weighted average, $\langle X\rangle$, is obtained by division with the convolved weights. Consequently, $\langle X\rangle$ is at kpc-scale resolution and can easily be compared to the kpc-scale \fdense and \sfedense measurements pixel-by-pixel.

The above formalism was introduced by \citet{Leroy2016} and is designed to connect high resolution to low resolution measurements such as conducted in this study, having the advantage of preserving the high resolution information and down-weighting empty regions. As such it was utilised by e.g. \citet{Gallagher2018b} who performed a similar comparison as the one presented in this work.
\citet{Sun2020a} computed the weighted averages in terms of Equation~\eqref{EQU:weighted_convolution} applying a top-hat kernel to the cloud-scale data and then computed the weighted averages in each of these apertures. 
Here, we follow the Gaussian convolution approach using Equation~\eqref{EQU:weighted_convolution} in order to make the weighted averages similarly comparable to the kpc-scale observations.
We highlight the difference between the two approaches in the Appendix~\ref{SEC:appendix:weighted_averages}.

We estimate the propagated uncertainties in the weighted averages via Monte Carlo computations. We start with the \sigmol, \vdis, \avir, \Pturb maps, add random Gaussian noise with amplitudes taken from the cloud-scale maps. Then we run the noise-added maps through the weighted averages procedure and repeat this process 100 times. Finally, we take the standard deviation in \sigmolavg, \vdisavg, \aviravg, \Pturbavg over all realisations as the uncertainty estimate.

%%%%%%%%%%%%%%%%%%%%%%%%%
\subsection{Data Binning}
\label{SEC:binning}
%%%%%%%%%%%%%%%%%%%%%%%%%

We detect integrated HCN intensity (analogously for HCO$^+$ and CS) with S/N $\ge 3$ only in the brightest regions of the galaxies. In order to recover the low S/N information hidden in the data we bin the HCN data by \sigmolavg, or equivalently \intCOavg (following \citealp{Gallagher2018a}). \sigmolavg is detected at high significance across much of the galaxy discs in all 25 targets. 

We bin each galaxy's data individually, choosing a fixed number of 20 bins, equally spaced in \sigmolavg, over the full data range of each galaxy. 
Adapting the binning to each galaxy individually allows us to recover more of the low S/N signal.
We choose the number of 20 bins because it increases the number of HCN detections at low \sigmolavg without averaging over too large intervals thus maximising the dynamic range in the $x$-axis variable (\sigmolavg).
%Moreover, in doing so, by binning every galaxy individually from minimum to maximum (and not adopting fixed bins for all galaxies), we keep the distribution of the data in the binning variable similar to the measured data such that the \sigmol distribution is conserved. 
In each bin, we compute the binned ratio - $\intHCN/\intCO$ or $\sigsfr/\intHCN$ - as the mean of the nominator's data in that bin divided by the mean of the denominator's data in that bin (as in \citealp{Schruba2011} and \citealp{Jimenez-Donaire2017}):
\begin{align}
    \left.\dfrac{\intHCN}{\intCO}\right\vert_\text{bin} & = \dfrac{\text{mean}(\intHCN)\vert_\text{bin}}{\text{mean}(\intCO)\vert_\text{bin}} \\ 
    \left.\dfrac{\sigsfr}{\intHCN}\right\vert_\text{bin} & = \dfrac{\text{mean}(\sigsfr)\vert_\text{bin}}{\text{mean}(\intHCN)\vert_\text{bin}}
\end{align}

This means that for each bin we take the ratio of the bin means and not the bin mean of the ratios. The binning process extends the dynamic range of significant HCN data and has the advantage of reducing the linear regression bias which is naturally induced by converting from linear to logarithmic scale (for more details see Appendix \ref{SEC:appendix:linear_regression}).

We propagate the measurement uncertainties from the individual integrated intensity (and SFR) data points which enter the binning using Gaussian error propagation. As we sample the integrated intensities at the beam size (one sample per beam FWHM), we do not need to account for oversampling in the error propagation. In doing so, for each bin the propagated uncertainty roughly decreases as $1/\sqrt{N}$, where $N$ is the number of points in the bin. However, the binned measurements can often still have low S/N. Considering binned data detected if the signal-to-noise ratio $\ge 3$ and censored (non-detected) if S/N $<3$, we can define upper and lower limits on the binned data. The binned integrated \cotwo intensities and SFR surface densities are significant (S/N $\ge 3$) across the whole galactic disc for the full sample of galaxies. Thus, the S/N is purely dominated by the HCN data. Therefore, we define upper limits (UL) in the binned HCN/\cotwo data via:
\begin{align}
    \text{UL}\vert_\text{bin} = \dfrac{3\cdot I_\text{HCN, unc}\vert_\text{bin}}{\intCO\vert_\text{bin}} \; ,
\end{align}
where $I_\text{HCN, unc}$ is the (propagated) uncertainty of the integrated HCN intensity in each bin. For SFR/HCN we compute lower limits (LL) via:
\begin{align}
    \text{LL}\vert_\text{bin} = \dfrac{\sigsfr\vert_\text{bin}}{3\cdot I_\text{HCN, unc}\vert_\text{bin}} \; .
\end{align}
Although UL and LL are regarded (by definition) non-significant, they are still an important part of the data distribution and we use them in our linear regression analysis (Section~\ref{SEC:fitting_and_corr}).

%%%%%%%%%%%%%%%%%%%%%%%%%%%%%%%%
\subsection{Linear Regression and Correlation}
\label{SEC:fitting_and_corr}
%%%%%%%%%%%%%%%%%%%%%%%%%%%%%%%%

To investigate the correlations we fit a linear regression model to the log-scale binned data, resulting from the data processing described above (Section~\ref{SEC:data_and_methods}). We perform the linear regression by making use of the \texttt{LinMix} package\footnote{\url{https://linmix.readthedocs.io/en/latest/index.html}} which is based on the Bayesian approach to linear regression proposed by \citet{Kelly2007}. In this approach, a likelihood function of the linear regression model is built and MCMC simulations are run using a Gibbs sampler exploring the posterior distribution of the regression parameters. Here, we force the MCMC simulation to take at least \num{10000} steps after convergence was reached, i.e. close to the global maximum of the posterior distribution where every iteration can be considered a random draw from the posterior. The model accounts for heteroscedastic uncertainties in the data on both coordinates, intrinsic scatter and censored data, i.e. upper (or lower) limits in the independent variable.
\footnote{Note that LinMix can also account for the covariance between uncertainties in the x- and y-axis coordinates. You may expect that the uncertainties of HCN/CO and \sigmolavg are correlated since both axis depend on the \cotwo measurements. However, the HCN/CO uncertainties are completely dominated by the \hcnone measurement uncertainties. Therefore, the uncertainties between the axes show no significant correlation and we neglect the covariance term in the fitting scheme.}
Due to its statistical nature in exploring the parameter space, it naturally provides trustworthy uncertainty estimates and credibility intervals of the regression parameters. Moreover, it computes the Pearson correlation coefficient $\rho$ (and the $p$-value) using both detected and censored data. We choose this linear regression method because it accounts for non-detections, determines meaningful fit uncertainties and leads to less biased regression parameter estimates (see Appendix~\ref{SEC:appendix:linear_regression}). 

We perform the linear regression by fitting the following linear function to the data in log-log scale:
\begin{align}
    \log_{10}\,Y = b_\mathrm{y,x} + m_\mathrm{y,x} \left[ \log_{10}\,\Xavg - x_\mathrm{off,x} \right] ,
\end{align}
where $Y$ are the kpc-scale measurements (HCN/CO or SFR/HCN) and \Xavg are the weighted averages of the cloud-scale molecular gas properties (\sigmol, \vdis, \Pturb)\footnote{We skip \avir here, because we do not find any significant correlation with \avir and thus do not perform the linear regression.} in their respective units. $b_\mathrm{y,x}$ and $m_\mathrm{y,x}$ are the intercept and slope of the fit line, where $y=\{f, S\}$, $x=\{\Sigma, \sigma, P\}$ indicate the corresponding kpc-scale (HCN/CO, SFR/HCN) and cloud-scale quantities (\sigmol, \vdis, \Pturb). We recenter the distribution in the $x$-axis coordinate to minimise the covariance between the slope and intercept, applying $x_\mathrm{off,x}\equiv \{2.5, 1.1, 6.5\}$ for $x=\{\Sigma, \sigma, P\}$ which is near the middle of the data range. Note, that this has no effect on the fitting scheme. In addition, we compute the scatter of the data about the best fit line as the standard deviation of the fit residuals, i.e. the standard deviation in the y-axis data after the fit line has been removed. Here, we only consider significant data ($\text{SNR}\ge 3$) and give the scatter in units of dex.
%%%%%%%%%%%%%%%%%%%
\section{Results}
\label{SEC:results}
%%%%%%%%%%%%%%%%%%%

\begin{figure}
    \includegraphics{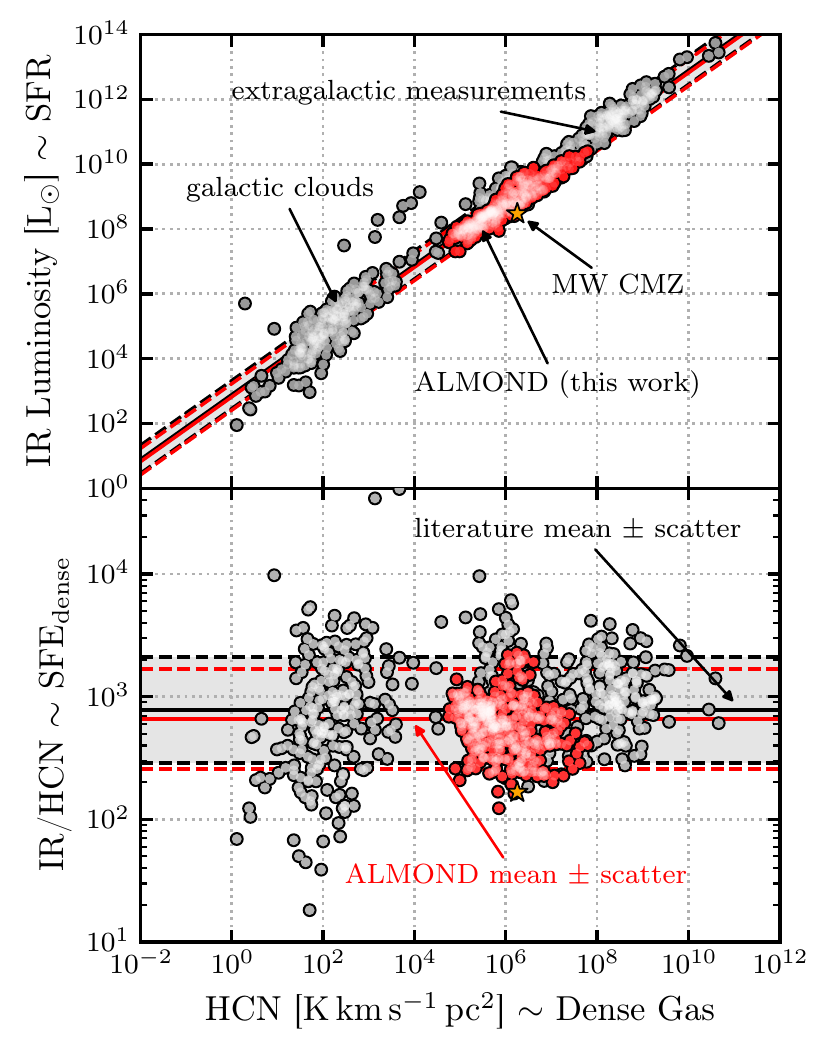}
    \caption{Relation between (total) infrared luminosity, tracing SFR, and \hcnone luminosity, tracing dense gas mass. We show our new ALMOND data, where $\text{S/N}\geq 5$ along with MW clouds \citep{Wu2010,Stephens2016}, the CMZ, GMCs in the SMC, LMC and other low metallicity environments \citep{Chin1997, Chin1998, Braine2017} as well as GMCs in other galaxies \citep{Brouillet2005, Buchbender2013, Chen2017}. Furthermore, we add other extragalactic observations, i.e. resolved nearby galaxy discs \citep{Kepley2014, Bigiel2015, Chen2015, Usero2015, Gallagher2018a} and whole galaxies \citep{Gao2004, Gao2007, Gracia-Carpio2008, Krips2008, Juneau2009, Garcia-Burillo2012, Privon2015}. The solid black line indicates the mean SFR/HCN of $10^{2.89}\SI{}{\Lsun}(\mathrm{K}\,\mathrm{km}\,\mathrm{s}^{-1})^{-1}\,\mathrm{pc}^{-2}$ from \citet{Jimenez-Donaire2019} over their literature compilation, with the dashed lines showing the scatter of $\pm\SI{0.37}{\dex}$. In addition, we show the mean ($10^{2.82}\SI{}{\Lsun}(\mathrm{K}\,\mathrm{km}\,\mathrm{s}^{-1})^{-1}\,\mathrm{pc}^{-2}$) and scatter ($\pm\SI{0.41}{\dex}$) computed from the significant ($\text{S/N}\geq 5$) ALMOND data.}
    \label{FIG:dense_gas_sf_relation}
\end{figure}

%%%%%%%%%%%%%%%%%%%%%%%%%%%%%%%%%%%%%%%%%%%%%%%%%%%%%%%%%%%%
% PLOTS (HCN/CO and SFR/HCN correlations)
%%%%%%%%%%%%%%%%%%%%%%%%%%%%%%%%%%%%%%%%%%%%%%%%%%%%%%%%%%%%

% HCN/CO and SFR/HCN vs cloud props; lowres
\begin{figure*}
    \includegraphics[width=\textwidth]{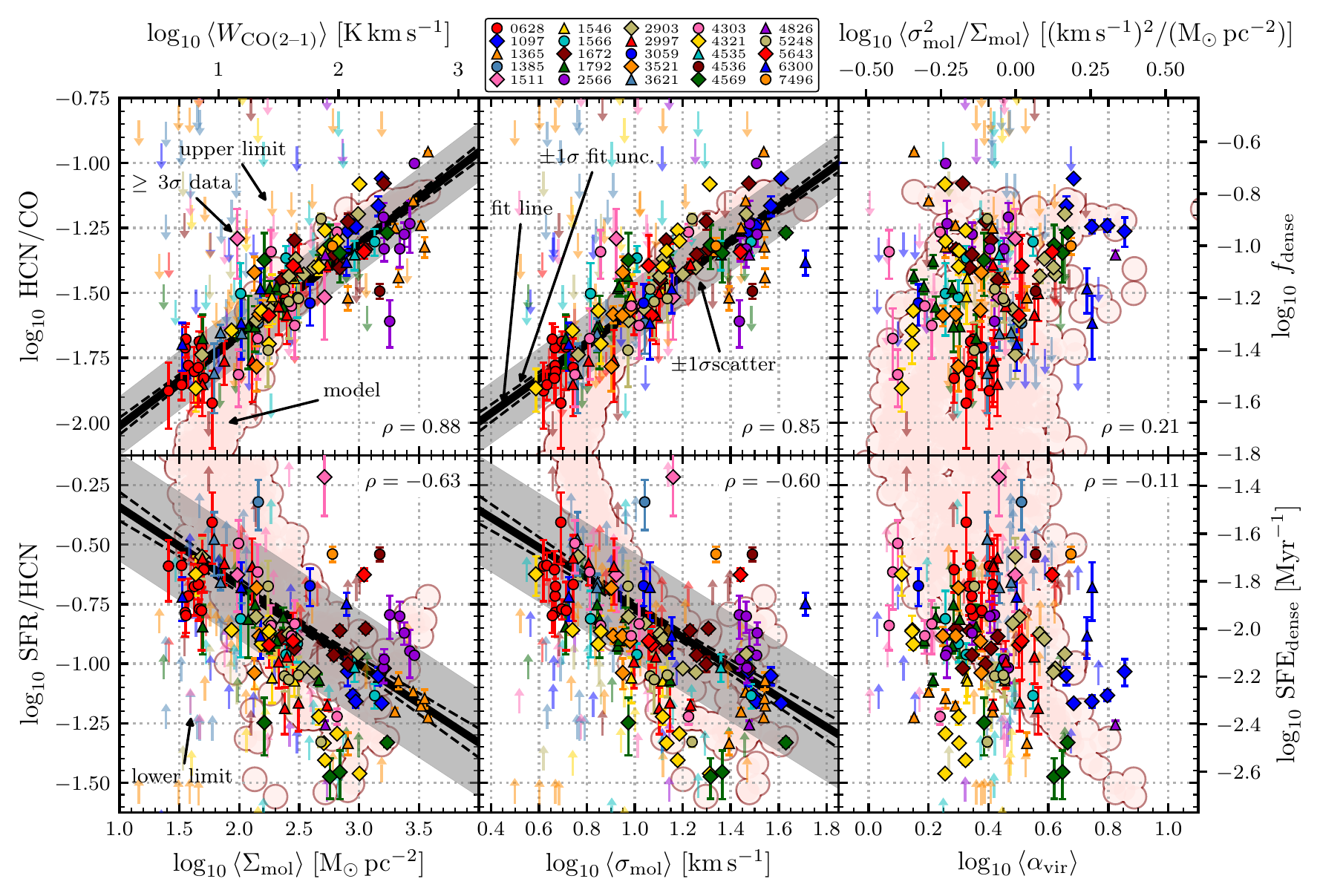}
    \caption{\textbf{HCN/CO vs. \Xavg and SFR/HCN vs. \Xavg (cloud-scale = \textbf{150}$\,$\textbf{pc}, kpc-scale = \textbf{2.1}$\,$\textbf{kpc})} \textit{Top:} HCN/CO as a proxy of dense gas fraction against molecular cloud properties (\sigmolavg, \vdisavg, \aviravg) as obtained from \cotwo data from left to right. The data are binned according to Section~\ref{SEC:binning}. Filled points specify significant data with $\text{SNR(HCN/CO)}\ge3$ and downward pointing arrows indicate $3\sigma$ upper limits on HCN/CO. The thick solid line denotes the best fit linear regression, i.e. the median realisation of the MCMC simulation. The dashed lines indicate the $1\sigma$ credibility interval of the MCMC realisations. The grey shaded area shows the scatter of the significant data about the fit line. For \aviravg we do not observe a correlation and thus do not fit a line to the data. \textit{Bottom:} Analogous to the upper panels, SFR/HCN as a proxy of the star formation efficiency of the dense gas vs. molecular cloud properties from left to right. Here, upward pointing arrows denote $3\sigma$ lower limits in SFR/HCN. Again, the linear regression to \aviravg was not determined due to lack of correlation. The light red shaded areas show the model prediction, equivalent to the red data in Figure~\ref{FIG:PDF_expectations_data}, but shifted by \SI{-1.0}{\dex} in HCN/CO and \SI{-0.6}{\dex} in SFR/HCN to visually overlap with the observational results.}
    \label{FIG:HCN_lowres}
\end{figure*}

% Fit parameters table; lowres
% HCN/CO and SFR/HCN CORRELATION FIT PARAMETERS TABLE (LOWRES; incl. centres and discs)

\begin{table*}
\resizebox{\textwidth}{!}{
\begin{threeparttable}[t]
\centering
\caption{HCN/CO and SFR/HCN Correlations}
\label{TAB:HCN_corr_table_lowres}
\begin{tabular}{ccccccccccc}
\hline\hline
Cloud-scale & \multirow{2}{*}{Environment} & \multicolumn{4}{c}{HCN/CO} && \multicolumn{4}{c}{SFR/HCN} \\\cline{3-6}\cline{8-11}
Property &  & Slope (unc.) & Interc. (unc.)\tnote{1} & Corr. $\rho$ ($p$) & Scatter && Slope (unc.) & Interc. (unc.)\tnote{1} & Corr. $\rho$ ($p$) & Scatter \\
\hline
           & centres + discs & 0.35 (0.02) & -1.49 (0.01) & 0.88 (0.0) & 0.11 &  & -0.33 (0.04) & -0.84 (0.02) & -0.63 (0.0) & 0.23 \\
\sigmolavg & centres         & 0.33 (0.05) & -1.42 (0.03) & 0.82 (0.0) & 0.11 && -0.20 (0.14) & -0.90 (0.08) & -0.31 (0.136) & 0.30 \\ 
           & discs           & 0.32 (0.02) & -1.50 (0.01) & 0.86 (0.0) & 0.14 && -0.35 (0.04) & -0.85 (0.02) & -0.66 (0.0) & 0.21 \\ 
 &  &  &  &  &  &  &  &  &  &  \\ 
         & centres + discs & 0.66 (0.04) & -1.5 (0.01) & 0.85 (0.0) & 0.12 &  & -0.63 (0.07) & -0.83 (0.02) & -0.60 (0.0) & 0.23 \\ 
\vdisavg & centres         & 0.51 (0.13) & -1.43 (0.04) & 0.69 (0.0) & 0.14 && -0.31 (0.27) & -0.89 (0.09) & -0.26 (0.203) & 0.31 \\ 
         & discs           & 0.64 (0.05) & -1.50 (0.01) & 0.83 (0.0) & 0.14 && -0.74 (0.08) & -0.86 (0.02) & -0.65 (0.0) & 0.20 \\ 
 &  &  &  &  &  &  &  &  &  &  \\ 
         & centres + discs & ... & ... & 0.21 (0.028) & ... &  & ... & ... & -0.11 (0.226) & ... \\ 
\aviravg & centres         & ... & ... & -0.12 (0.572) & ... && ... & ... & 0.19 (0.363) & ... \\ 
         & discs           & ... & ... & 0.25 (0.011) & ... && ... & ... & -0.23 (0.019) & ... \\ 
 &  &  &  &  &  &  &  &  &  &  \\ 
          & centres + discs & 0.17 (0.01) & -1.49 (0.01) & 0.88 (0.0) & 0.11 &  & -0.15 (0.02) & -0.83 (0.02) & -0.62 (0.0) & 0.22 \\ 
\Pturbavg & centres         & 0.15 (0.03) & -1.41 (0.03) & 0.75 (0.0) & 0.12 && -0.09 (0.07) & -0.90 (0.08) & -0.29 (0.160) & 0.31 \\ 
          & discs           & 0.16 (0.01) & -1.50 (0.01) & 0.89 (0.0) & 0.14 && -0.17 (0.02) & -0.84 (0.02) & -0.67 (0.0) & 0.20 \\ 
\hline\hline
\end{tabular}
\begin{tablenotes}
\small
\item \textbf{Notes.} Fit parameters resulting from the linear regression of HCN/CO (tracing \fdense) and SFR/HCN (tracing \sfedense) both at \SI{2.1}{\kilo\parsec} scale vs. molecular cloud properties (\sigmol, \vdis, \avir, \Pturb) at \SI{150}{\parsec} scale. Column 2 indicates the environment considered for the fit, where centre + disc means the whole galaxy as in Figure~\ref{FIG:HCN_lowres}. Centre and disc are defined as introduced in Section~\ref{SEC:centres_vs_discs} and are shown in Figure~\ref{FIG:HCN_lowres_centres_vs_discs}. Columns 3 and 4 list the slope and intercept with corresponding uncertainty estimates as determined by the linear regression tool. Column 5 shows the Pearson correlation coefficient $\rho$ and its corresponding $p$-value. Column 6 displays the $y$-axis scatter of the data about the best fit line measured in units of dex. Due to lack of correlation between HCN/CO, SFR/HCN and the virial parameter, we do not show linear regression results for \aviravg, but only list the correlation coefficients and $p$-values based on the significant data points. Note, that for the other cloud-scale properties, the correlations coefficient (and the $p$-value) are determined using both the censored and the significant data.
\item[1] Note that the intercept is measured at ca. the median of the respective cloud-scale property as described in Section~\ref{SEC:fitting_and_corr}. 
\end{tablenotes}
\end{threeparttable}}
\end{table*}

%%%%%%%%%%%%%%%%%%%%%%%%%%%%%%%%%%%%%%%%%%%%%%%%%%

We analyse the dependence of the ratios $\intHCN/\intCO$ (hereafter HCN/CO) and $\sigsfr/\intHCN$ (hereafter SFR/HCN) as a function of the cloud-scale molecular gas properties \sigmol, \vdis, \avir and \Pturb.
HCN/CO is used as a proxy for the dense gas fraction (\fdense) and SFR/HCN as a proxy for the star formation efficiency of the dense gas ($\sfedense=\text{SFR}/\mdense$), both at \SI{2.1}{\kilo\parsec} physical scale, albeit with some important caveats (see Sections~\ref{SEC:line_emissivity} and~\ref{SEC:fdense}),
The cloud-scale properties are inferred from the \cotwo measurements at a fixed physical scale of \SI{150}{\parsec}. 
We use the \cotwo intensity to trace \sigmol and the \cotwo line width to trace \vdis. 
We trace \avir and \Pturb via $\vdis^2/\sigmol$ and $\vdis^2\sigmol$, respectively (Section~\ref{SEC:cloud_scale_props}).
Figure~\ref{FIG:HCN_lowres} shows the observed relationships. The upper panels show the HCN/CO correlations with the three molecular cloud properties (\sigmol, \vdis, \avir) from left to right, which are discussed in Section~\ref{SEC:HCN_CO_correlations}. 
Similarly, the lower panels display the SFR/HCN correlations discussed in Section~\ref{SEC:SFR_HCN_correlations}.
For each relation we perform linear regression fitting to the data in logarithmic scale as described in Section ~\ref{SEC:fitting_and_corr}. Moreover, we determine the Pearson correlation and corresponding $p$-value and compute the scatter in the fit residuals.

In addition, we examine the impact of resolution in Appendix~\ref{SEC:appendix:res_configs} and find consistent results across all adopted resolutions, i.e. varying cloud-scale and kpc-scale from \SIrange{75}{150}{\parsec} and \SIrange{1.0}{2.1}{\kilo\parsec}, respectively. Moreover, we show the same relationships taking \hcopone or \cstwo as a tracer of the denser molecular gas (see Appendix~\ref{SEC:appendix:HCOP_correlations} and~\ref{SEC:appendix:CS_correlations})

%%%%%%%%%%%%%%%%%%%%%%%%%%%%%%%%%%%%%%%
\subsection{Dense Gas Star Formation Relation}
\label{SEC:dense_gas_sf_law}
%%%%%%%%%%%%%%%%%%%%%%%%%%%%%%%%%%%%%%%

In Figure \ref{FIG:dense_gas_sf_relation}, we show the relation between HCN luminosity and total infrared luminosity, measured at the native resolution of the HCN observations, as has been reported in many previous works \citep[e.g.][]{Lada2003, Gao2004, Jimenez-Donaire2019}.
We inferred the total IR (TIR) luminosity from the SFR maps using the following equation \citep{Murphy2011}:
\begin{align}
    \dfrac{\sigsfr}{\SI{}{\Msun\per\year\per\square\kilo\parsec}} = \num{1.48e-10} \, \dfrac{\Sigma_{\rm TIR}}{\SI{}{\Lsun\per\square\kilo\parsec}}
\end{align}
Overall, our HCN and SFR measurements are in agreement with previous works confirming the, to zeroth order, linear relation between HCN inferred dense gas mass and IR inferred SFR. Certainly, our data are on average $\SI{0.07}{\dex}$ lower then the mean value of $\sfedense=\SI{776}{\Lsun}(\mathrm{K}\,\mathrm{km}\,\mathrm{s}^{-1})^{-1}\,\mathrm{pc}^{-2}$ reported by \citet{Jimenez-Donaire2019} and in fact consistent with the low \sfedense found in the milky way central molecular zone (CMZ).

%%%%%%%%%%%%%%%%%%%%%%%%%%%%%%%%
\subsection{HCN/CO vs. Molecular Cloud Properties}
\label{SEC:HCN_CO_correlations}
%%%%%%%%%%%%%%%%%%%%%%%%%%%%%%%%

%%%%%%%%%%%%%%%%%%%%%%%%%%%%%%%%%%%%%%%%%%%%%%%%%%%%%%%%%%%%%
\subsubsection{HCN/CO vs. Molecular Gas Surface Mass Density}
\label{SEC:HCN_CO_vs_sigmol}
%%%%%%%%%%%%%%%%%%%%%%%%%%%%%%%%%%%%%%%%%%%%%%%%%%%%%%%%%%%%%

Assuming that cloud-scale surface density traces mean volume density, we expect a positive correlation between the surface density of the molecular cloud (\sigmol) and the HCN/CO line ratio as laid out in Section~\ref{SEC:models_HCN_CO_corr}. 
The upper left panel of Figure~\ref{FIG:HCN_lowres} shows the observed relationship between HCN/CO and \sigmolavg (significant data points and upper limits). 
The underlying red shaded region shows the model expectations which are in good agreement with the data if shifted by \SI{-1.0}{\dex} in HCN/CO. 
At lower \sigmolavg, the model produces a steeper relation than the data.
This discrepancy is expected and can be attributed to the simplified model, which does not account for systematic variations of the HCN emission as a function of the cloud density (see Section~\ref{SEC:models_HCN_CO_corr}).
Our model does not take into account the \cotwo or \hcnone light-to-mass conversion factors \aCOtwo and \aHCN, respectively. Hence, the employed shift would imply a ratio between the conversion factors of $\aHCN/\aCOtwo\sim 10$.
In agreement with the model expectations and expanding the results by \citet{Gallagher2018b}, we find a strong positive correlation between HCN/CO and \sigmolavg (see Figure~\ref{FIG:HCN_lowres}) with Pearson correlation coefficient $\rho=0.88$ ($p$-values smaller than \num{e-5}) and a linear regression slope of $m_\mathrm{f,\Sigma}=0.35\pm0.02$. 
We find small scatter of \SI{0.11}{\dex} about the fit line pointing towards a tight correlation.

For sub-samples of galaxies, where higher resolutions (i.e. \SI{120}{\parsec}, \SI{75}{\parsec} cloud-scale and \SI{1.5}{\kilo\parsec}, \SI{1.0}{\kilo\parsec} averaging-scale) can be accessed, we find comparable correlations with $\rho=$ \numrange{0.88}{0.97}, $m_\mathrm{f,\Sigma}=$ \numrange{0.35}{0.49} (see Appendix~\ref{SEC:appendix:HCN_CO_corr}). In general, we find that the derived relationship can change significantly depending on which galaxies are included in the sample. However, for a fixed sample of galaxies the correlations are consistent for different resolutions, where smaller scales seem to show steeper slopes (a more detailed discussion is found in Appendix~\ref{SEC:appendix:HCN_CO_corr}).

%%%%%%%%%%%%%%%%%%%%%%%%%%%%%%%%%%%%%%%%%%%%%%
\subsubsection{HCN/CO vs. Velocity Dispersion}
\label{SEC:HCN_CO_vs_vdis}
%%%%%%%%%%%%%%%%%%%%%%%%%%%%%%%%%%%%%%%%%%%%%%

Similar to the HCN/CO vs. \sigmol correlation, turbulent cloud models predict a positive correlation between HCN/CO and \vdis assuming the effective line width traces the turbulent Mach number (see Section~\ref{SEC:models_HCN_CO_corr}). 
Consistent with the model expectations, we report a positive correlation between HCN/CO and \vdisavg with Pearson correlation coefficient $\rho=0.85$ and small $p$-value $<\num{e-5}$. 
The regression slope is $m_\mathrm{f,\sigma}=0.66\pm0.04$ and we find small scatter of \SI{0.12}{\dex} indicating a strong and tight correlation. 
Variations in the correlation at different resolutions (see Appendix ~\ref{SEC:appendix:HCN_CO_corr}) are consistent for the same sample of galaxies and follow similar systematics as seen for HCN/CO vs. \sigmolavg which is expected due to the strong correlation between \sigmol and \vdis (see e.g. \citealp{Sun2020b,Rosolowsky2021}).

Tracing the velocity dispersion via the line width is appropriate for the discs of galaxies but may lead to biased estimates in the galactic centres \citep[e.g.][]{Henshaw2016}. 
In Section~\ref{SEC:centres_vs_discs}, we additionally inspect the correlations for the central regions (defined as the central pixel of each galaxy, i.e. the inner $\sim$ \SI{2.1}{\kilo\parsec}) and the discs separately (the fit parameters are listed in Table~\ref{TAB:HCN_corr_table_lowres}). We find that the correlations as obtained from the central regions are slightly offset by $<\SI{0.1}{\dex}$ from the correlations associated to the discs suggesting that the kpc-scale centres are not statistically distinct to the discs.

%%%%%%%%%%%%%%%%%%%%%%%%%%%%%%%%%%%%%%%%%%%
\subsubsection{HCN/CO vs. Virial Parameter}
%%%%%%%%%%%%%%%%%%%%%%%%%%%%%%%%%%%%%%%%%%%
As discussed in Section~\ref{SEC:models_HCN_CO_corr}, the connection between HCN/CO and the virial parameter is complex. In the simple KM theory, \avir does not affect the PDF and thus keeps HCN/CO unchanged. However, the empirical \avir (Equation~\eqref{EQU:virial_parameter}), which assumes a fixed cloud size, correlates with \vdis and anti-correlates with \sigmol such that, given the observed cloud-scale properties, variations in $\avir\propto\vdis^2/\sigmol$ might be correlated with HCN/CO as shown in Figure~\ref{FIG:PDF_expectations_data}.

In accordance with the model picture, we find a weak positive ($\rho=0.21$, $p=0.028$), but no significant correlation between HCN/CO and $\vdis^2/\sigmol$ tracing the virial parameter. Here, the correlation coefficient was computed using only the significant data points (i.e., where $\text{SNR}\ge 3$, hence not including censored data as for \sigmol or \vdis), because the fitting algorithm does not converge.

We consistently find positive correlation coefficients, spanning $\rho=$ \numrange{0.21}{0.77}, at different resolutions which supports a positive correlation between HCN/CO and $\vdis^2/\sigmol$, especially for individual galaxies (e.g. NGC 2903 or NGC 4321, which are also the ones with the highest S/N) and at smaller scales (\SI{75}{\parsec} cloud-scale and \SI{1.0}{\kilo\parsec} averaging-scale).
However, including the complete sample of 25 galaxies, our data do not confidently suggest any correlation between HCN/CO and $\vdis^2/\sigmol\propto\avir$.

%%%%%%%%%%%%%%%%%%%%%%%%%%%%%%%%%%%%%%%%%%%%%%%%%%%%%%
\subsubsection{HCN/CO vs. Internal Turbulent Pressure}
%%%%%%%%%%%%%%%%%%%%%%%%%%%%%%%%%%%%%%%%%%%%%%%%%%%%%%

The internal turbulent pressure, or equivalently the kinetic energy density, measures the turbulence of the gas, $\vdis^2$, weighted by the amount of molecular gas, \sigmol, so that $\Pturb\propto\sigmol\vdis^2$ (see Equation~\eqref{EQU:turbulent_pressure}). We have shown in Sections~\ref{SEC:HCN_CO_vs_sigmol} and~\ref{SEC:HCN_CO_vs_vdis} that HCN/CO positively correlates with \sigmolavg and \vdisavg. Thus, also agreeing with model predictions, we expect a positive correlation between HCN/CO and \Pturbavg.
The HCN/CO vs \Pturb relation plot is not shown in Figure~\ref{FIG:HCN_lowres} because it directly follows from and is almost identical to the \sigmol and \vdis relations. 
Though, the linear regression results are listed in Table~\ref{TAB:HCN_corr_table} and the plot is shown in the Appendix~\ref{SEC:appendix:HCN_CO_corr}.

As expected, we find a strong positive correlation between HCN/CO and \Pturbavg with correlation coefficient $\rho=0.88$ and $p<\num{e-5}$ which are very similar to the correlation coefficients found for \sigmolavg ($\rho=0.88$) and \vdisavg ($\rho=0.85$). Though, the regression slope is small ($m_\mathrm{f,P}=0.17\pm0.01$ due to the huge dynamic range in \Pturbavg spanning five orders of magnitude. The scatter in the correlation is small (\SI{0.11}{\dex}) indicating a tight correlation. Variations in the correlations as a function of resolution configurations show similar trends as for \sigmolavg (Section~\ref{SEC:HCN_CO_vs_sigmol}) and \vdisavg  (Section~\ref{SEC:HCN_CO_vs_vdis}).

%%%%%%%%%%%%%%%%%%%%%%%%%%%%%%%%%
\subsection{SFR/HCN vs. Molecular Cloud Properties}
\label{SEC:SFR_HCN_correlations}
%%%%%%%%%%%%%%%%%%%%%%%%%%%%%%%%%

%%%%%%%%%%%%%%%%%%%%%%%%%%%%%%%%%%%%%%%%%%%%%%%%%%%%%%%%%%%%%%
\subsubsection{SFR/HCN vs. Molecular Gas Surface Mass Density}
\label{SEC:SFR_HCN_vs_surface_mass_density}
%%%%%%%%%%%%%%%%%%%%%%%%%%%%%%%%%%%%%%%%%%%%%%%%%%%%%%%%%%%%%%

Based on simple turbulent models of star formation (e.g. KM theory; Section~\ref{SEC:models}) we expect a negative correlation between SFR/HCN and \sigmolavg. 
The main driver of the negative correlation is that with increasing mean density of the cloud, HCN traces more of the bulk molecular gas thus decreasing SFR/HCN (Section~\ref{SEC:models_SFR_HCN_corr}).
The lower left panel of Figure~\ref{FIG:HCN_lowres} shows the relationship between SFR/HCN and \sigmolavg. The underlying model predictions (red area) is in good agreement with the data if shifted by \SI{0.6}{\dex} in SFR/HCN.
In accordance with the model expectations, we find a negative correlation between SFR/HCN and \sigmolavg with Pearson correlation coefficient $\rho=-0.63$ and $p$-value smaller than \num{e-3}. The regression slope is $m_\mathrm{S,\Sigma}=-0.33\pm0.04$ indicating a sub-linear anti-correlation, where the scatter is \SI{0.23}{\dex}. Note, however, that the scatter is larger at higher \sigmolavg and can be up to $\sim\SI{0.5}{\dex}$ at $\sigmolavg\sim\SI{e3}{\Msun\per\square\parsec}$. In comparison with the HCN/CO correlations (Section~\ref{SEC:HCN_CO_vs_sigmol}) the SFR/HCN correlation with \sigmolavg is weaker, but still significant. Furthermore, the scatter is roughly twice as large compared to the HCN/CO realtion as also indicated by the the model. The stronger scatter can be explained by the non-monotonic relation between SFR/HCN and \nmean. We find consistent results among different resolutions (for fixed galaxy sample) with the same trend of steeper correlation at smaller scales (see Appendix~\ref{SEC:appendix:SFR_HCN_corr} for more details).

%%%%%%%%%%%%%%%%%%%%%%%%%%%%%%%%%%%%%%%%%%%%%%%
\subsubsection{SFR/HCN vs. Velocity Dispersion}
\label{SEC:SFR_HCN_vs_vdis}
%%%%%%%%%%%%%%%%%%%%%%%%%%%%%%%%%%%%%%%%%%%%%%%

As described in Section~\ref{SEC:models_SFR_HCN_corr}, turbulent cloud models can predict a negative correlation between SFR/HCN and the turbulence of the molecular gas due to the widening of the density PDF resulting in a lower SFR/HCN ratio. 
We find a negative correlation between SFR/HCN and \vdisavg with Pearson correlation coefficient $\rho=-0.60$ and $p$-value smaller than \num{e-3}. We report a regression slope of $m_\mathrm{S,\sigma}=-0.63\pm0.07$ with moderate scatter \SI{0.23}{\dex}. Similar to the \sigmolavg relation, the scatter is larger at higher \vdisavg. 

The correlation coefficients are very similar to the ones found for SFR/HCN vs. \sigmolavg, as expected due to the strong correlation between \sigmol and \vdis. The measured correlations vary with resolution and sample, where the steepness of the correlation tends to increase with the resolution, i.e. with decreasing physical scale (see Appendix~\ref{SEC:appendix:SFR_HCN_corr}).

%%%%%%%%%%%%%%%%%%%%%%%%%%%%%%%%%%%%%%%%%%%%
\subsubsection{SFR/HCN vs. Virial Parameter}
%%%%%%%%%%%%%%%%%%%%%%%%%%%%%%%%%%%%%%%%%%%%

Naively, one could expect that a cloud with lower virial parameter and thus higher gravitational boundedness could form stars more efficiently, suggesting an anti-correlation between SFR/HCN and \avir. 
Moreover, assuming \avir to have only little effect on the PDF, based on Equation~\eqref{EQU:PDF_nSF}, increasing \avir would shift the star formation density threshold (\nSF) to higher densities hence decreasing SFR. 
In this consideration, we would expect an anti-correlation between SFR/HCN and \avir. 
In the model description adopted here (Section~\ref{SEC:expectations}), it is less obvious to explore the effect of \avir on the log-normal PDF, the SFR and hence SFR/HCN.
Yet, by assuming that \avir traces $\vdis^2/\sigmol$, we explore variations of HCN/CO as a function of empirically based $\vdis^2/\sigmol$ values (red area in Figure~\ref{FIG:HCN_lowres}) and predicted a small positive correlation ($\rho=-0.68$) with significant scatter.

In our data we find no correlation ($\rho=-0.11$ , $p=0.226$) between SFR/HCN and $\langle\vdis^2/\sigmol\rangle$ tracing \aviravg, suggesting that SFR/HCN and \aviravg are uncorrelated. However, for the sub-sample that includes the five closes galaxies, we find $\rho=-0.53$ and $p=0.003$ indicating a moderate negative correlation accordance with the model predictions at least for some galaxies (Appendix~\ref{SEC:appendix:SFR_HCN_corr}).

%%%%%%%%%%%%%%%%%%%%%%%%%%%%%%%%%%%%%%%%%%%%%%%%%%%%%%%
\subsubsection{SFR/HCN vs. Internal Turbulent Pressure}
%%%%%%%%%%%%%%%%%%%%%%%%%%%%%%%%%%%%%%%%%%%%%%%%%%%%%%%

Following the same reasoning as in Section~\ref{SEC:HCN_CO_correlations}, the effect of the turbulent pressure (\Pturb) on SFR/HCN can be inferred from the expected correlations of SFR/HCN with \sigmol and \vdis, using $\Pturb\propto\sigmol\vdis^2$. Hence, we expect a negative correlation between SFR/HCN and \Pturbavg due to the negative correlation of SFR/HCN with both \sigmolavg and \vdisavg.
We report a negative correlation finding a Pearson correlation coefficient of $\rho=-0.62$ with $p$-value  $<\num{e-3}$. Due to the huge dynamic range of \Pturbavg the regression slope is shallow ($m_\mathrm{S,P}=-0.15\pm0.02$. The scatter about the fit line is \SI{0.22}{\dex} very similar to the scatter seen in the \sigmolavg and \vdisavg relations.
Similar to the SFR/HCN vs. \sigmolavg and SFR/HCN vs. \vdisavg correlations, we find a steeper correlation with increasing resolution, but consistent results among the same sample of galaxies (Appendix~\ref{SEC:appendix:SFR_HCN_corr}).

%%%%%%%%%%%%%%%%%%%%%%%%%%%%%%%%%%%%%%%%%%%%%%%%%%%%%%%
\subsection{\texorpdfstring{HCO$^+$}{HCO+} and CS}
%%%%%%%%%%%%%%%%%%%%%%%%%%%%%%%%%%%%%%%%%%%%%%%%%%%%%%%

Analogously to \hcnone (Sections \ref{SEC:HCN_CO_correlations} and~\ref{SEC:SFR_HCN_correlations}), we perform the same analysis using \hcopone as well as \cstwo as a tracer of the denser molecular medium. These molecular lines have expected excitation densities comparable to \hcnone.
Thereofore, we expect to find similar (anti-) correlations. 
Accordingly, we study how HCO$^+$/CO, CS/CO and SFR/HCO$^+$, SFR/CS vary with the cloud-scale molecular gas properties. The detailed results are shown in the Appendix \ref{SEC:appendix:HCOP_correlations} and~\ref{SEC:appendix:CS_correlations}. 

We find that both HCO$^+$/CO and CS/CO positively correlate with \sigmolavg, \vdisavg and \Pturbavg with Pearson correlation coefficients $\sim 0.8$ and negligible $p$-values $<\num{e-5}$. In general, we find very similar slopes for the HCO$^+$/CO and CS/CO relations as for the HCN/CO relations showing that HCN, HCO$^+$ and CS are likewise sensitive to variations of the cloud-scale molecular gas properties. The scatter in the HCO$^+$/CO data is slightly larger which can be explained by the slightly larger HCO$^+$ measurement uncertainties. The CS/CO relations are shifted to lower values due to the lower CS brightness compared to HCN or HCO$^+$. We also observe larger scatter due to the larger CS measurement uncertainties. These results show, that not only HCN/CO, but also HCO$^+$/CO and CS/CO at kpc-scale are good proxies of the average density structure of the molecular gas.

As for SFR/HCN, we find that both SFR/HCO$^+$ and SFR/CS anti-correlate with \sigmolavg, \vdisavg and \Pturbavg with $\rho\sim 0.5$ ($p<\num{e-3}$). This suggests that HCN, HCO$^+$ and CS are a similarly tracing the star forming gas and that the ratios with SFR are likewise affected by variations of the cloud-scale molecular gas properties.
%%%%%%%%%%%%%%%%%%%
\section{Correlation with Local Environment}
\label{SEC:analysis}
%%%%%%%%%%%%%%%%%%%

In the following, we study how the observed correlations may depend on the environment of the galaxies, where we separate the central kpc-scale regions from the discs (Section~\ref{SEC:centres_vs_discs}). 
%In the Appendix, we analyse the ratio of the internal and external pressure of the molecular gas at cloud-scales as a potential candidate for explaining the intrinsic scatter in the correlations (Section \ref{SEC:appendix:pressure}). 
We perform the analysis focusing on the same resolution configuration, i.e. \SI{150}{\parsec} cloud-scale and \SI{2.1}{\kilo\parsec} kpc-scale, as in Section~\ref{SEC:results}.

%%%%%%%%%%%%%%%%%%%%%%%%%%%%%%%%%%%%%%%%%%%%
% PLOT (Correlations for centres vs. Discs)

\begin{figure*}
    \includegraphics[width=\textwidth]{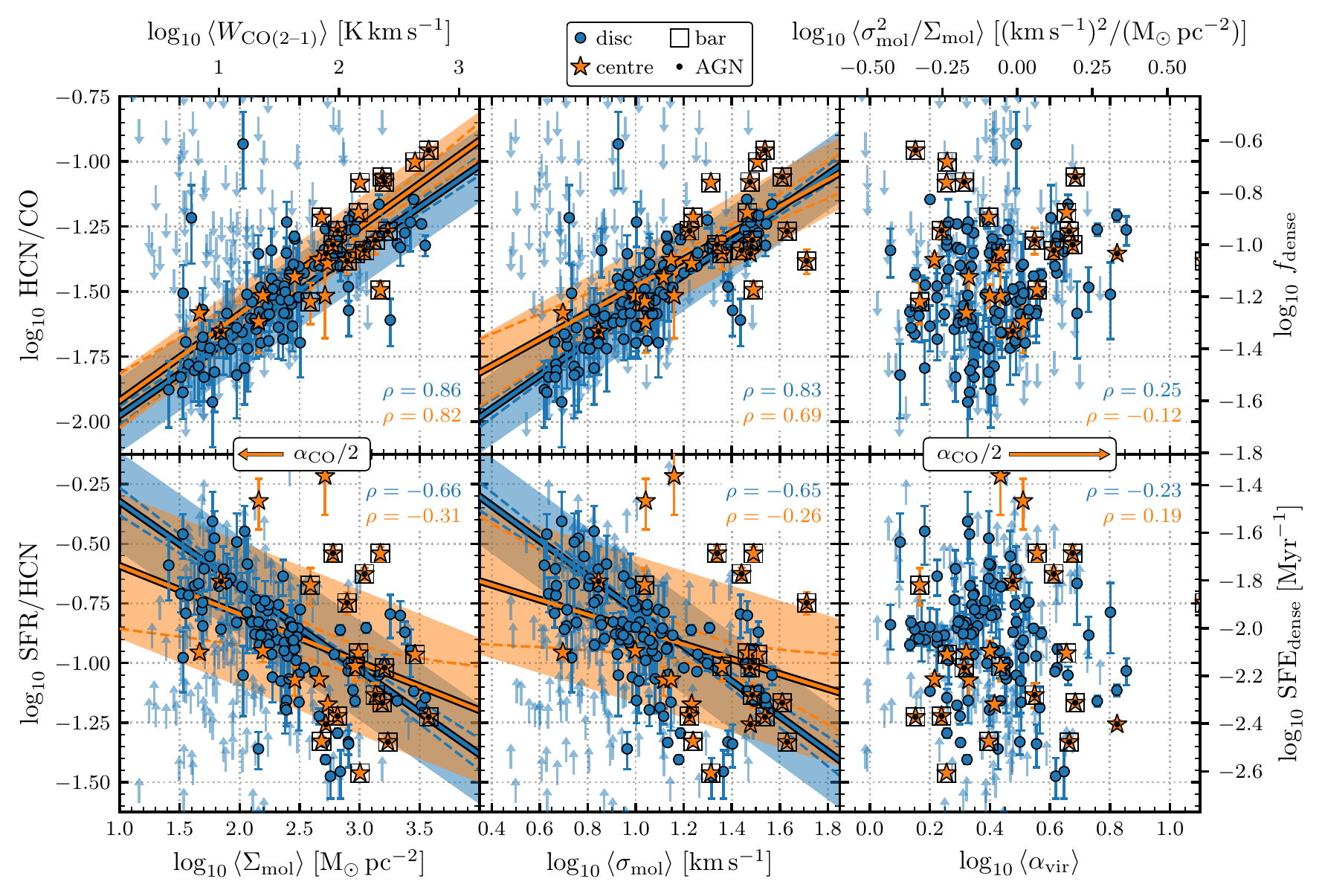}
    \caption{\textbf{HCN/CO vs. \Xavg and SFR/HCN vs. \Xavg (cloud-scale = \textbf{150}$\,$\textbf{pc}, kpc-scale = \textbf{2.1}$\,$\textbf{kpc})} HCN/CO \textit{(top)} and SFR/HCN \textit{(bottom)} against molecular cloud properties - \sigmolavg, \vdisavg, \aviravg - from left to right, separately fitted for galaxy discs (blue circles) and central regions (orange stars). The shaded ares indicate the scatter and the dotted lines the $1\,\sigma$ credibility areas of the linear regression realisations (see Table~\ref{TAB:HCN_corr_table_lowres} for details on the fit results). The central regions are taken as the single pixel at the galaxy centre, i.e. the inner \SI{2.1}{\kilo\parsec}. The remaining data points is referred to as "disc" and processed as in Section~\ref{SEC:binning}. For the central regions we indicate if the galaxies are barred (black squares) and/or contain an AGN (black circle). The length of the orange arrow labelled with $\aCO/2$ indicates the shift of the data points in the \sigmol and \avir plots if \aCO would decrease by a factor of two.}
    \label{FIG:HCN_lowres_centres_vs_discs}
\end{figure*}

%%%%%%%%%%%%%%%%%%%%%%%%%%%%%%%%%%%%%%%
\subsection{Central Regions vs. Discs}
\label{SEC:centres_vs_discs}
%%%%%%%%%%%%%%%%%%%%%%%%%%%%%%%%%%%%%%%

The central regions of galaxies (M51, \citealp{Querejeta2019}; NGC 253, \citealp{Jiang2020}; NGC 6946, \citealp{Eibensteiner2022}) as well as the galactic CMZ \citep{Longmore2013,Kruijssen2014b,Barnes2017} are typically much denser and less efficient at forming stars from the dense gas, making them a particularly interesting environment to study as they form an interesting contrast to the discs. Therefore, we study the same relations as in Section~\ref{SEC:results} separately for the central regions and the discs. We define the central region (also referred to as "centre" throughout this section) as the single kpc-scale (i.e. \SI{2.1}{\kilo\parsec}) pixel at the centre of each galaxy. Note that the physical size of the galaxy centres are typically a factor of $\sim 3$ smaller (median size of the centre, i.e. small bulge or nucleus of the PHANGS galaxies is $\sim\SI{600}{\parsec}$, \citealp{Querejeta2021b}) compared to the \SIrange{1}{2}{\kilo\parsec} size adopted here. Therefore, we may underestimate the difference between the centres and the discs in our analysis. Given that we are sampling the maps with one sample per beam, the centre is one single pixel and consequently we do not bin the centres data. For the remaining pixels (i.e. all pixels except the centre) we perform the binning procedure as described in Section~\ref{SEC:binning}, but use 18 instead of 20 bins which results in similar bin sizes for the discs data compared to the binning of the full data. Finally, we separately fit linear functions to the data for the discs and the centres, analogous to the procedure used in Section~\ref{SEC:results} (see Figure~\ref{FIG:HCN_lowres_centres_vs_discs}).

In agreement with other studies, we find that, on average, centres appear to have higher HCN/CO by about \SI{0.17}{\dex} (KS $p$-value\footnote{The two-sample Kolmogorov-Smirnov (KS) test quantifies the significance of the difference between the distributions of two samples \citep{Hodges1958}. Here, we test the probability $p_\text{KS}$ again the null hypothesis that e.g. centres have lower HCN/CO than discs.}: $p_\text{KS}=0.001$) and lower SFR/HCN by about \SI{0.14}{\dex} ($p_\text{KS}=0.011$) across our sample of 25 nearby galaxies (see Figure\,\ref{FIG:centres_vs_discs_histograms}). 
Nonetheless, centres also have higher \sigmol and \vdis, and, hence, are found to follow similar HCN/CO and SFR/HCN relations as are observed in the discs; i.e. in agreement with the model expectations and the correlations found in Section~\ref{SEC:results}. 
This suggests that the physical connection between molecular cloud properties, density distribution and star formation is, to first order and on kpc-scales, valid independent of the local environment. 

In detail, the HCN/CO against \sigmolavg or \vdisavg relations show very similar linear regression slopes for the centres compared to the discs (Figure~\ref{FIG:HCN_lowres_centres_vs_discs}). 
It is worth noting that we do see a minor offset between HCN/CO vs \sigmolavg for centres and discs of about $\sim\SI{0.1}{\dex}$ (measured as the difference in the intercepts of the fit lines at $\sigmol=\SI{e2.5}{\Msun\per\square\parsec}$).
% the HCN/CO vs \sigmolavg fit line of the centres is offset from the fit line of the discs by about $\sim\SI{0.1}{\dex}$ if measured as the difference in the intercepts of the fit lines at $\sigmol=\SI{e2.5}{\Msun\per\square\parsec}$. 
On the one hand, this may suggest that there are other physical parameters at play which systematically affect \fdense and hence HCN/CO at fixed \sigmolavg and \vdisavg. These parameters could be connected to the galaxy's environment such as the dynamical equilibrium pressure or shear \citep[see e.g.][]{Federrath2016,Kruijssen2019}. On the other hand, offsets in \fdense or \sigmol may be connected to systematic variations of the \aCO and \aHCN conversion factors (see Section~\ref{SEC:conversion_factors}). Overall, although the centres are slightly (to within $1-2\,\sigma$) offset to higher HCN/CO values, they follow the same trends with the cloud-scale molecular gas properties. Thus, also in the centres, HCN/CO appears to be a good first-order tracer of mean molecular gas density.

For the SFR/HCN correlations we do not find a significant offset between the centres and the discs as is observed for the HCN/CO correlations. However, we find a flatter slope and significantly larger scatter for the centres ($\sim\SI{0.3}{\dex}$) compared to the discs ($\sim\SI{0.2}{\dex}$), especially at high \sigmolavg or \vdisavg. This increasing scatter is also seen in the model predictions (Figure~\ref{FIG:PDF_expectations_data}) and is caused by the decrease of the free-fall time at large cloud densities which results in an increase of SFR/HCN at large \sigmol. Therefore, the KM theory can predict both a lower and a higher SFR/HCN in the centres of galaxies depending on the turbulence of the molecular clouds. Certainly, there are alternative explanations for large variations of SFR and \sfedense in galaxy centres. One idea is that star formation in galaxy centres is episodic due to stellar feedback cycles \citep[e.g.][]{Krumholz2015}. In addition, the accretion of dense gas to the galaxy centre may vary, leading to SFR fluctuations \citep[][]{Seo2019,Sormani2020b,Moon2022}.

%%%%%%%%%%%%%%%%%%%%%%%%%%%%%%%%%%%%%%%
\subsection{Impact of Bars and AGN}
\label{SEC:bar_agn}
%%%%%%%%%%%%%%%%%%%%%%%%%%%%%%%%%%%%%%%

\begin{figure}
    \includegraphics{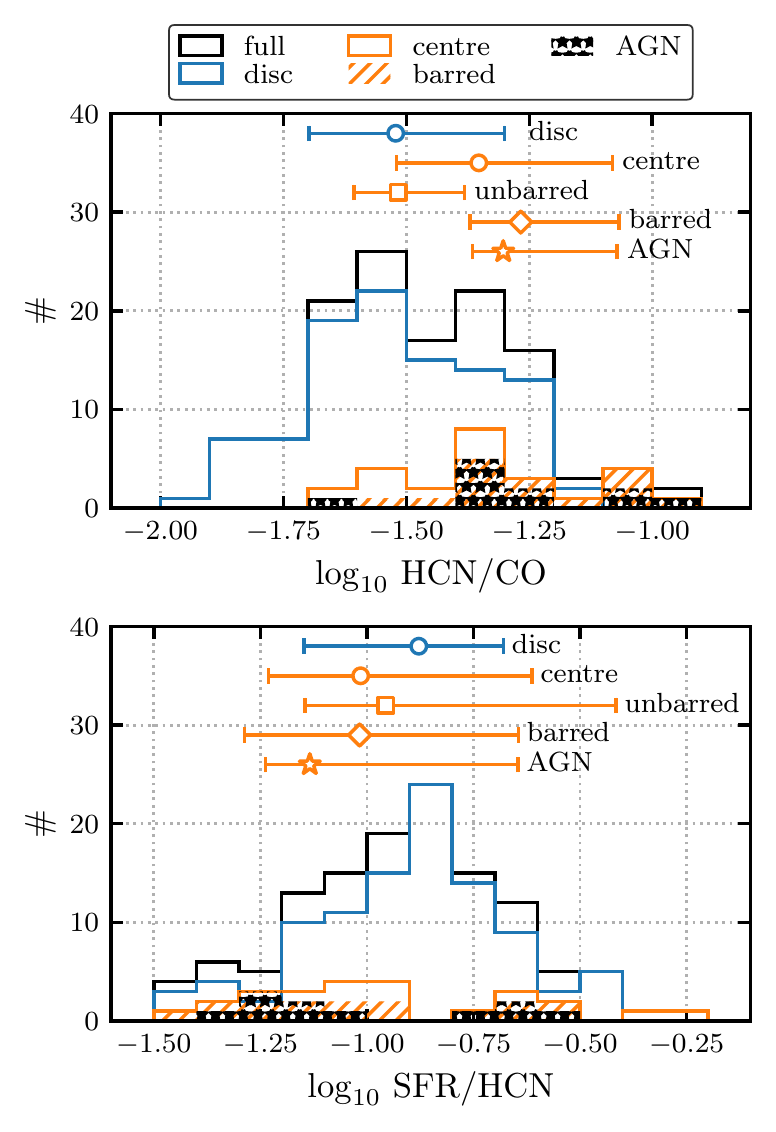}
    \caption{Histograms of HCN/CO \textit{(top)} and SFR/HCN \textit{(bottom)} at \SI{2.1}{\kilo\parsec} scale in different environments. The full data is shown in black. Centres and discs are colored in orange and blue in analogy with Figure~\ref{FIG:HCN_lowres_centres_vs_discs}. In addition, for the centres, we denote galaxies with a bar (diagonal hatching) or and AGN (starred hatching). The markers and lines above the histogram indicate the median and 16 to 84 percentiles of the respective data.}
    \label{FIG:centres_vs_discs_histograms}
\end{figure}

In addition to separating the centre from the disc, we want to study the impact of a bar or an AGN on the kpc-scale dense gas quantities in the centres of galaxies (the classifications are listed in Table~\ref{TAB:galaxy_sample}). 
\citet{Sun2020b} analysed the molecular gas properties at \SI{150}{\parsec}-scale in a larger sample of 70 PHANGS galaxies and found that gas in centres of barred galaxies have higher surface density \sigmol and velocity dispersion \vdis compared to gas in centres of unbarred galaxies (as noted above, the defined sizes of the centres in \citet{Sun2020b} are typically smaller than the central regions studied here). 
In this work, we also find that centres of barred galaxies tend to show higher HCN/CO by about \SI{0.25}{\dex} ($p_\text{KS}=0.0002$) (see Figures \ref{FIG:HCN_lowres_centres_vs_discs} and~\ref{FIG:centres_vs_discs_histograms}). 
SFR/HCN is only insignificantly lower ($p_\text{KS}=0.436$) in barred galaxies by about \SI{0.06}{\dex}.
Moreover, we find that molecular gas in centres of unbarred galaxies is similar in terms of HCN/CO and SFR/HCN to the values found in discs (a result reported for the molecular cloud properties by \citet{Sun2018}).

Moreover, we examine how an AGN may affect the (dense) molecular gas in the central region of galaxies. Our sample contains eleven AGN galaxies (14 without AGN). Note that there is a significant overlap between AGN and barred galaxies, so we cannot easily discriminate the impact of bars and AGN. 
On average, the AGN seems to boost HCN/CO in the centres of galaxies. We find \SI{0.12}{\dex} higher median HCN/CO ($p_\text{KS}=0.040$) in the centres of AGN galaxies compared to the centres which do not harbour an AGN. 
These results suggest that centres of AGN galaxies have higher molecular gas surface densities and turbulence, which, following the correlations found in this work, lead to higher HCN/CO. 
It is less clear how AGNs affect SFR/HCN, which is only insignificantly ($p_\text{KS}=0.208$) lower by \SI{0.17}{\dex}. 
Also, in some AGN galaxies, we observe higher SFR/HCN in the central regions. 
This could be explained by the increase of SFR/HCN at very high \sigmol as seen in models, or point at more complex gas dynamics in centres.

%%%%%%%%%%%%%%%%%%%%%%%%%%%%%%%%%%%%%%%%%%%%
% PLOT (Correlations for centres vs. Discs)

%%%%%%%%%%%%%%%%%%%%%%%%%%%%%%%%%%%%%%%
\subsection{Conversion Factors}
\label{SEC:conversion_factors}
%%%%%%%%%%%%%%%%%%%%%%%%%%%%%%%%%%%%%%%

In Section~\ref{SEC:molecular_gas_surface_density} we discussed how the CO-to-\htwo conversion factor \aCO can vary with local environment. Most notable, \aCO has been observed to be lower in the centres of galaxies compared to the disc which is linked to the high surface density, turbulence and temperature yielding a brighter CO emission \citep[see e.g. ][]{Watanabe2011,Shetty2011a,Shetty2011b,Papadopoulos2012,Bolatto2013,Sandstrom2013,Israel2020,Teng2022}. We note that \aCO can be $3-10$ times lower in galaxy centres compared to the default MW value that is also adopted here \citep{Israel2020}. \citet{Sandstrom2013} report a factor of $\sim 2$ lower \aCO in the central kpc regions compared to the average disc value in their sample of 26 nearby, star-forming galaxies. Therefore, we indicate how lowering \aCO by a factor of two in the central regions of galaxies affects the data points in the correlations studied here. In the first instance, we only consider changes in the cloud-scale properties and leave the $y$-axis coordinate unchanged. As denoted by the orange arrows in Figure~\ref{FIG:HCN_lowres_centres_vs_discs}), reducing \aCO by \SI{0.3}{\dex} decreases \sigmol and increases \avir by \SI{0.3}{\dex}. This has only little effect on the correlations with \sigmolavg, but would increase the offset in the correlations between the centres and the discs. Though, it would slightly increase the strength of the correlations with \aviravg due to making the clouds in the centres much less bound, such that we would find $\rho=0.33$ for HCN/CO vs \aviravg and $\rho=-0.12$ for SFR/HCN vs \aviravg. However, if we would account for the power-law extension of the log-normal PDF, bound clouds would always have higher dense gas fraction thus counteracting the shift to higher virial parameter values for the centres. In the end, variations with \avir remain complex and we cannot infer a clear conclusion whether HCN/CO or SFR/HCN varies significantly with \avir.

We investigate how decreasing \aCO for the centres may affect the $y$-axis coordinates if HCN/CO and SFR/HCN are converted to \fdense and \sfedense, respectively. In contrast to \aCO, there is very little information on environmental variations of \aHCN in the literature. We could assume that \aHCN varies similarly as \aCO, which might be justified because \aHCN becomes optically thick towards centres \citep[see e.g.][]{Jimenez-Donaire2019} yielding stronger HCN emission. Then, \fdense would be unaffected, while \sfedense would increase in the centres of galaxies thus decreasing the correlation with the cloud-scale molecular as properties. We could also assume that \aCO varies much more than \aHCN and thus neglect \aHCN variations. In this case, \sfedense would be unaffected, while \fdense would increase in the centres of galaxies which increases the observed correlation, but also significantly enhances the offset between centres and galaxies. The center-disc offset could only be dissolved if \aHCN is lowered even more than \aCO in the centres thus yielding a lower \fdense. 

Overall, variations of \aCO and \aHCN will eventually change the slope and strength of the correlations, but only at the \SI{0.3}{\dex} level, which is not sufficient to change the direction of the relations. Primarily, the correlations are driven by the discs, which are much less affected by variations of the conversion factors than the centres. We thus, highlight that our findings show significant systematic variations of HCN/CO and SFR/HCN with cloud-scale gas properties. Independent of whether HCN/CO and SFR/HCN can be accurately translated to \fdense and \sfedense, respectively, they are very useful tools to trace the mean density structure of molecular gas. 
%%%%%%%%%%%%%%%%%%%
\section{Summary \& Discussion}
\label{SEC:summary}
%%%%%%%%%%%%%%%%%%%

In this work, we investigate the connection of the density-sensitive kpc-scale (\SI{2.1}{\kilo\parsec}) HCN/CO and SFR/HCN ratios with various structural and dynamical properties (\sigmol, \vdis, \avir, \Pturb) of the cloud-scale (\SI{150}{\parsec}) molecular gas across 25 nearby galaxies.
In the literature, HCN/CO and SFR/HCN are often synonymous with the dense gas fraction and dense gas star formation efficiency, respectively.
This is based on the assumption that CO and HCN emission is originating from molecular gas differing within different (often fixed) density regimes.
However, observations \citep[e.g.][]{Pety2017,Kauffmann2017,Barnes2020,Evans2020} and simulations \citep[e.g.][]{Shirley2015,Mangum2015,Leroy2017,Onus2018,Jones2021} highlight that there is still a significant uncertainty in the exact density thresholds and their mass conversion factors.
In this study, we focus on the quantities HCN/CO and SFR/HCN and are careful to draw conclusions from the less certain physical quantities, i.e. \fdense and \sfedense.

In Section~\ref{SEC:expectations} we lay out qualitative predictions about the direction of the studied correlations based on single free-fall time turbulent cloud models (e.g. the KM theory; \citealp{Krumholz2005}). We find that molecular cloud properties affect the density distribution of the molecular gas such that, within this simplified model description, HCN/CO is expected to correlate and SFR/HCN to anti-correlate with molecular cloud properties like the mean density, traced by the surface density, or the Mach number, traced by the velocity dispersion of the molecular gas. The underlying physical mechanisms are that the mean density shifts the density PDF, while the Mach number affects the width of the PDF which in return affects the line emissivity of molecular lines like \cotwo and \hcnone as well as the star formation rate.

We compare the cloud scale properties to the kpc-scale HCN/CO and SFR/HCN via intensity-weighted averaging (Section~\ref{SEC:weighted_avg}). To quantitatively analyse the correlations, we fit a linear regression model to the data in log-log scale in order to determine a first order power-law dependence. We measure the strength of the correlation by computing the Pearson correlation coefficient and the corresponding $p$-value (Section~\ref{SEC:results}). Moreover, we study the correlation with local environment by separately analysing the central kpc-scale regions to contrast with the discs (Section \ref{SEC:analysis}). In the following we summarise and interpret our main findings:

\begin{enumerate}[labelindent=*,style=multiline,leftmargin=*,label=\arabic*., itemsep=8pt]

    \item We report systematic variations of HCN/CO with cloud-scale molecular gas properties (Figure~\ref{FIG:HCN_lowres} and Section~\ref{SEC:HCN_CO_correlations}). Building up on the works of \citet{Gallagher2018a,Gallagher2018b}, we find a strong positive correlation ($\rho\approx 0.9$) between HCN/CO and the cloud-scale surface density \sigmol as traced by the \cotwo line intensity adopting a fixed line-to-mass conversion factor $\aCO=4.3\, \text{M}_\odot \,\text{pc}^{-2} \,(\text{K}\,\text{km}\,\text{s}^{-1})^{-1}$ \citep{Bolatto2013} and a fixed \cotwo-to-\coone line ratio $R_{21}=0.64$ \citep{denBrok2021}. The results are in agreement with the model predictions, where the mean density (assumed to be traced by \sigmol) affects the median of the density PDF without altering its shape such that higher \nmean leads to higher HCN/CO. This is a powerful indication that both HCN/CO and cloud-scale CO trace density. Moreover, we observe a strong positive correlation ($\rho\approx 0.9$) between HCN/CO and the cloud-scale velocity dispersion as traced by the \cotwo line width in agreement with our simplified model, in which the Mach number (traced by \vdis) affects the width of the density PDF such that higher \Mach leads to higher HCN/CO. These correlations also imply that HCN/CO positively correlates with the cloud-scale internal turbulent pressure as traced via $\Pturb\propto\sigmol\vdis^2$. Furthermore, we find a weak ($\rho\approx 0.2$, $p$-value $< 0.03$) positive correlation between HCN/CO and the virial parameter as measured via $\avir\propto\vdis^2/\sigmol$ which is supported by models if \nmean and \Mach are traced by the cloud-scale CO intensity and line width, respectively.
    
    \item We report that SFR/HCN systematically varies with cloud-scale molecular gas properties (Figure~\ref{FIG:HCN_lowres} and Section~\ref{SEC:SFR_HCN_correlations}) finding a negative correlation ($\rho\approx 0.6$) between SFR/HCN and the cloud-scale \sigmol and \vdis. These results are in agreement with turbulent cloud models, in which stars are assumed to form from the dense gas above some threshold density $\nSF\propto\nmean\avir\Mach^2$. Our findings show that, although SFR linearly correlates with HCN over several orders of magnitude, SFR/HCN varies systematically as a function of the cloud-scale molecular gas properties, thus disclaiming the constant \sfedense hypothesis put forward by \citet{Gao2004}. Extending the works of \citet{Longmore2013,Kruijssen2014b,Bigiel2016,Barnes2017,Gallagher2018a,Gallagher2018b,Jimenez-Donaire2019,Querejeta2019,Jiang2020,Eibensteiner2022} who showed that the amount of dense gas is not enough to set the star formation rate, we conclude that SFR/HCN is significantly affected by the density distribution of molecular clouds which, based on turbulent cloud models, affects both the emissivity of dense gas tracers like HCN and the star formation rate and hence SFR/HCN. Moreover, we find no universal evidence for a correlation between SFR/HCN and $\vdis^2/\sigmol$ tracing \avir ($\rho\approx -0.1$, $p$-value $\sim 0.2$). For some galaxies (e.g. NGC 2903) we find indications of a negative correlation between SFR/HCN and \avir ($\rho\approx -0.5$, $p$-value $< 0.01$) This trend is supported by the model predictions (Figure~\ref{FIG:PDF_expectations_data}) and would point towards less bound clouds being less efficient in forming stars from a fixed fraction of dense gas.
    
    \item Using HCO$^+$ or CS as a tracer of the dense molecular gas, we find the same correlations with the cloud-scale molecular gas properties as seen with HCN. This is a powerful indicator that not only HCN, but also other tracers with critical densities in excess of that of low-J CO lines like HCO$^+$ or CS, observed at kpc-scale, are sensitive to the density structure of the cloud-scale molecular gas. 
    
    \item Separating the central $\sim\SI{}{\kilo\parsec}$ regions from the rest of the galaxy discs. 
    We find that centres have significantly higher HCN/CO and lower SFR/HCN compared to discs (Figure~\ref{FIG:centres_vs_discs_histograms} and Section~\ref{SEC:centres_vs_discs}).
    Nonetheless, both environments follow similar HCN/CO and SFR/HCN trends against the cloud-scale properties (Figure~\ref{FIG:HCN_lowres_centres_vs_discs}). 
    This suggests that the physical connection between molecular cloud properties, density distribution and star formation is independent of the local environment and extends from low density, less turbulent clouds as predominantly found in the disc to high density and turbulent clouds as found in the centres of galaxies.
    We also studied the impact of bars and AGN on the central regions of galaxies, finding typically higher HCN/CO and lower SFR/HCN for barred and AGN galaxies compared to their complements (unbarred and without AGN), respectively. This suggest that bars and AGNs boost HCN/CO and lower SFR/HCN in the centres of galaxies. Differences are though small $\sim$ \SIrange{0.1}{0.2}{\dex} and only significant for HCN/CO.
    Throughout this work we assumed a constant \aCO conversion factor.
    We study whether these scaling relations change when we assume that centres have systematically lower \aCO than discs which has been reported in the literature \citep{Sandstrom2013}.
    Adopting $\aCO/2$ for the central regions, we find no significant effect on either the HCN/CO or the SFR/HCN relations with the cloud-scale properties.

\end{enumerate}

Our findings demonstrate that density, cloud-scale molecular gas properties and star formation appear interrelated in a coherent way and one that agrees reasonably well with current models. 
Our results also strongly reinforce the view that HCN/CO and similar line ratios (e.g. HCO$^+$/CO or CS/CO) are sensitive measures of the density distribution of the molecular gas and thus powerful tools in extragalatic studies. 
Regardless of physical interpretation, we observe clear correlations between molecular cloud properties and line ratios sampling different physical densities. 
These should represent significant observational constraints on any theory attempting to relate star formation, gas density, and the ISM in galaxies.
Many previous studies \citep[e.g.][]{Chin1997,Chin1998, Gao2004, Brouillet2005, Lada2010, Wu2010, Rosolowsky2011, Garcia-Burillo2012, Buchbender2013, Longmore2013, Kepley2014, Chen2015, Usero2015, Bigiel2016, Chen2017, Shimajiri2017, Gallagher2018a, Jimenez-Donaire2019, Beslic2021} show that HCN luminosity (tracing dense gas mass) and SFR are strongly correlated probing scales ranging from nearby galactic cloud to entire galaxy spanning $\sim 8$ orders of magnitude. 
Therefore, \citet{Shimajiri2017} propose a quasi-universal \sfedense. 
Our results support this picture. 
However, all previous works as well as our results show a $\sim\SI{1}{\dex}$ scatter in \sfedense. 
Here, we show that this scatter is not random, but that SFR/HCN correlates with the properties of the molecular gas, i.e. \sigmol and \vdis, at \SI{150}{\parsec} scale.
It is still much of an open question what drives \sfedense in galaxy centres, where we observe typically lower \sfedense but also large scatter. 
Ultimately, we need high resolution (cloud-scale), high sensitivity spectroscopic mapping of a large sample of galaxies in order to resolve and study the effect of local environment on the dense molecular gas and star formation. This work also motivates to further investigate how spiral arms, bars and AGN may affect the density distribution of molecular gas in galaxy centres.

\section*{Acknowledgements}
We would like to thank the referee for their constructive feedback that
helped improve the paper.
This work was carried out as part of the PHANGS collaboration.
ATB, JP, JdB and FB would like to acknowledge funding from the European Research Council (ERC) under the European Union’s Horizon 2020 research and innovation programme (grant agreement No.726384/Empire).
The work of AKL and MJG on the early parts of this work was partially supported by the National Science Foundation under Grants No.~1615105, 1615109,and 1653300. 
MC gratefully acknowledges funding from the Deutsche Forschungsgemeinschaft (DFG) through an Emmy Noether Research Group (grant number CH2137/1-1). COOL Research DAO is a Decentralized Autonomous Organization supporting research in astrophysics aimed at uncovering our cosmic origins.
MC and JMDK gratefully acknowledge funding from the Deutsche Forschungsgemeinschaft (DFG) through an Emmy Noether Research Group (grant number KR4801/1-1) and the DFG Sachbeihilfe (grant number KR4801/2-1), as well as from the European Research Council (ERC) under the European Union's Horizon 2020 research and innovation programme via the ERC Starting Grant MUSTANG (grant agreement number 714907).
CE gratefully acknowledges funding from the Deutsche Forschungsgemeinschaft (DFG) Sachbeihilfe, grant number BI1546/3-1.
RSK and SCOG acknowledge  support  from  the  Deutsche  Forschungsgemeinschaft (DFG) in the Collaborative Research Centre (SFB 881, ID 138713538)  ``The Milky Way System'' (subprojects A1, B1, B2, and B8) and from the Heidelberg Cluster of Excellence (EXC 2181, ID 390900948) ``STRUCTURES: A unifying approach to emergent phenomena in the physical world, mathematics, and complex data'', funded by the German Excellence Strategy. RSK also thanks for funding form the European Research Council in the ERC Synergy Grant ``ECOGAL -- Understanding our Galactic ecosystem: From the disk of the Milky Way to the formation sites of stars and planets'' (ID 855130). RSK and SCOG also benefit from computing resources provided by the State of Baden-W\"urttemberg through bwHPC and DFG through grant INST 35/1134-1 FUGG, and from the data storage facility SDS@hd supported through grant INST 35/1314-1 FUGG, and they thank resources provided by the Leibniz Computing Centre (LRZ) for project pr74nu.
MQ acknowledges support from the Spanish grant PID2019-106027GA-C44, funded by MCIN/AEI/10.13039/501100011033.
ER acknowledges the support of the Natural Sciences and Engineering Research Council of Canada (NSERC), funding reference number RGPIN-2022-03499.
The work of JS is partially supported by the Natural Sciences and Engineering Research Council of Canada (NSERC) through the Canadian Institute for Theoretical Astrophysics (CITA) National Fellowship.
Y.-H.T. acknowledges funding support from NRAO Student Observing Support Grant SOSPADA-012 and from the National Science Foundation (NSF) under grant No. 2108081.
MCS acknowledges financial support from the European Research Council via the ERC Synergy Grant ``ECOGAL - Understanding our Galactic ecosystem: from the disk of the Milky Way to the formation sites of stars and planets'' (grant 855130).
TGW and ES acknowledge funding from the European Research Council (ERC) under the European Union’s Horizon 2020 research and innovation programme (grant agreement No. 694343).
AU acknowledges support from the Spanish grants PGC2018-094671-B-I00, funded by MCIN/AEI/10.13039/501100011033 and by ``ERDF A way of making Europe'', and PID2019-108765GB-I00, funded by MCIN/AEI/10.13039/501100011033. 
K.G. is supported by the Australian Research Council through the Discovery Early Career Researcher Award (DECRA) Fellowship DE220100766 funded by the Australian Government. 
K.G. is supported by the Australian Research Council Centre of Excellence for All Sky Astrophysics in 3 Dimensions (ASTRO~3D), through project number CE170100013. 
JeP acknowledges support by the French Agence Nationale de la Recherche through the DAOISM grant ANR-21-CE31-0010, and by the Programme National ``Physique et Chimie du Milieu Interstellaire'' (PCMI) of CNRS/INSU with INC/INP, co-funded by CEA and CNES.

This paper makes use of the following ALMA data, which have been processed as part of the ALMOND and PHANGS--ALMA surveys: \\
\noindent ADS/JAO.ALMA\#2012.1.00650.S, \linebreak % (PHANGS-ALMA N628/M74)
%ADS/JAO.ALMA\#2013.1.00803.S, \linebreak % (PHANGS-ALMA N5128/CenA)
ADS/JAO.ALMA\#2013.1.01161.S, \linebreak % (PHANGS-ALMA N1365 + N5236/M83)
%ADS/JAO.ALMA\#2015.1.00121.S, \linebreak % (PHANGS-ALMA N5236/M83)
%ADS/JAO.ALMA\#2015.1.00782.S, \linebreak % (PHANGS-ALMA N1313 + N7793)
ADS/JAO.ALMA\#2015.1.00925.S, \linebreak % (PHANGS-ALMA pilot low mass)
ADS/JAO.ALMA\#2015.1.00956.S, \linebreak % (PHANGS-ALMA pilot high mass)
%ADS/JAO.ALMA\#2016.1.00386.S, \linebreak % (PHANGS-ALMA N5236/M83)
ADS/JAO.ALMA\#2017.1.00230.S, \linebreak % (ALMOND 7m+tp)
ADS/JAO.ALMA\#2017.1.00392.S, \linebreak % (PHANGS-ALMA low mass follow-up)
ADS/JAO.ALMA\#2017.1.00766.S, \linebreak % (PHANGS-ALMA early-type)
ADS/JAO.ALMA\#2017.1.00815.S, \linebreak % (ALMOND NGC 4321)
ADS/JAO.ALMA\#2017.1.00886.L, \linebreak % (PHANGS-ALMA large program)
ADS/JAO.ALMA\#2018.1.01171.S, \linebreak % (ALMOND 7m+tp)
%ADS/JAO.ALMA\#2018.1.01321.S, \linebreak % (PHANGS-ALMA N253, N300, Circinus)
ADS/JAO.ALMA\#2018.1.01651.S, \linebreak % (PHANGS-ALMA main sample follow-up)
ADS/JAO.ALMA\#2018.A.00062.S. \linebreak % (PHANGS-ALMA ACA-only nearby)
ADS/JAO.ALMA\#2019.2.00134.S, \linebreak % (ALMOND 7m+tp)
ADS/JAO.ALMA\#2021.1.00740.S, \linebreak % (ALMOND NGC 2903)
ALMA is a partnership of ESO (representing its member states), NSF (USA), and NINS (Japan), together with NRC (Canada), NSC and ASIAA (Taiwan), and KASI (Republic of Korea), in cooperation with the Republic of Chile. The Joint ALMA Observatory is operated by ESO, AUI/NRAO, and NAOJ. The National Radio Astronomy Observatory (NRAO) is a facility of the National Science Foundation operated under cooperative agreement by Associated Universities, Inc.

This work makes use of data products from the \textit{Wide-field Infrared Survey Explorer (WISE)}, which is a joint project of the University of California, Los Angeles, and the Jet Propulsion Laboratory/California Institute of Technology, funded by NASA.

This work is based in part on observations made with the \textit{Galaxy Evolution Explorer (GALEX)}. \textit{GALEX} is a NASA Small Explorer, whose mission was developed in cooperation with the Centre National d'Etudes Spatiales (CNES) of France and the Korean Ministry of Science and Technology. \textit{GALEX} is operated for NASA by the California Institute of Technology under NASA contract NAS5-98034.

%We acknowledge the usage of the HyperLeda database\footnote{\url{http://leda.univ-lyon1.fr}} \citep{Makarov_etal_2014} \citep{Maka} and the SAO/NASA Astrophysics Data System\footnote{\url{http://www.adsabs.harvard.edu}}.

\textit{Facilities:} ALMA, WISE, GALEX

\textit{Software:}
\texttt{NumPy} \citep{NumPy2020},
\texttt{SciPy} \citep{SciPy2020},
\texttt{Astropy} \citep{Astropy2018},
\texttt{pandas} \citep{Pandas_1.3.4},
\texttt{Matplotlib} \citep{Matplotlib2007},
\texttt{Colorcet} \citep{Kovesi2015},
\texttt{LinMix} \citep{Kelly2007}

%%%%%%%%%%%%%%%%%%%%%%%%%%%%%%%%%%%%%%%%%%%%%%%%%%
\section*{Data Availability}
The data used within this paper will be provided on reasonable request to the corresponding author.



\bibliographystyle{mnras}
\bibliography{bibliography} % if your bibtex file is called example.bib

%\begin{thebibliography}{99}
%\end{thebibliography}

%%%%%%%%%%%%%%%%%%%%%%%%%%%%%%%%%%%%%%%%%%%%%%%%%%

%%%%%%%%%%%%%%%%% APPENDICES %%%%%%%%%%%%%%%%%%%%%

%\clearpage
\appendix
%%%%%%%%%%%%%%%%%%%%%%%%%%%%%%%%%%%%%%%
\section{Cloud-scale Molecular Gas Properties}
\label{SEC:appendix:cloud_props}
%%%%%%%%%%%%%%%%%%%%%%%%%%%%%%%%%%%%%%%

\begin{figure}
    \includegraphics{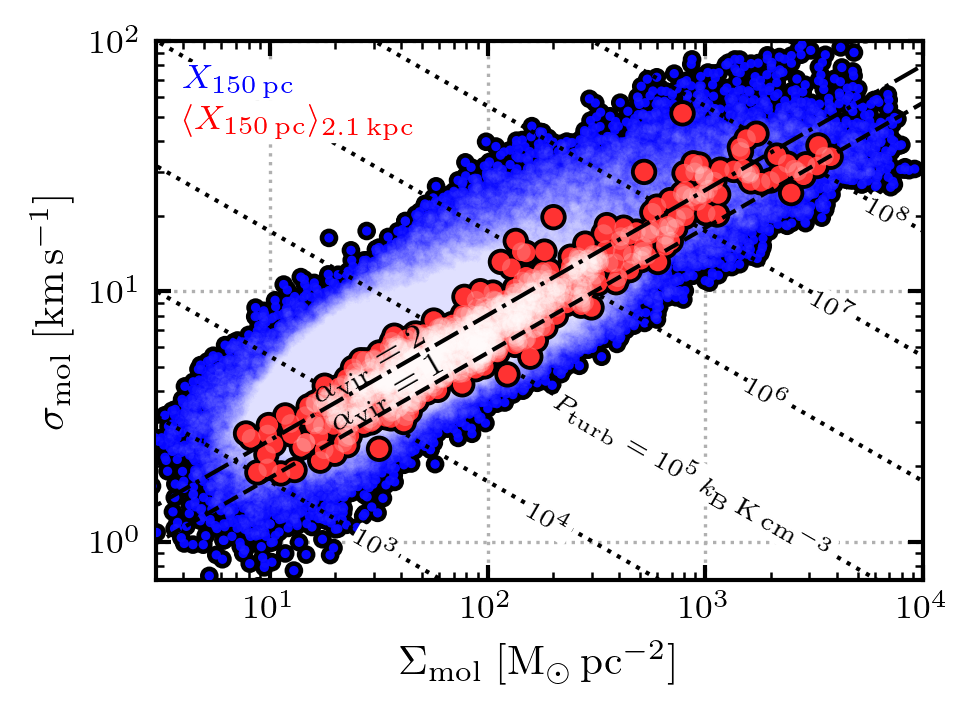}
    \caption{Molecular gas velocity dispersion (\vdis) against surface density (\sigmol) at \SI{150}{\parsec} scale across 22 nearby galaxies. The blue points denote the original \SI{150}{\parsec} resolution measurements, while the red points are the intensity-weighted averages obtained at \SI{2.1}{\kilo\parsec} apertures. The loci of constant virial paramter (\avir) and internal turbulent pressure (\Pturb) are obtained assuming fixed cloud size, i.e. $\avir\propto\vdis^2/\sigmol$ (Equation \ref{EQU:virial_parameter}), $\Pturb\propto\sigmol\vdis^2$ (Equation \ref{EQU:turbulent_pressure}, and are only valid for the original \SI{150}{\parsec} measurements.}
    \label{FIG:cloud_properties}
\end{figure}

Figure \ref{FIG:cloud_properties} displays the velocity dispersion of the molecular gas (\vdis) against its surface density (\sigmol) for all individual sitelines across the full sample of 22 galaxies at \SI{150}{\parsec} resolution (blue data points) similar to figure 1 in \citealp{Sun2020b}. \vdis and \sigmol are inferred from the CO(2-1) observations as described in Sections \ref{SEC:molecular_gas_surface_density} and \ref{SEC:velocity_dispersion}, respectively. The plot also shows loci of constant virial parameter (\avir) and internal turbulent pressure (\Pturb) as obtained from the CO(2-1) observations as described in Sections \ref{SEC:virial_parameter} and \ref{SEC:Pturb}, respectively, such that $\avir\propto\vdis^2/\sigmol$ and $\Pturb\propto\sigmol\vdis^2$ at fixed scale (here: \SI{150}{\parsec}). Moreover, we indicate the intensity-weighted averages (red points) of the \SI{150}{\parsec} measurements at \SI{2.1}{\kilo\parsec} averaging scale following Section \ref{SEC:weighted_avg}. We find that the distribution of the weighted averages in the \vdis--\sigmol plane resembles the distribution of the (original) high resolution measurements very well, providing similar dynamic range in both \vdis and \sigmol. However, the weighted averages show significantly lower dynamic range in \avir. Note that the loci of constant \avir and \Pturb are not valid for the weighted averages, because we take the weighted averages of the cloud-scale properties individually for each quantity, such that $\aviravg\not\propto\vdisavg^2/\sigmolavg$ and $\Pturbavg\not\propto\sigmolavg\vdisavg^2$.

%%%%%%%%%%%%%%%%%%%%%%%%%%%%%%%%%%%%%%
\section{Spectral Stacking}
\label{SEC:appendix:stacking}
%%%%%%%%%%%%%%%%%%%%%%%%%%%%%%%%%%%%%%

In order to recover more emission, in particular outside of galaxy centers, we perform spectral stacking of the \hcnone, \hcopone and \cstwo cubes as in \citet{Schruba2011, Jimenez-Donaire2017, Jimenez-Donaire2019, Beslic2021}. 
The basic idea is that the spectral axis is matched with a know velocity field from a high significance prior, i.e. here \cotwo. 
After shuffling the velocities, we average the spectra in bins defined by the galactocentric radius ($r_\mathrm{gal}$).
We select five bins up to $r_\mathrm{gal}=\SI{5}{\kilo\parsec}$ with bin widths of \SI{2}{\kilo\parsec}. 
In Figure~\ref{FIG:ngc4321_stacking}, we show the resulting stacked spectra (bottom panels) and stacked integrated intensities (a).
%A complete atlas of all galaxies is presented online.
A complete atlas of the remaining 24 ALMOND galaxies is presented in Appendix~\ref{SEC:appendix:supplements}.
The spectral stacking results demonstrate that, despite the low detection rate at the pixel level across much of the molecular gas discs, we are able to recover significant emission of \hcnone \hcopone and \cstwo outside of galaxy centres via stacking at the expense of spatial information. 
We detect significant HCN emission out to \SI{6}{\kilo\parsec} in more than a third (9/25) of the galaxies compared to only \SI{3}{\percent} for individual sightlines (Table~\ref{TAB:HCN_detection_fraction}), which demonstrates that stacking can successfully unveil HCN emission across most of the molecular gas discs.
In particular, these results motivate the binning approach described in Section~\ref{SEC:binning}, where we average the HCN data in bins of \intCOavg.
The two approaches, binning and stacking, yield very similar results within $\sim$ \SI{10}{\percent}, on average, and without bias \citep{Gallagher2018b}.

\begin{table}
\resizebox{\columnwidth}{!}{
\begin{threeparttable}[t]
\centering
\caption{HCN detection fraction across the 25 ALMOND galaxies}
\label{TAB:HCN_detection_fraction}
\begin{tabular}{ccccccc}
\hline\hline
\multirow{2}{*}{$r_{\rm gal}$ [kpc]} && \multicolumn{2}{c}{Sightlines} &&  \multicolumn{2}{c}{Stacking} \\\cline{3-4}\cline{6-7}
 && $N_{\rm det}/N_{\rm tot}$ & $N_{\rm frac}$ [\%] && $N_{\rm det}/N_{\rm tot}$ & $N_{\rm frac}$ [\%]\\
\hline
$0-2$  && 79/171 & 46.3 && 25/25 & 100 \\
$2-4$  && 78/473 & 16.6 && 21/25 & 84 \\
$4-6$  && 49/601 & 8.1  && 9/25 & 36 \\
$6-8$  && 19/696 & 2.8  && 5/25 & 20 \\
$8-10$ && 6/705  & 0.9  && 2/25  & 8 \\
\hline\hline
\end{tabular}
\begin{tablenotes}
\small
\item \textbf{Notes.} \hcnone detection fraction as a function galactocentric radius. $N_{\rm det}$ is the number of detected spectra for individual lines-of-sight (left), or the radially stacked spectra (right), where the S/N of the integrated intensity $> 3\,\sigma$. $N_{\rm tot}$ is the total number of spectra inside the radial bin. $N_{\rm frac}=N_{\rm det}/N_{\rm tot}$ depicts the detection fraction.
\end{tablenotes}
\end{threeparttable}
}
\end{table}

%%%%%%%%%%%%%%%%%%%%%%%%%%%%%%%%%%%%%%
\section{Weighted Averages}
\label{SEC:appendix:weighted_averages}
%%%%%%%%%%%%%%%%%%%%%%%%%%%%%%%%%%%%%%

\begin{figure*}
    \includegraphics{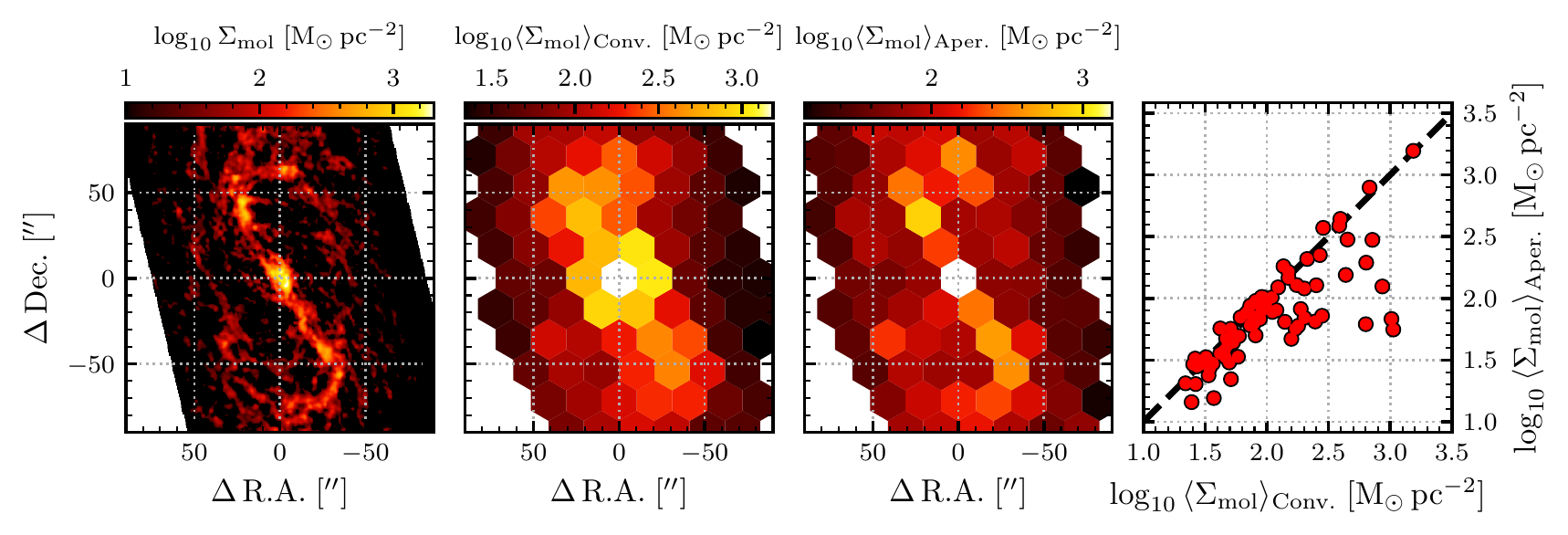}
    \caption{Comparison between different approaches of computing intensity-weighted averages. \textit{Left:} Molecular gas surface density (\sigmol) map of NGC 2903 at \SI{75}{\parsec} resolution. \textit{Centre left:} Intensity-weighted average \sigmol map at \SI{1}{\kilo\parsec} scale based on the Gaussian convolution as defined by Equation \ref{EQU:weighted_convolution}. This method is employed in this work to compare with the native kpc-scale observations, i.e. the HCN data. \textit{Centre right:} Intensity-weighted averages based on sharp apertures defined as the hexagonal shaped pixels. This method has been applied by e.g. \citet{Sun2020b}. \textit{Right:} Pixel-by-pixel comparison between the aperture and convolution based approaches.}
    \label{FIG:weighted_averages_methods}
\end{figure*}

In Section~\ref{SEC:weighted_avg}, we explain the idea of computed intensity-weighted averages from the high-resolution CO data in order to compare with the coarse-scale dense gas observations using the following equation:
\begin{align}
    \langle X \rangle_\text{Conv.} = \dfrac{(X\cdot \intCO)\ast\Omega}{\intCO\ast\Omega} \; ,
    \label{EQU:weighted_average_conv}
\end{align}
where $X$ is the high-resolution quantity (e.g. \sigmol) and $\Omega$ is the convolution kernel to go from the high to the coarse resolution. \citet{Sun2020a} computed the weighted averages inside sharp apertures, such that:
\begin{align}
    \langle X \rangle_\text{Aper.} = \dfrac{\sum\limits_{i\,\in\,\text{Aper.}} X_i \cdot I_{\text{CO(2--1)},i}}{\sum\limits_{i\,\in\,\text{Aper.}} I_{\text{CO(2--1)},i}} \; .
\label{EQU:weighted_average_aper}
\end{align}
We compare the two methods for the galaxy NGC 2903 in Figure~\ref{FIG:weighted_averages_methods}. While both methods lead to very similar results in the centre or along the bar, there are large discrepancies for the adjacent pixels, where the aperture method produces much lower values. The aperture approach is not affected by any Gaussian kernel dilution and thus useful if the aperture based weighted averages are used to study individually or for comparison with other aperture based weighted averages. However, comparison with observations performed at or convolved to the averaging scale should only be done using the convolution based method, which is symmetrically affected by beam dilution.

%%%%%%%%%%%%%%%%%%%%%%%%%%%%%%%%%%%%%%
\section{Linear Regression}
\label{SEC:appendix:linear_regression}
%%%%%%%%%%%%%%%%%%%%%%%%%%%%%%%%%%%%%%

Linear regression of astronomical data is far from trivial and it is crucial to apply a linear fitting routine which is tailored to the science question and the noise properties of the data appropriately. Here, we ask the question of how the cloud-scale molecular gas properties ($x$ data) affect the dense gas fraction and star formation efficiency ($y$ data). Statistically speaking, the $x$ data can be considered as the independent variable and the $y$ data as the dependent variable, such that we seek to constrain $y(x)=b+m\cdot x$, where $b,m=\mathrm{const}$. In principle one could also ask the inverted question, i.e. how $x$ depends on $y$ and thus constrain $x(y)=b^\prime +m^\prime\cdot y$ ($b^\prime,m^\prime=\mathrm{const}$). However, based on the formulated science question and given that the $x$ data are detected significantly throughout most of the discs of all galaxies, as opposed to the $y$ data, where about \SI{50}{\percent} of the data points are censored (here we consider the fully processed, binned data which enters the fitting routine), it is well-grounded to consider $x$ as the independent variable.

We detect HCN significantly (S/N $\ge 3$) only for about \SI{50}{\percent} of the binned data points. Hence, we have many censored data points, which result in upper limits (HCN/CO) or lower limits (SFR/HCN). Although these data are not significant, it is still valuable information: we know with high certainty (\SI{99.7}{\percent}) that the emission of that data point can not be larger than $3\sigma$ thus providing an upper limit. This information should be taken into account in the fitting routine to better constrain the assumed correlation and linear dependence. In addition, conversion to log-log scale can generate a bias in the estimated linear regression if censored data are not taken into account. Moreover, the true correlation most likely does not perfectly follow a linear correlation. Also, there is not necessarily a physical model which predicts a linear dependence (power-law in linear scale) between the $x$ and $y$ data. Thus, we need to account for an intrinsic scatter in the correlation. Even more so, it is important to account for the intrinsic scatter and the data uncertainties separately, in order to get reasonable regression uncertainties (s. \citealp{Kelly2007}).

Given our science question and the properties of our data, we want to use a linear regression tool which constrains the linear correlation of the dependent variable $y$ as a function of the independent variable $x$, while taking into account measurement uncertainties in both variables, intrinsic scatter about the regression and censored $y$ data. All of these requirements are met by the Python regression tool \textit{LinMix} which implements the Bayesian approach to linear regression introduced by \citealp{Kelly2007}. The tool assumes that the true data distribution is sampled from a superposition of Gaussians in $x$ and $y$. It performs a Markov chain Monte Carlo (MCMC) simulation using the Gibbs sampler to explore the posterior distribution, i.e. the true distribution of the regression parameters. \textit{LinMix} is capable of computing the Pearson correlation coefficient using both the significant and the censored data. Due to its statistical approach, the tool naturally finds trustworthy constraints on the regression parameters (intercept and slope) and also gives credibility areas, which we use to illustrate the uncertainty of the linear fits.

\begin{figure}
    \includegraphics{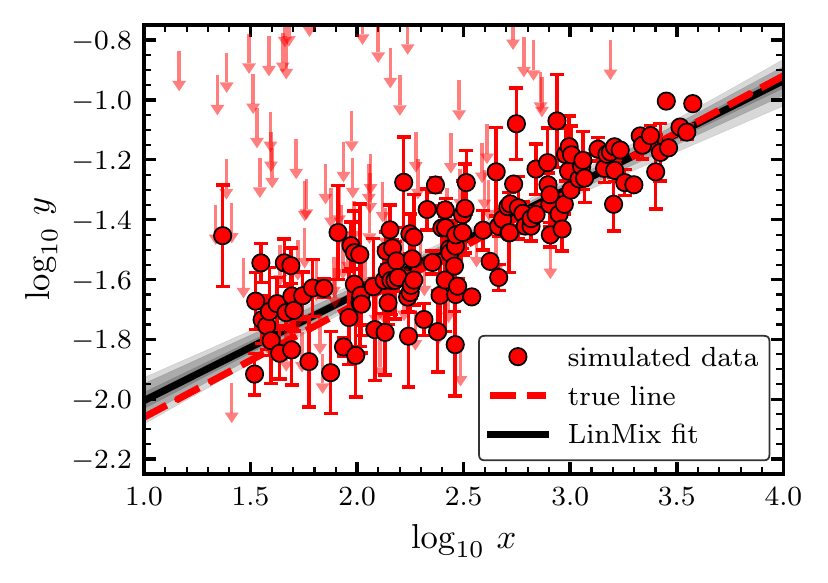}
    \caption{Bias estimation of the linear regression results. Adapted to the HCN/CO ($y$ data) vs. \sigmolavg ($x$ data) correlation, we use the \sigmolavg data and create perfectly correlated (linear relation) $y$ data, indicated by the red dotted line. Then we add Gaussian noise using the measurement uncertainties in HCN/CO, \sigmolavg and also add intrinsic scatter based on the estimated intrinsic scatter of the measured data. Finally, we apply the \textit{LinMix} fitting routine to determine the best fit linear regression (black solid line) and the $\{1, 2, 3\}$-sigma credibility regions (grey shaded areas).}
    \label{FIG:linear_regression_bias}
\end{figure}

In astronomy it is very common to determine the power-law of two astronomical quantities by converting the data from linear to logarithmic scale and fitting a line through the data. However, this procedure has some drawbacks. First, conversion to logarithmic scale is only valid for positive data, and negative data (i.e. negative intensities which arise from the data reduction and represent mostly noise) is removed. As a consequence the log-scale data is biased towards positive values and thus biases the linear regression. We can mitigate this bias by using a linear regression tool which can handle censored data and thus takes the insignificant and negative data into account. Next, conversion to logarithmic scale produces asymmetric uncertainties, i.e. if the uncertainties are symmetric in linear scale, they appear shorter in the positive and larger in the negative direction. Again, this will bias the linear regression if the regression tool assumes symmetric uncertainties, because it either overestimates the uncertainties in positive direction or underestimates the uncertainties in negative direction. We note that the fitting routine applied here is affected by this bias. Though, we are not aware of any regression tool which can take into account asymmetric uncertainties in addition to handling censored data. Moreover, we estimated the expected log-scale induced bias with the following simulation: To estimate the bias of the linear regression we start with the measured $x$-data and produce perfectly correlated $y$-data in logarithmic scale. Then we convert to linear scale and add Gaussian noise with amplitudes matching the measurement uncertainties. We also add Gaussian intrinsic scatter with amplitude as obtained from the linear regression of the observed data. Finally we convert back to logarithmic scale and run the fitting algorithm. Figure \ref{FIG:linear_regression_bias} shows the result customized to the HCN/CO vs. \intCOavg correlation. We find that the determined linear regression slope is in fact biased towards lower values by about $\sim$ \SI{10}{\percent}. In general, repeating this procedure for the other correlations, we find that the determined slopes are probably $\sim$ \SI{10}{\percent} flatter compared to the true correlation, if it were perfectly correlated.

%%%%%%%%%%%%%%%%%%%%%%%%%%%%%%%%%%%%%%%
\section{Line-of-sight Correlations}
\label{SEC:appendix:los}
%%%%%%%%%%%%%%%%%%%%%%%%%%%%%%%%%%%%%%%

\begin{figure*}
    \includegraphics{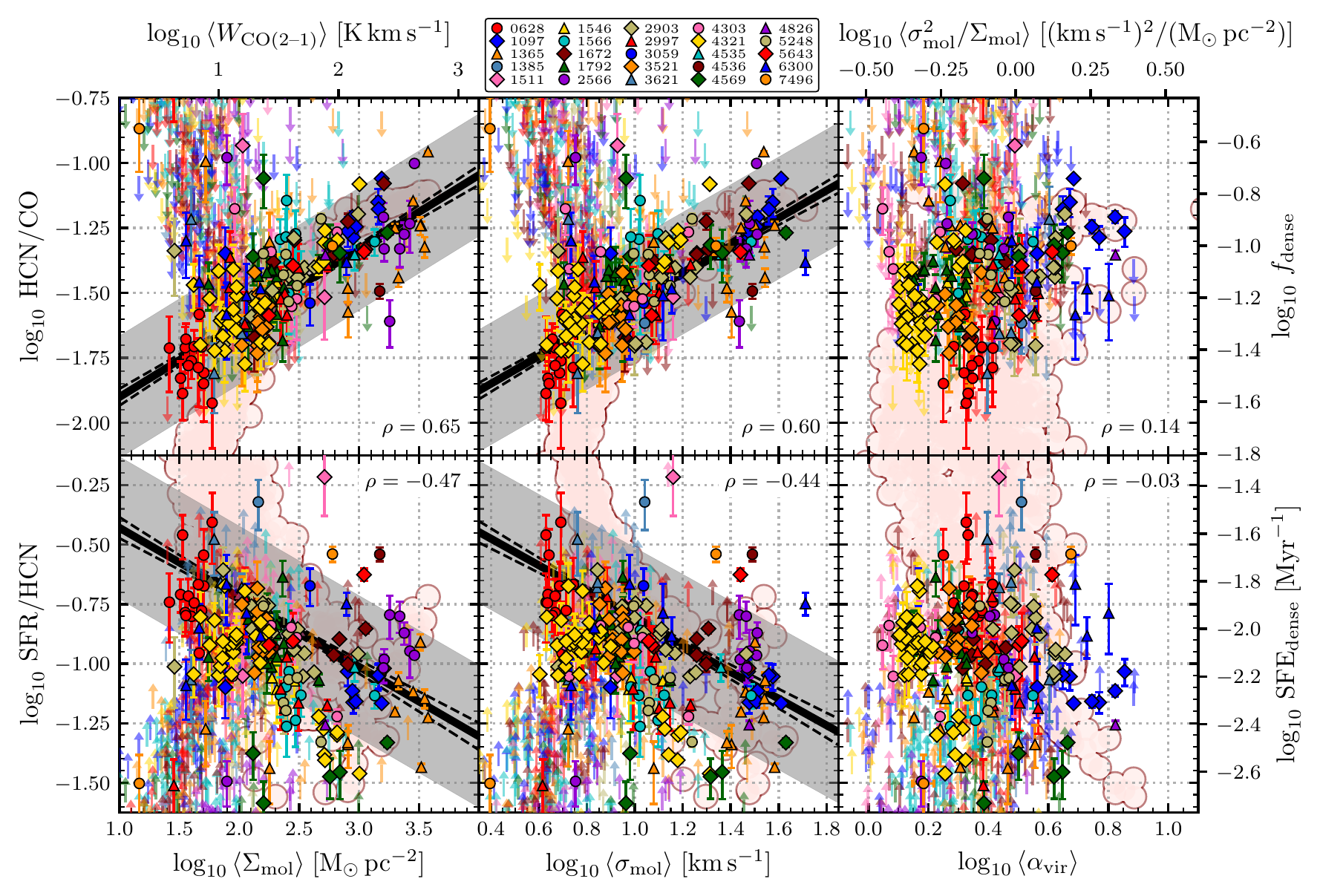}
    \caption{Analogous to Figure~\ref{FIG:HCN_lowres}, but for individual line-of-sight measurements, i.e. without binning the data.}
    \label{FIG:HCN_lowres_los}
\end{figure*}

In Section~\ref{SEC:binning}, we explain how we bin the data via \intCOavg to recover more emission, especially in the low \intCOavg regime.
We show in Appendix~\ref{SEC:appendix:stacking}, that averaging data via a high significant prior, i.e. \cotwo, is effectively unveiling more emission in the CO-emitting regions.
As a consequence, the binning method allows us to constrain the relations between HCN/CO, SFR/HCN and the cloud-scale properties with higher significance and with a higher weighting of the significant measurements.
However, binning is expected to reduce the scatter in the binned quantities, i.e. HCN/CO and SFR/HCN, thus potentially reducing the scatter and increasing the measured correlation.
Therefore, we also present the HCN/CO and SFR/HCN correlations with \sigmolavg, \vdisavg, \aviravg using the individual line-of-sight (LOS) measurements (Figure~\ref{FIG:HCN_lowres_los}.
We perform the linear regression on the LOS measurements analogous to Section~\ref{SEC:results}, i.e. taking into account measurement uncertainties, intrinsic scatter and censored data.
Qualitatively, we find the same results for the LOS data as for the binned data, i.e. a positive (negative) correlation between HCN/CO (SFR/HCN) with the cloud-scale molecular gas properties.
Certainly, we find lower correlations and higher scatter (Table~\ref{TAB:HCN_corr_table_los}). The lower correlation is however partly due to the higher statistical weight of the censored data (higher fraction of censored data taken into account in the fit).

% HCN/CO and SFR/HCN CORRELATION FIT PARAMETERS TABLE (LOWRES; incl. centres and discs)

\begin{table*}
\resizebox{\textwidth}{!}{
\begin{threeparttable}[t]
\centering
\caption{HCN/CO and SFR/HCN Correlations}
\label{TAB:HCN_corr_table_los}
\begin{tabular}{ccccccccccc}
\hline\hline
Cloud-scale & \multirow{2}{*}{Data} & \multicolumn{4}{c}{HCN/CO} && \multicolumn{4}{c}{SFR/HCN} \\\cline{3-6}\cline{8-11}
Property &  & Slope (unc.) & Interc. (unc.)\tnote{1} & Corr. $\rho$ ($p$) & Scatter && Slope (unc.) & Interc. (unc.)\tnote{1} & Corr. $\rho$ ($p$) & Scatter \\
\hline
\multirow{2}{*}{\sigmolavg} & sightlines & 0.28 (0.02) & -1.47 (0.01) & 0.65 (0.0) & 0.24 && -0.29 (0.03) & -0.87 (0.02) & -0.47 (0.0) & 0.30 \\
 & binned & 0.35 (0.02) & -1.49 (0.01) & 0.88 (0.0) & 0.11 && -0.33 (0.04) & -0.84 (0.02) & -0.63 (0.0) & 0.23 \\
 &  &  &  &  &  &  &  &  &  &  \\
\multirow{2}{*}{\vdisavg}   & sightlines & 0.53 (0.04) & -1.48 (0.01) & 0.60 (0.0) & 0.23 && -0.56 (0.06) & -0.87 (0.02) & -0.44 (0.0) & 0.30 \\
 & binned & 0.66 (0.04) & -1.5 (0.01) & 0.85 (0.0) & 0.12 && -0.63 (0.07) & -0.83 (0.02) & -0.60 (0.0) & 0.23 \\ 
 &  &  &  &  &  &  &  &  &  &  \\ 
\multirow{2}{*}{\aviravg}   & sightlines & ... & ... & 0.14 (0.059) & ... && ... & ... & -0.03 (0.643) & ... \\
 & binned & ... & ... & 0.21 (0.028) & ... && ... & ... & -0.11 (0.226) & ... \\
 &  &  &  &  &  &  &  &  &  &  \\ 
\multirow{2}{*}{\Pturbavg}  & sightlines & 0.13 (0.01) & -1.48 (0.01) & 0.66 (0.0) & 0.23 && -0.14 (0.02) & -0.86 (0.02) & -0.48 (0.0) & 0.30\\
 & binned & 0.17 (0.01) & -1.49 (0.01) & 0.88 (0.0) & 0.11 && -0.15 (0.02) & -0.83 (0.02) & -0.62 (0.0) & 0.22 \\
\hline\hline
\end{tabular}
\begin{tablenotes}
\small
\item \textbf{Notes.} Linear regression parameters analogous to Table~\ref{TAB:HCN_corr_table} using lowres resolution configuration (HCN/CO, SFR/HCN at \SI{2.1}{\kilo\parsec} scale; molecular cloud properties (\sigmol, \vdis, \avir, \Pturb) at \SI{150}{\parsec} scale). The table shows results obtained from individual line-of-sight measurements corresponding to Figure~\ref{FIG:HCN_lowres_los}, as well as from binned data corresponding to Figure~\ref{FIG:HCN_lowres}.
%Fit parameters resulting from the linear regression of HCN/CO (tracing \fdense) and SFR/HCN (tracing \sfedense) both at \SI{2.1}{\kilo\parsec} scale vs. molecular cloud properties (\sigmol, \vdis, \avir, \Pturb) at \SI{150}{\parsec} scale. Column 2 indicates the environment considered for the fit, where centre + disc means the whole galaxy as in Figure~\ref{FIG:HCN_lowres}. Centre and disc are defined as introduced in Section~\ref{SEC:centres_vs_discs} and are shown in Figure~\ref{FIG:HCN_lowres_centres_vs_discs}. Columns 3 and 4 list the slope and intercept with corresponding uncertainty estimates as determined by the linear regression tool. Column 5 shows the Pearson correlation coefficient $\rho$ and its corresponding $p$-value. Column 6 displays the $y$-axis scatter of the data about the best fit line measured in units of dex. Due to lack of correlation between HCN/CO, SFR/HCN and the virial parameter, we do not show linear regression results for \aviravg, but only list the correlation coefficients and $p$-values based on the significant data points. Note, that for the other cloud-scale properties, the correlations coefficient (and the $p$-value) are determined using both the censored and the significant data.
\item[1] Note that the intercept is measured at ca. the median of the respective cloud-scale property as described in Section~\ref{SEC:fitting_and_corr}. 
\end{tablenotes}
\end{threeparttable}}
\end{table*}

%%%%%%%%%%%%%%%%%%%%%%%%%%%%%%%%%%%%%%%
\section{Variation with Resolution}
\label{SEC:appendix:res_configs}
%%%%%%%%%%%%%%%%%%%%%%%%%%%%%%%%%%%%%%%

We study the HCN/CO and SFR/HCN correlations as a function of the cloud-scale and large-scale resolutions choosing three cloud-scale physical resolutions (\SI{75}{\parsec}, \SI{120}{\parsec}, \SI{150}{\parsec}) associated with the CO(2-1) data and three large-scale physical resolutions (\SI{1.0}{\kilo\parsec}, \SI{1.5}{\kilo\parsec}, \SI{2.1}{\kilo\parsec}) associated with the HCN data defined as the highest available common resolutions for galaxies inside \SI{11.6}{\kilo\parsec} ("highres"; 3 galaxies), \SI{15.3}{\kilo\parsec} ("midres"; 9 galaxies), \SI{23.4}{\kilo\parsec} ("lowres"; 22 galaxies), respectively. In addition we measure the correlations at the native angular resolutions of the CO(2-1) and HCN observations ("natres"; 22 galaxies). This defines the finest resolution configuration available but accesses different physical scales. The adopted resolution configurations are listed in Table \ref{TAB:res_configs}.

The resolution configurations introduced above include different galaxy samples. In order to investigate the dependence of the correlations on the adopted resolutions for fixed samples of galaxies we introduce sub-samples of the natres, lowres and midres configurations marked by the suffixed "midtar" and "hightar". Midtar and hightar denote the sub-sample of galaxies which are included in the midres and highres sample, respectively. For instance lowres-hightar denotes the lowres resolution configuration (\SI{150}{\parsec} cloud-scale, \SI{2.1}{\kilo\parsec} large-scale), but only includes the sub-sample of three galaxies which are also included in highres. Figure \ref{FIG:HCN_corr_plots} shows a compilation of the HCN/CO and SFR/HCN correlations for the different resolution configurations. Complementary, Table \ref{TAB:HCN_corr_table} lists the linear regression results for all adopted resolution configurations.

Overall, we report similar HCN/CO and SFR/HCN correlations with the cloud-scale molecular gas properties across all resolution configurations, where the linear regression parameters are in agreement with each other if the galaxy sample is fixed. For varying samples of galaxies we observe significant deviations in the linear regression slope in some cases indicating galaxy-to-galaxy variations in the HCN/CO and SFR/HCN relations.

\begin{table*}
\begin{threeparttable}[t]
\centering
\caption{Resolutions}
\label{TAB:res_configs}
\begin{tabular}{cccccc}
\hline\hline
\multirow{3}{*}{Sample} & \multirow{3}{*}{Galaxies} & \multicolumn{4}{c}{Resolution} \\\cline{3-6}
 & & lowres & midres & highres & natres \\
 & & $\langle\SI{150}{\parsec}\rangle_{\SI{2.1}{\kilo\parsec}}$ & $\langle\SI{120}{\parsec}\rangle_{\SI{1.5}{\kilo\parsec}}$ & $\langle\SI{75}{\parsec}\rangle_{\SI{1.0}{\kilo\parsec}}$ & $\langle\sim\ang{;;1}\rangle_{\sim\ang{;;20}}$\\
\hline
 & & & & & \\
\multirow{5}{*}{full} & NGC 0628, NGC 1097, NGC 1365, NGC 1385, NGC 1511, & \multirow{5}{*}{\cmark} & \multirow{5}{*}{\xmark} & \multirow{5}{*}{\xmark} & \multirow{5}{*}{\cmark} \\
 & NGC 1546, NGC 1566, NGC 1672, NGC 1792, NGC 2566, & & & & \\
 & NGC 2903, NGC 2997, NGC 3059, NGC 3521, NGC 3621, & & & & \\
 & NGC 4303, NGC 4321, NGC 4535, NGC 4536, NGC 4569, & & & & \\
 & NGC 4826, NGC 5248, NGC 5643, NGC 6300, NGC 7496\; & & & & \\
  & & & & & \\
\multirow{3}{*}{midtar} & NGC 0628, NGC 1097, NGC 1511, NGC 2903, NGC 2997, & \multirow{3}{*}{\cmark} & \multirow{3}{*}{\cmark} & \multirow{3}{*}{\xmark} & \multirow{3}{*}{\cmark} \\
 & NGC 3521, NGC 3621, NGC 4321, NGC 4826, NGC 5248, & & & & \\
 & NGC 5643, NGC 6300  & & & & \\
 & & & & & \\
hightar & NGC 0628, NGC 2903, NGC 3621, NGC 4826, NGC 6300 & \cmark & \cmark & \cmark & \cmark \\
 & & & & & \\
\hline\hline
\end{tabular}
\begin{tablenotes}
\small
\item \textbf{Notes.} Column 2 shows the galaxies included in the respective (sub-) samples resulting from the accessible galaxies at given resolutions. The full sample can reach \SI{150}{\parsec} cloud-scale and \SI{2.1}{\kilo\parsec} kpc-scale resolution. For the midtar and hightar samples the accessible resolutions are \SI{120}{\parsec} cloud-scale, \SI{1.5}{\kilo\parsec} kpc-scale and \SI{75}{\parsec} cloud-scale, \SI{1.0}{\kilo\parsec} kpc-scale, respectively.
\end{tablenotes}
\end{threeparttable}
\end{table*}

%%%%%%%%%%%%%%%%%%%%%%%%%%%%%%%%%%%%%%%%%%%%%%%%%%%%%%%%%%%%%
\subsection{HCN/CO vs. Molecular Cloud Properties}
\label{SEC:appendix:HCN_CO_corr}
%%%%%%%%%%%%%%%%%%%%%%%%%%%%%%%%%%%%%%%%%%%%%%%%%%%%%%%%%%%%%

For the physically homogenised resolution configurations we consistently find strong positive correlations between HCN/CO and \sigmolavg, \vdisavg and \Pturbavg (see Figure \ref{FIG:HCN_corr_plots} (top) and Table \ref{TAB:HCN_corr_table} (left)) with Pearson correlation coefficients ranging from $\rho=$ \numrange{0.70}{0.82}, \numrange{0.61}{0.79} and \numrange{0.60}{0.79}, respectively, with p-values all smaller than \num{e-5}. For any given correlation (e.g. HCN/CO vs \sigmolavg), the regression slopes vary among different physical resolution configurations and samples of galaxies, spanning $m_\mathrm{f,\Sigma}=$ \numrange{0.35}{0.54}, $m_\mathrm{f,\sigma}=$ \numrange{0.51}{0.93} and $m_\mathrm{f,P}=$ \numrange{0.12}{0.24}. Though, the linear regression parameters are in agreement within the $1\sigma$ uncertainties for fixed galaxy sample, meaning resolution does not significantly affect the observed relation between HCN/CO and the molecular cloud properties. For instance, for the HCN/CO vs \sigmolavg correlation we find $\rho=0.70$, $0.73$, $0.81$ and $m_\mathrm{f,\Sigma}=0.35\pm 0.09$, $0.37\pm 0.09$, $0.51\pm 0.10$ for lowres-hightar, midres-hightar and highres, respectively, all in agreement within the $1\sigma$ uncertainty limits. In contrast, for fixed resolution but varying sample we observe slopes deviating more than $1\sigma$, e.g. midres and midres-hightar lead to $m_\mathrm{f,\Sigma}=0.54\pm 0.06$ and $0.37\pm 0.09$. This points towards a galaxy-to-galaxy variation of the studied HCN/CO relations. However, these variations are not huge, because within the $2\sigma$ uncertainty range all resolution configuration are again consistent. In general, we find the trend of increasing correlation and steeper slopes for decreasing scale, i.e. at higher resolution, suggesting a small but systematic resolution dependence of the correlations. For the correlation of HCN/CO with the virial parameter (\aviravg) we find much lower correlation coefficients spanning $\rho=$ \numrange{0.17}{0.59} and p-values from \numrange{e-3}{0.10} suggesting a weak positive correlation between HCN/CO and \aviravg. However, the stronger positive correlation seen in the hightar configurations is mainly produced by one galaxy, i.e. NGC 2903, and is not confidently seen in the other targets. Note also that the dynamic range in \aviravg is barely \SI{1}{\dex} so that we might be insensitive to any potentially existing correlation with \avir. In the end, we have no convincing evidence for a correlation between HCN/CO and \aviravg.

Above all, studying the HCN/CO correlations with molecular cloud properties at different resolutions leads to consistent results which confidently demonstrates a positive correlation between HCN/CO and \sigmolavg, \vdisavg, \Pturbavg, with the trend of increasing correlation with increasing resolution (decreasing scale). The correlation of HCN/CO with \aviravg remains less clear. But consistently positive correlation coefficients point towards weak positive correlation between HCN/CO and \aviravg.

%%%%%%%%%%%%%%%%%%%%%%%%%%%%%%%%%%%%%%%%%%%%%%%%%%%%%%%%%%%%%
\subsection{SFR/HCN vs. Molecular Cloud Properties}
\label{SEC:appendix:SFR_HCN_corr}
%%%%%%%%%%%%%%%%%%%%%%%%%%%%%%%%%%%%%%%%%%%%%%%%%%%%%%%%%%%%%

\begin{figure*}
    \includegraphics[width=\textwidth]{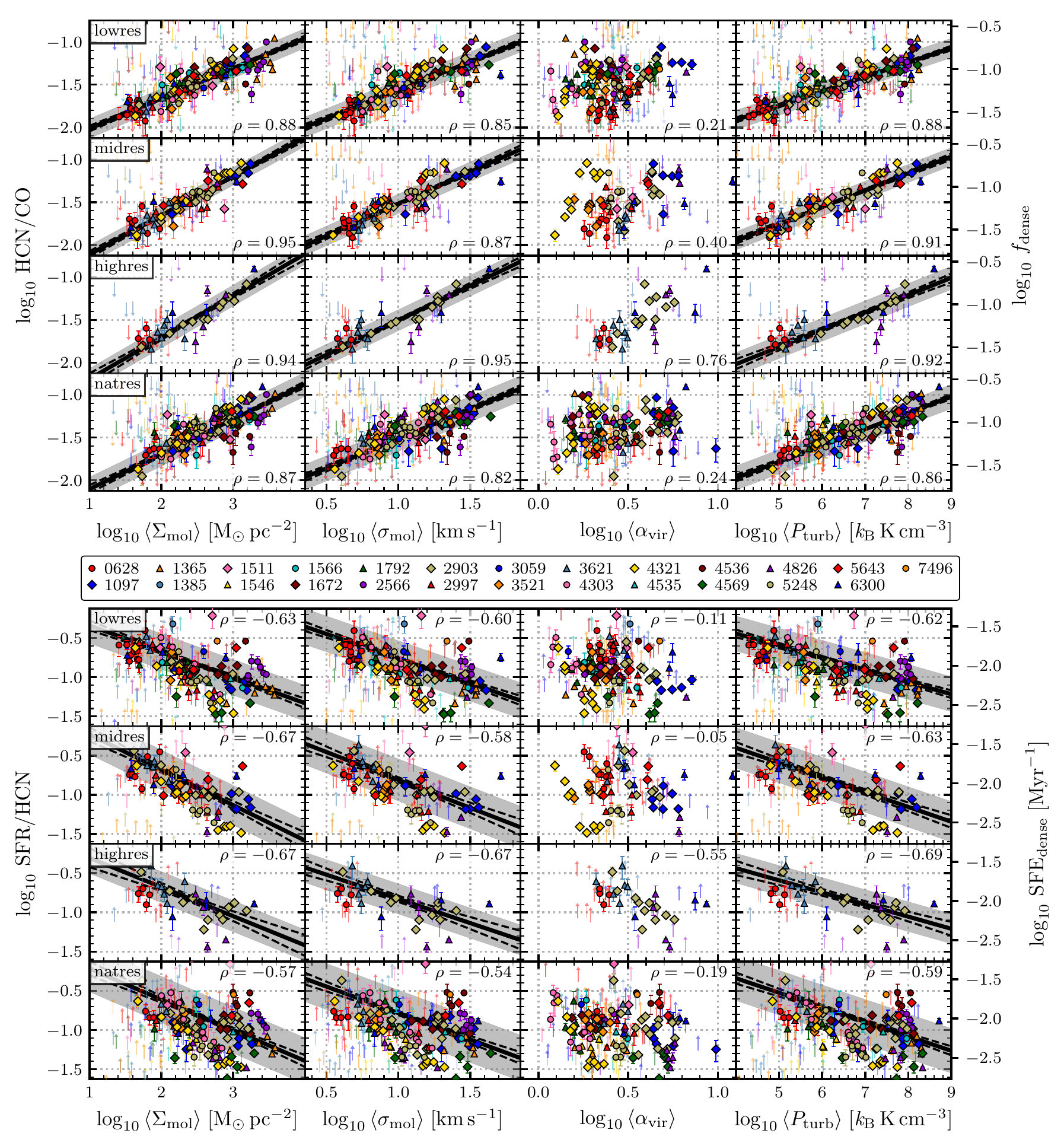}
    \caption{HCN/CO \textit{(top)} vs. \Xavg and SFR/HCN \textit{(bottom)} vs. \Xavg at different resolutions listed in Table~\ref{TAB:res_configs}. The solid line shows the best fit line where the dotted line is the $1\,\sigma$ uncertainty. The grey shaded area indicates the scatter of the significant data about the fit line.}
    \label{FIG:HCN_corr_plots}
\end{figure*}

We consistently find negative correlations between SFR/HCN and the cloud-scale properties \sigmolavg, \vdisavg, \Pturbavg across all adopted resolution configurations (see Figure \ref{FIG:HCN_corr_plots} (bottom) and Table \ref{TAB:HCN_corr_table} (right), where Pearson correlation coefficients range from \numrange{-0.45}{-0.63} (\sigmolavg), \numrange{-0.33}{-0.56} (\vdisavg) and \numrange{-0.32}{-0.59} (\Pturbavg) and slopes span $m_\mathrm{S,\Sigma}=$ \numrange{-0.23}{-0.49}, $m_\mathrm{S,\sigma}=$ \numrange{-0.27}{-0.78} and $m_\mathrm{S,P}=$ \numrange{-0.06}{-0.21} for the physically homogenised resolutions. Compared to the HCN/CO correlations, the strength of the SFR/HCN correlation is about $0.2$ lower and the intrinsic scatter about the median regression line is $2-3$ times as large, indicating a weaker correlation and suggesting potentially other physical processes in setting SFR/HCN. Still, we find strong evidence for a negative correlation between SFR/HCN and the aforementioned cloud properties at all resolutions. Moreover, the lack of correlation between SFR/HCN and \aviravg found at the lowest resolution (lowres) is also supported at higher resolution. In fact, the correlation coefficients are $|\rho|<0.2$ at maximum with p-values as large as $0.98$ indicating a very weak negative or no correlation with the virial parameter. The dependence on resolution follows similar systematics as of HCN/CO meaning the correlation increases and the slope steepens for increasing resolution, i.e. decreasing physical scale.

Overall, based on different resolution configurations we find strong evidence for a negative correlation between SFR/HCN tracing \sfedense and molecular cloud properties \sigmolavg, \vdisavg and \Pturbavg, where the correlation and steepness of the slope seems to increase with increasing resolution. Furthermore, we find no correlation between SFR/HCN and \aviravg. The opposite sign in the correlations compared to HCN/CO points towards an anti-correlation between SFR/HCN and HCN/CO and thus \sfedense and \sfedense.

%%%%%%%%%%%%%%%%%%%%%%%%%%%%%%%%%%%%%%%
\section{\texorpdfstring{HCO$^+$}{HCO+}/CO and SFR/\texorpdfstring{HCO$^+$}{HCO+} Correlations}
\label{SEC:appendix:HCOP_correlations}
%%%%%%%%%%%%%%%%%%%%%%%%%%%%%%%%%%%%%%%
In analogy to the HCN/CO and the SFR/HCN correlations we show the results of the determined HCO$^+$/CO and SFR/HCO$^+$ correlations in Figure \ref{FIG:HCOP_corr_plots} and in Table \ref{TAB:HCOP_corr_table}. First and foremost, we find the same correlations and anti-correlations between the HCO$^+$ spectroscopic measurements with the molecular cloud properties as of HCN with similar correlation coefficients, slopes and scatter. Thus, at \SIrange{1}{2}{\kilo\parsec} resolution, HCN(1-0) and HCO$^+$(1-0) are sensitive to the same density variations.

%%%%%%%%%%%%%%%%%%%%%%%%%%%%%%%%%%%%%%%
\section{CS/CO and SFR/CS Correlations}
\label{SEC:appendix:CS_correlations}
\begin{figure*}
    \includegraphics{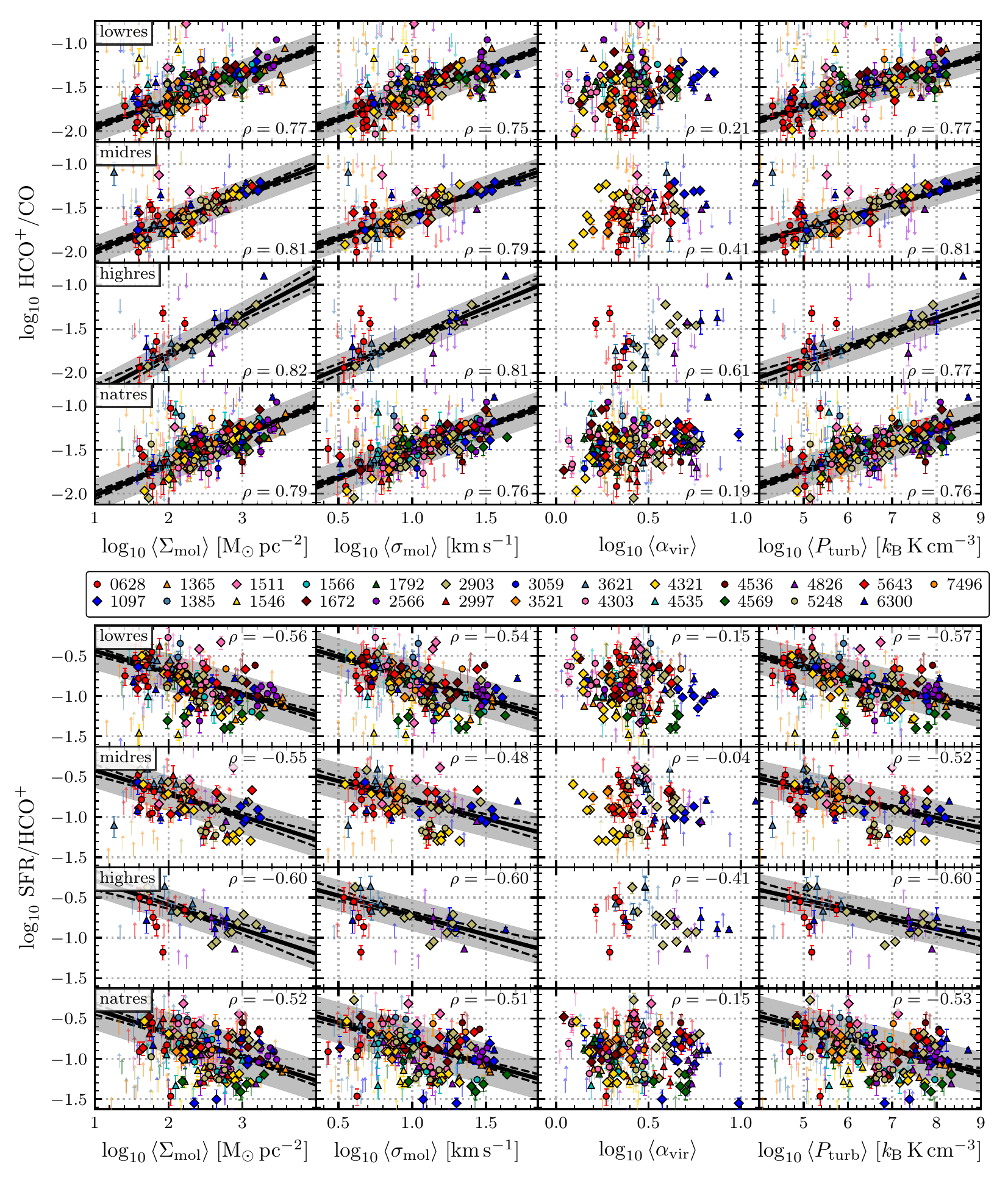}
    \caption{HCO$^+$/CO \textit{(top)} vs. \Xavg and SFR/HCO$^+$ \textit{(bottom)} vs. \Xavg at different resolutions listed in Table~\ref{TAB:res_configs}.}
    \label{FIG:HCOP_corr_plots}
\end{figure*}

%%%%%%%%%%%%%%%%%%%%%%%%%%%%%%%%%%%%%%%
In analogy to the HCN/CO and the SFR/HCN correlations we show the results of the determined CS/CO and SFR/CS correlations in Figure \ref{FIG:HCOP_corr_plots} and in Table \ref{TAB:HCOP_corr_table}. Despite the much lower signal-to-noise of the CS data we recover the same trends with cloud-scale molecular gas properties as seen for HCN or HCO$^+$, though with larger uncertainties.

%%%%%%%%%%%%%%%%%%%%%%%%%%%%%%%%%%%%%%%
\section{Individual Galaxies}
\label{SEC:appendix:galaxies}
%%%%%%%%%%%%%%%%%%%%%%%%%%%%%%%%%%%%%%%
In Figure~\ref{FIG:HCN_corr_plots_galaxies}, we show the same HCN/CO against \sigmolavg correlations as in Figure~\ref{FIG:HCN_lowres} (left panels), but for each galaxy individually. 

%%%%%%%%%%%%%%%%%%%%%%%%%%%%%%%%%%%%%%%
%% HCN res configs compilation plot

% HCN/CO and SFR/HCN CORRELATION FIT PARAMETERS TABLE

\begin{table*}
\resizebox{\textwidth}{!}{
\begin{threeparttable}[h]
\centering
\caption{HCN/CO and SFR/HCN Correlations}
\label{TAB:HCN_corr_table}
\begin{tabular}{ccccccccccc}
\hline\hline
Cloud-scale & \multirow{2}{*}{Resolution} & \multicolumn{4}{c}{HCN/CO} && \multicolumn{4}{c}{SFR/HCN} \\\cline{3-6}\cline{8-11}
Property &  & Slope (unc.) & Interc. (unc.) & Corr. $\rho$ ($p$) & Scatter && Slope (unc.) & Interc. (unc.) & Corr. $\rho$ ($p$) & Scatter\\
\hline
\sigmolavg & natres & 0.41 (0.03) & -1.49 (0.01) & 0.87 (0.0) & 0.14 &  & -0.39 (0.05) & -0.81 (0.03) & -0.57 (0.0) & 0.29 \\ 
\sigmolavg & natres-midtar & 0.5 (0.03) & -1.46 (0.01) & 0.94 (0.0) & 0.12 &  & -0.46 (0.07) & -0.85 (0.03) & -0.60 (0.0) & 0.29 \\ 
\sigmolavg & natres-hightar & 0.5 (0.04) & -1.47 (0.02) & 0.95 (0.0) & 0.10 &  & -0.44 (0.1) & -0.86 (0.05) & -0.66 (0.0) & 0.24 \\ 
\sigmolavg & lowres & 0.35 (0.02) & -1.49 (0.01) & 0.88 (0.0) & 0.11 &  & -0.33 (0.04) & -0.84 (0.02) & -0.63 (0.0) & 0.23 \\ 
\sigmolavg & lowres-midtar & 0.43 (0.02) & -1.44 (0.01) & 0.95 (0.0) & 0.10 &  & -0.42 (0.05) & -0.93 (0.03) & -0.7 (0.0) & 0.20 \\ 
\sigmolavg & lowres-hightar & 0.39 (0.03) & -1.43 (0.02) & 0.97 (0.0) & 0.07 &  & -0.31 (0.06) & -0.90 (0.04) & -0.73 (0.0) & 0.13 \\ 
\sigmolavg & midres & 0.46 (0.03) & -1.43 (0.01) & 0.95 (0.0) & 0.10 &  & -0.45 (0.06) & -0.91 (0.03) & -0.67 (0.0) & 0.25 \\ 
\sigmolavg & midres-hightar & 0.41 (0.04) & -1.43 (0.02) & 0.92 (0.0) & 0.10 &  & -0.37 (0.08) & -0.89 (0.05) & -0.66 (0.0) & 0.19 \\ 
\sigmolavg & highres & 0.49 (0.04) & -1.46 (0.02) & 0.94 (0.0) & 0.12 &  & -0.37 (0.08) & -0.86 (0.04) & -0.67 (0.0) & 0.20 \\ 
 &  &  &  &  &  &  &  &  &  &  \\ 
\vdisavg & natres & 0.69 (0.05) & -1.45 (0.01) & 0.82 (0.0) & 0.15 &  & -0.67 (0.09) & -0.85 (0.03) & -0.54 (0.0) & 0.29 \\ 
\vdisavg & natres-midtar & 0.80 (0.06) & -1.40 (0.02) & 0.88 (0.0) & 0.14 &  & -0.74 (0.13) & -0.90 (0.04) & -0.56 (0.0) & 0.29 \\ 
\vdisavg & natres-hightar & 0.81 (0.09) & -1.38 (0.02) & 0.9 (0.0) & 0.13 &  & -0.71 (0.17) & -0.93 (0.05) & -0.61 (0.0) & 0.26 \\ 
\vdisavg & lowres & 0.66 (0.04) & -1.5 (0.01) & 0.85 (0.0) & 0.12 &  & -0.63 (0.07) & -0.83 (0.02) & -0.60 (0.0) & 0.23 \\ 
\vdisavg & lowres-midtar & 0.69 (0.05) & -1.49 (0.01) & 0.88 (0.0) & 0.10 &  & -0.65 (0.10) & -0.87 (0.03) & -0.61 (0.0) & 0.22 \\ 
\vdisavg & lowres-hightar & 0.58 (0.07) & -1.52 (0.02) & 0.9 (0.0) & 0.08 &  & -0.46 (0.1) & -0.83 (0.03) & -0.68 (0.0) & 0.13 \\ 
\vdisavg & midres & 0.75 (0.06) & -1.44 (0.02) & 0.87 (0.0) & 0.11 &  & -0.71 (0.11) & -0.88 (0.03) & -0.58 (0.0) & 0.27 \\ 
\vdisavg & midres-hightar & 0.63 (0.07) & -1.48 (0.02) & 0.89 (0.0) & 0.10 &  & -0.56 (0.13) & -0.84 (0.04) & -0.64 (0.0) & 0.19 \\ 
\vdisavg & highres & 0.81 (0.07) & -1.42 (0.02) & 0.95 (0.0) & 0.12 &  & -0.62 (0.13) & -0.89 (0.04) & -0.67 (0.0) & 0.19 \\ 
 &  &  &  &  &  &  &  &  &  &  \\ 
\aviravg & natres & ... & ... & 0.24 (0.01) & ... &  & ... & ... & -0.19 (0.037) & ... \\ 
\aviravg & natres-midtar & ... & ... & 0.41 (0.001) & ... &  & ... & ... & -0.15 (0.233) & ... \\ 
\aviravg & natres-hightar & ... & ... & 0.73 (0.0) & ... &  & ... & ... & -0.53 (0.003) & ... \\ 
\aviravg & lowres & ... & ... & 0.21 (0.028) & ... &  & ... & ... & -0.11 (0.226) & ... \\ 
\aviravg & lowres-midtar & ... & ... & 0.46 (0.0) & ... &  & ... & ... & -0.12 (0.325) & ... \\ 
\aviravg & lowres-hightar & ... & ... & 0.76 (0.0) & ... &  & ... & ... & -0.53 (0.005) & ... \\ 
\aviravg & midres & ... & ... & 0.4 (0.001) & ... &  & ... & ... & -0.05 (0.666) & ... \\ 
\aviravg & midres-hightar & ... & ... & 0.77 (0.0) & ... &  & ... & ... & -0.54 (0.005) & ... \\ 
\aviravg & highres & ... & ... & 0.76 (0.0) & ... &  & ... & ... & -0.55 (0.002) & ... \\ 
 &  &  &  &  &  &  &  &  &  &  \\ 
\Pturbavg & natres & 0.19 (0.01) & -1.5 (0.01) & 0.86 (0.0) & 0.14 &  & -0.19 (0.02) & -0.80 (0.03) & -0.59 (0.0) & 0.28 \\ 
\Pturbavg & natres-midtar & 0.22 (0.01) & -1.49 (0.02) & 0.92 (0.0) & 0.12 &  & -0.2 (0.03) & -0.82 (0.04) & -0.6 (0.0) & 0.28 \\ 
\Pturbavg & natres-hightar & 0.22 (0.02) & -1.52 (0.02) & 0.95 (0.0) & 0.10 &  & -0.19 (0.04) & -0.82 (0.05) & -0.68 (0.0) & 0.23 \\ 
\Pturbavg & lowres & 0.17 (0.01) & -1.49 (0.01) & 0.88 (0.0) & 0.11 &  & -0.15 (0.02) & -0.83 (0.02) & -0.62 (0.0) & 0.22 \\ 
\Pturbavg & lowres-midtar & 0.18 (0.01) & -1.47 (0.01) & 0.92 (0.0) & 0.10 &  & -0.17 (0.02) & -0.88 (0.03) & -0.64 (0.0) & 0.21 \\ 
\Pturbavg & lowres-hightar & 0.15 (0.01) & -1.5 (0.02) & 0.94 (0.0) & 0.08 &  & -0.12 (0.02) & -0.84 (0.03) & -0.71 (0.0) & 0.13 \\ 
\Pturbavg & midres & 0.2 (0.01) & -1.46 (0.01) & 0.91 (0.0) & 0.11 &  & -0.19 (0.03) & -0.87 (0.03) & -0.63 (0.0) & 0.26 \\ 
\Pturbavg & midres-hightar & 0.16 (0.02) & -1.49 (0.02) & 0.89 (0.0) & 0.11 &  & -0.15 (0.03) & -0.84 (0.04) & -0.67 (0.0) & 0.19 \\ 
\Pturbavg & highres & 0.20 (0.02) & -1.51 (0.02) & 0.92 (0.0) & 0.14 &  & -0.15 (0.03) & -0.82 (0.04) & -0.69 (0.0) & 0.19 \\ 
\hline\hline
\end{tabular}
\begin{tablenotes}
\small
\item \textbf{Notes.} HCN/CO (tracing \fdense) and SFR/HCN (tracing \sfedense) vs. molecular cloud properties - \sigmol, \vdis, \avir, \Pturb - correlations for all adopted resolution configurations. Columns 3 and 4 list the slope and intercept with its uncertainty estimates as determined by the linear regression. Column 5 shows the Pearson correlation coefficient $\rho$ and its corresponding p-value. Column 6 displays the scatter, i.e. the standard deviation of the fit residuals of the significant ($\text{SNR}>3$) data. Due to lack of correlation between HCN/CO or SFR/HCN and the virial parameter, we do not show linear regression results, but only list the correlation coefficient and p-value based on the significant data points. Note, that for the other cloud-scale properties, the correlation coefficient (and the p-value) are determined using both the censored and the significant data.
\end{tablenotes}
\end{threeparttable}}
\end{table*}

%%%%%%%%%%%%%%%%%%%%%%%%%%%%%%%%%%%%%%%
%% HCO+ res configs compilation plot

% HCOP/CO and SFR/HCOP CORRELATION FIT PARAMETERS TABLE

\begin{table*}
\resizebox{\textwidth}{!}{
\begin{threeparttable}[h]
\centering
\caption{HCO$^+$/CO and SFR/HCO$^+$ Correlations}
\label{TAB:HCOP_corr_table}
\begin{tabular}{ccccccccccc}
\hline\hline
Cloud-scale & \multirow{2}{*}{Res. Config.} & \multicolumn{4}{c}{HCO$^+$/CO} && \multicolumn{4}{c}{SFR/HCO$^+$} \\\cline{3-6}\cline{8-11}
Property &  & Slope (unc.) & Interc. (unc.) & Corr. $\rho$ ($p$) & Scatter && Slope (unc.) & Interc. (unc.) & Corr. $\rho$ ($p$) & Scatter\\
\hline
\sigmolavg & natres & 0.34 (0.03) & -1.51 (0.01) & 0.79 (0.0) & 0.18 &  & -0.3 (0.04) & -0.83 (0.02) & -0.52 (0.0) & 0.26 \\ 
\sigmolavg & natres-midtar & 0.39 (0.04) & -1.52 (0.02) & 0.82 (0.0) & 0.19 &  & -0.32 (0.06) & -0.83 (0.03) & -0.51 (0.0) & 0.26 \\ 
\sigmolavg & natres-hightar & 0.39 (0.05) & -1.55 (0.03) & 0.82 (0.0) & 0.17 &  & -0.3 (0.09) & -0.79 (0.04) & -0.53 (0.0) & 0.24 \\ 
\sigmolavg & lowres & 0.3 (0.02) & -1.51 (0.01) & 0.77 (0.0) & 0.17 &  & -0.27 (0.03) & -0.84 (0.02) & -0.56 (0.0) & 0.23 \\ 
\sigmolavg & lowres-midtar & 0.33 (0.03) & -1.51 (0.02) & 0.84 (0.0) & 0.15 &  & -0.31 (0.05) & -0.87 (0.03) & -0.6 (0.0) & 0.20 \\ 
\sigmolavg & lowres-hightar & 0.3 (0.05) & -1.53 (0.03) & 0.76 (0.0) & 0.12 &  & -0.21 (0.06) & -0.80 (0.04) & -0.58 (0.0) & 0.14 \\ 
\sigmolavg & midres & 0.32 (0.03) & -1.51 (0.02) & 0.81 (0.0) & 0.16 &  & -0.29 (0.05) & -0.85 (0.03) & -0.55 (0.0) & 0.22 \\ 
\sigmolavg & midres-hightar & 0.22 (0.07) & -1.56 (0.05) & 0.52 (0.0) & 0.20 &  & -0.15 (0.07) & -0.78 (0.05) & -0.39 (0.01) & 0.19 \\ 
\sigmolavg & highres & 0.44 (0.06) & -1.57 (0.03) & 0.82 (0.0) & 0.16 &  & -0.30 (0.08) & -0.74 (0.04) & -0.60 (0.0) & 0.20 \\ 
 &  &  &  &  &  &  &  &  &  &  \\ 
\vdisavg & natres & 0.59 (0.05) & -1.47 (0.01) & 0.76 (0.0) & 0.19 &  & -0.51 (0.07) & -0.86 (0.02) & -0.51 (0.0) & 0.25 \\ 
\vdisavg & natres-midtar & 0.63 (0.06) & -1.47 (0.02) & 0.79 (0.0) & 0.20 &  & -0.51 (0.10) & -0.86 (0.03) & -0.48 (0.0) & 0.26 \\ 
\vdisavg & natres-hightar & 0.6 (0.1) & -1.49 (0.03) & 0.76 (0.0) & 0.18 &  & -0.43 (0.14) & -0.84 (0.05) & -0.46 (0.0) & 0.25 \\ 
\vdisavg & lowres & 0.58 (0.05) & -1.52 (0.01) & 0.75 (0.0) & 0.17 &  & -0.53 (0.07) & -0.83 (0.02) & -0.54 (0.0) & 0.23 \\ 
\vdisavg & lowres-midtar & 0.58 (0.05) & -1.55 (0.01) & 0.83 (0.0) & 0.15 &  & -0.50 (0.09) & -0.83 (0.03) & -0.54 (0.0) & 0.20 \\ 
\vdisavg & lowres-hightar & 0.48 (0.08) & -1.6 (0.02) & 0.79 (0.0) & 0.11 &  & -0.35 (0.09) & -0.76 (0.03) & -0.62 (0.0) & 0.13 \\ 
\vdisavg & midres & 0.56 (0.05) & -1.51 (0.02) & 0.79 (0.0) & 0.15 &  & -0.47 (0.1) & -0.84 (0.03) & -0.48 (0.0) & 0.22 \\ 
\vdisavg & midres-hightar & 0.42 (0.10) & -1.57 (0.04) & 0.62 (0.0) & 0.19 &  & -0.27 (0.11) & -0.76 (0.04) & -0.42 (0.006) & 0.18 \\ 
\vdisavg & highres & 0.7 (0.10) & -1.54 (0.03) & 0.81 (0.0) & 0.16 &  & -0.48 (0.12) & -0.77 (0.04) & -0.60 (0.0) & 0.19 \\ 
 &  &  &  &  &  &  &  &  &  &  \\ 
\aviravg & natres & ... & ... & 0.19 (0.023) & ... &  & ... & ... & -0.15 (0.069) & ... \\ 
\aviravg & natres-midtar & ... & ... & 0.36 (0.001) & ... &  & ... & ... & -0.17 (0.115) & ... \\ 
\aviravg & natres-hightar & ... & ... & 0.34 (0.044) & ... &  & ... & ... & -0.26 (0.134) & ... \\ 
\aviravg & lowres & ... & ... & 0.21 (0.013) & ... &  & ... & ... & -0.15 (0.069) & ... \\ 
\aviravg & lowres-midtar & ... & ... & 0.44 (0.0) & ... &  & ... & ... & -0.14 (0.22) & ... \\ 
\aviravg & lowres-hightar & ... & ... & 0.78 (0.0) & ... &  & ... & ... & -0.53 (0.003) & ... \\ 
\aviravg & midres & ... & ... & 0.41 (0.0) & ... &  & ... & ... & -0.04 (0.738) & ... \\ 
\aviravg & midres-hightar & ... & ... & 0.61 (0.001) & ... &  & ... & ... & -0.3 (0.122) & ... \\ 
\aviravg & highres & ... & ... & 0.61 (0.001) & ... &  & ... & ... & -0.41 (0.034) & ... \\ 
 &  &  &  &  &  &  &  &  &  &  \\ 
\Pturbavg & natres & 0.16 (0.01) & -1.51 (0.01) & 0.76 (0.0) & 0.19 &  & -0.14 (0.02) & -0.82 (0.02) & -0.53 (0.0) & 0.25 \\ 
\Pturbavg & natres-midtar & 0.17 (0.02) & -1.54 (0.02) & 0.8 (0.0) & 0.19 &  & -0.14 (0.03) & -0.81 (0.03) & -0.51 (0.0) & 0.26 \\ 
\Pturbavg & natres-hightar & 0.16 (0.03) & -1.59 (0.03) & 0.78 (0.0) & 0.18 &  & -0.13 (0.04) & -0.77 (0.04) & -0.54 (0.0) & 0.24 \\ 
\Pturbavg & lowres & 0.14 (0.01) & -1.51 (0.01) & 0.77 (0.0) & 0.17 &  & -0.13 (0.02) & -0.83 (0.02) & -0.57 (0.0) & 0.23 \\ 
\Pturbavg & lowres-midtar & 0.15 (0.01) & -1.54 (0.01) & 0.84 (0.0) & 0.15 &  & -0.13 (0.02) & -0.84 (0.02) & -0.59 (0.0) & 0.20 \\ 
\Pturbavg & lowres-hightar & 0.13 (0.02) & -1.58 (0.02) & 0.83 (0.0) & 0.10 &  & -0.09 (0.02) & -0.77 (0.03) & -0.67 (0.0) & 0.13 \\ 
\Pturbavg & midres & 0.14 (0.01) & -1.53 (0.02) & 0.81 (0.0) & 0.15 &  & -0.12 (0.02) & -0.83 (0.03) & -0.52 (0.0) & 0.22 \\ 
\Pturbavg & midres-hightar & 0.1 (0.03) & -1.59 (0.04) & 0.58 (0.0) & 0.19 &  & -0.07 (0.03) & -0.76 (0.04) & -0.44 (0.004) & 0.18 \\ 
\Pturbavg & highres & 0.17 (0.03) & -1.63 (0.04) & 0.77 (0.0) & 0.17 &  & -0.12 (0.03) & -0.71 (0.04) & -0.6 (0.0) & 0.20 \\ 
\hline\hline
\end{tabular}
\begin{tablenotes}
\small
\item \textbf{Notes.} Analog to Table~\ref{TAB:HCN_corr_table} but for \hcopone.
\end{tablenotes}
\end{threeparttable}}
\end{table*}

%%%%%%%%%%%%%%%%%%%%%%%%%%%%%%%%%%%%%%%
%% CS res configs compilation plot
\begin{figure*}
    \includegraphics{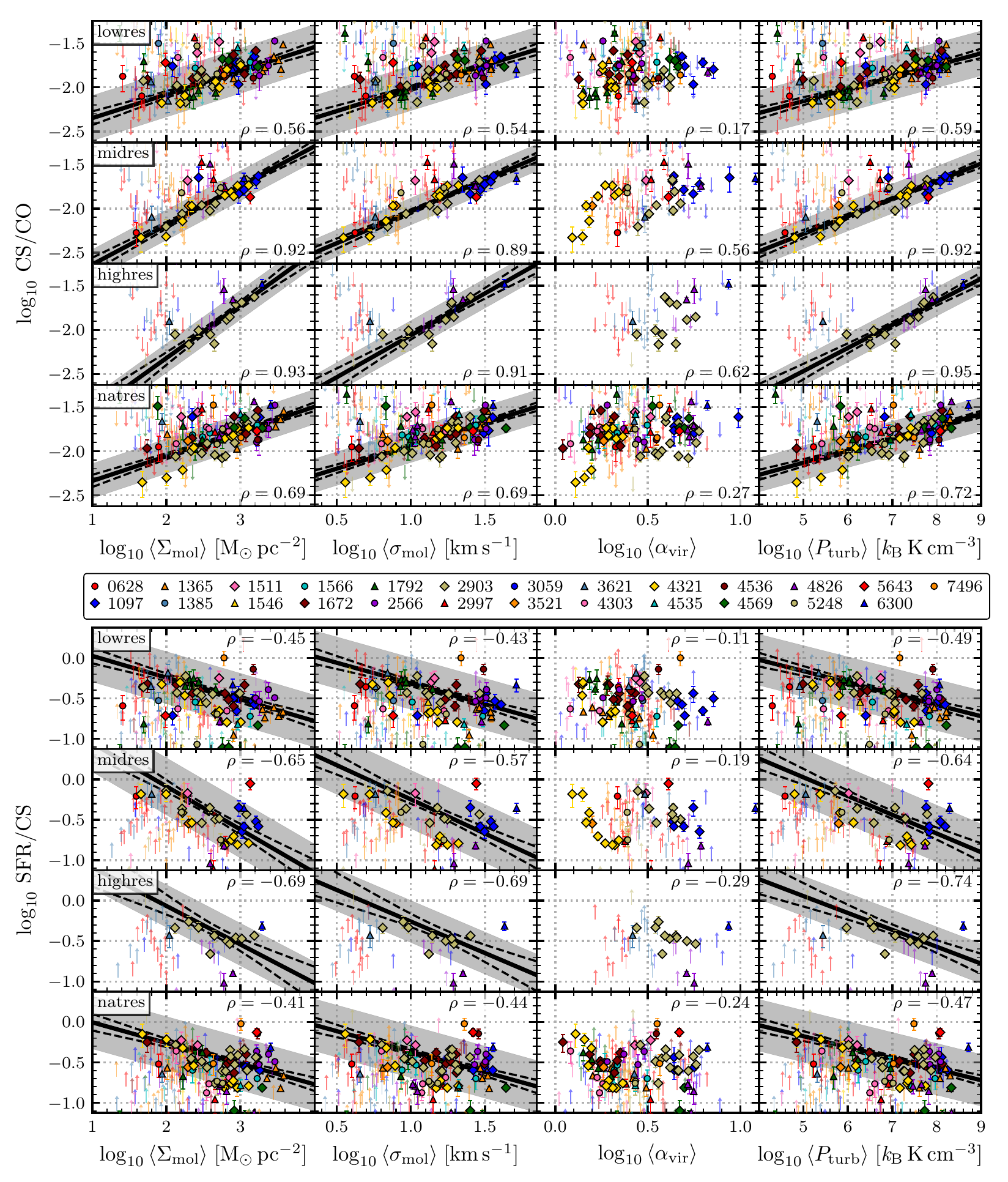}
    \caption{CS/CO \textit{(top)} vs. \Xavg and SFR/CS \textit{(bottom)} vs. \Xavg at different resolutions listed in Table~\ref{TAB:res_configs}.}
    \label{FIG:CS_corr_plots}
\end{figure*}

% CS/CO and SFR/CS CORRELATION FIT PARAMETERS TABLE

\begin{table*}
\resizebox{\textwidth}{!}{
\begin{threeparttable}[h]
\centering
\caption{CS/CO and SFR/CS Correlations}
\label{TAB:CS_corr_table}
\begin{tabular}{ccccccccccc}
\hline\hline
\multirow{2}{*}{MC Prop.} & \multirow{2}{*}{Res. Config.} & \multicolumn{4}{c}{CS/CO} && \multicolumn{4}{c}{SFR/CS} \\\cline{3-6}\cline{8-11}
 &  & Slope (unc.) & Interc. (unc.) & Corr. $\rho$ ($p$) & Scatter && Slope (unc.) & Interc. (unc.) & Corr. $\rho$ ($p$) & Scatter \\
 \hline
 \sigmolavg & natres & 0.28 (0.04) & -1.92 (0.02) & 0.69 (0.0) & 0.20 &  & -0.26 (0.06) & -0.39 (0.03) & -0.41 (0.0) & 0.33 \\ 
\sigmolavg & natres-midtar & 0.42 (0.05) & -1.95 (0.03) & 0.82 (0.0) & 0.18 &  & -0.4 (0.09) & -0.37 (0.04) & -0.54 (0.0) & 0.32 \\ 
\sigmolavg & natres-hightar & 0.45 (0.08) & -1.96 (0.04) & 0.84 (0.0) & 0.17 &  & -0.36 (0.12) & -0.38 (0.06) & -0.58 (0.0) & 0.26 \\ 
\sigmolavg & lowres & 0.27 (0.05) & -1.95 (0.03) & 0.56 (0.0) & 0.26 &  & -0.27 (0.06) & -0.38 (0.03) & -0.45 (0.0) & 0.32 \\ 
\sigmolavg & lowres-midtar & 0.45 (0.1) & -1.96 (0.05) & 0.60 (0.0) & 0.32 &  & -0.45 (0.11) & -0.41 (0.05) & -0.55 (0.0) & 0.35 \\ 
\sigmolavg & lowres-hightar & 0.36 (0.08) & -1.92 (0.04) & 0.81 (0.0) & 0.17 &  & -0.32 (0.11) & -0.41 (0.06) & -0.62 (0.0) & 0.21 \\ 
\sigmolavg & midres & 0.45 (0.05) & -1.96 (0.02) & 0.92 (0.0) & 0.17 &  & -0.53 (0.10) & -0.35 (0.05) & -0.65 (0.0) & 0.35 \\ 
\sigmolavg & midres-hightar & 0.50 (0.07) & -1.95 (0.03) & 0.97 (0.0) & 0.15 &  & -0.51 (0.16) & -0.37 (0.07) & -0.71 (0.0) & 0.26 \\ 
\sigmolavg & highres & 0.60 (0.09) & -2.02 (0.04) & 0.93 (0.0) & 0.17 &  & -0.48 (0.16) & -0.30 (0.07) & -0.69 (0.0) & 0.26 \\ 
 &  &  &  &  &  &  &  &  &  &  \\ 
\vdisavg & natres & 0.52 (0.07) & -1.89 (0.02) & 0.69 (0.0) & 0.19 &  & -0.51 (0.11) & -0.42 (0.03) & -0.44 (0.0) & 0.32 \\ 
\vdisavg & natres-midtar & 0.71 (0.1) & -1.90 (0.03) & 0.79 (0.0) & 0.18 &  & -0.67 (0.16) & -0.42 (0.04) & -0.53 (0.0) & 0.32 \\ 
\vdisavg & natres-hightar & 0.71 (0.16) & -1.88 (0.04) & 0.76 (0.0) & 0.19 &  & -0.52 (0.22) & -0.44 (0.06) & -0.51 (0.0) & 0.27 \\ 
\vdisavg & lowres & 0.52 (0.09) & -1.96 (0.03) & 0.54 (0.0) & 0.25 &  & -0.51 (0.11) & -0.37 (0.03) & -0.43 (0.0) & 0.32 \\ 
\vdisavg & lowres-midtar & 0.73 (0.16) & -2.02 (0.05) & 0.56 (0.0) & 0.32 &  & -0.69 (0.19) & -0.35 (0.05) & -0.47 (0.0) & 0.36 \\ 
\vdisavg & lowres-hightar & 0.53 (0.13) & -2. (0.04) & 0.76 (0.0) & 0.16 &  & -0.46 (0.18) & -0.33 (0.06) & -0.58 (0.0) & 0.21 \\ 
\vdisavg & midres & 0.72 (0.08) & -1.96 (0.02) & 0.89 (0.0) & 0.15 &  & -0.85 (0.17) & -0.33 (0.05) & -0.57 (0.0) & 0.36 \\ 
\vdisavg & midres-hightar & 0.76 (0.11) & -2.00 (0.03) & 0.95 (0.0) & 0.13 &  & -0.76 (0.25) & -0.29 (0.08) & -0.68 (0.0) & 0.26 \\ 
\vdisavg & highres & 0.95 (0.15) & -1.97 (0.04) & 0.91 (0.0) & 0.17 &  & -0.78 (0.25) & -0.34 (0.07) & -0.69 (0.0) & 0.26 \\ 
 &  &  &  &  &  &  &  &  &  &  \\ 
\aviravg & natres & ... & ... & 0.27 (0.014) & ... &  & ... & ... & -0.24 (0.032) & ... \\ 
\aviravg & natres-midtar & ... & ... & 0.51 (0.0) & ... &  & ... & ... & -0.22 (0.155) & ... \\ 
\aviravg & natres-hightar & ... & ... & 0.38 (0.087) & ... &  & ... & ... & -0.16 (0.475) & ... \\ 
\aviravg & lowres & ... & ... & 0.17 (0.14) & ... &  & ... & ... & -0.11 (0.361) & ... \\ 
\aviravg & lowres-midtar & ... & ... & 0.26 (0.128) & ... &  & ... & ... & 0.01 (0.936) & ... \\ 
\aviravg & lowres-hightar & ... & ... & 0.58 (0.031) & ... &  & ... & ... & -0.27 (0.354) & ... \\ 
\aviravg & midres & ... & ... & 0.56 (0.0) & ... &  & ... & ... & -0.19 (0.271) & ... \\ 
\aviravg & midres-hightar & ... & ... & 0.77 (0.003) & ... &  & ... & ... & -0.44 (0.155) & ... \\ 
\aviravg & highres & ... & ... & 0.62 (0.019) & ... &  & ... & ... & -0.29 (0.311) & ... \\ 
 &  &  &  &  &  &  &  &  &  &  \\ 
\Pturbavg & natres & 0.14 (0.02) & -1.92 (0.02) & 0.72 (0.0) & 0.19 &  & -0.14 (0.03) & -0.38 (0.03) & -0.47 (0.0) & 0.32 \\ 
\Pturbavg & natres-midtar & 0.19 (0.02) & -1.98 (0.03) & 0.82 (0.0) & 0.19 &  & -0.18 (0.04) & -0.35 (0.05) & -0.56 (0.0) & 0.32 \\ 
\Pturbavg & natres-hightar & 0.20 (0.04) & -2. (0.04) & 0.85 (0.0) & 0.17 &  & -0.16 (0.05) & -0.36 (0.06) & -0.63 (0.0) & 0.25 \\ 
\Pturbavg & lowres & 0.14 (0.02) & -1.95 (0.03) & 0.59 (0.0) & 0.25 &  & -0.14 (0.03) & -0.37 (0.03) & -0.49 (0.0) & 0.32 \\ 
\Pturbavg & lowres-midtar & 0.2 (0.04) & -2. (0.05) & 0.61 (0.0) & 0.32 &  & -0.2 (0.05) & -0.37 (0.05) & -0.55 (0.0) & 0.35 \\ 
\Pturbavg & lowres-hightar & 0.15 (0.03) & -1.98 (0.03) & 0.85 (0.0) & 0.16 &  & -0.13 (0.04) & -0.36 (0.05) & -0.66 (0.0) & 0.20 \\ 
\Pturbavg & midres & 0.2 (0.02) & -1.99 (0.02) & 0.92 (0.0) & 0.17 &  & -0.23 (0.05) & -0.32 (0.05) & -0.64 (0.0) & 0.35 \\ 
\Pturbavg & midres-hightar & 0.21 (0.03) & -2.02 (0.03) & 0.96 (0.0) & 0.14 &  & -0.20 (0.06) & -0.3 (0.07) & -0.75 (0.0) & 0.24 \\ 
\Pturbavg & highres & 0.26 (0.04) & -2.07 (0.04) & 0.95 (0.0) & 0.16 &  & -0.21 (0.06) & -0.26 (0.07) & -0.74 (0.0) & 0.25 \\ 
\hline\hline
\end{tabular}
\begin{tablenotes}
\small
\item \textbf{Notes.} Analog to Table~\ref{TAB:HCN_corr_table} but for \cstwo.
\end{tablenotes}
\end{threeparttable}}
\end{table*}

%%%%%%%%%%%%%%%%%%%%%%%%%%%%%%%%%%%%%%%
%% individual galaxies (HCN/CO vs Sigmol)
\begin{figure*}
    \includegraphics[width=\textwidth]{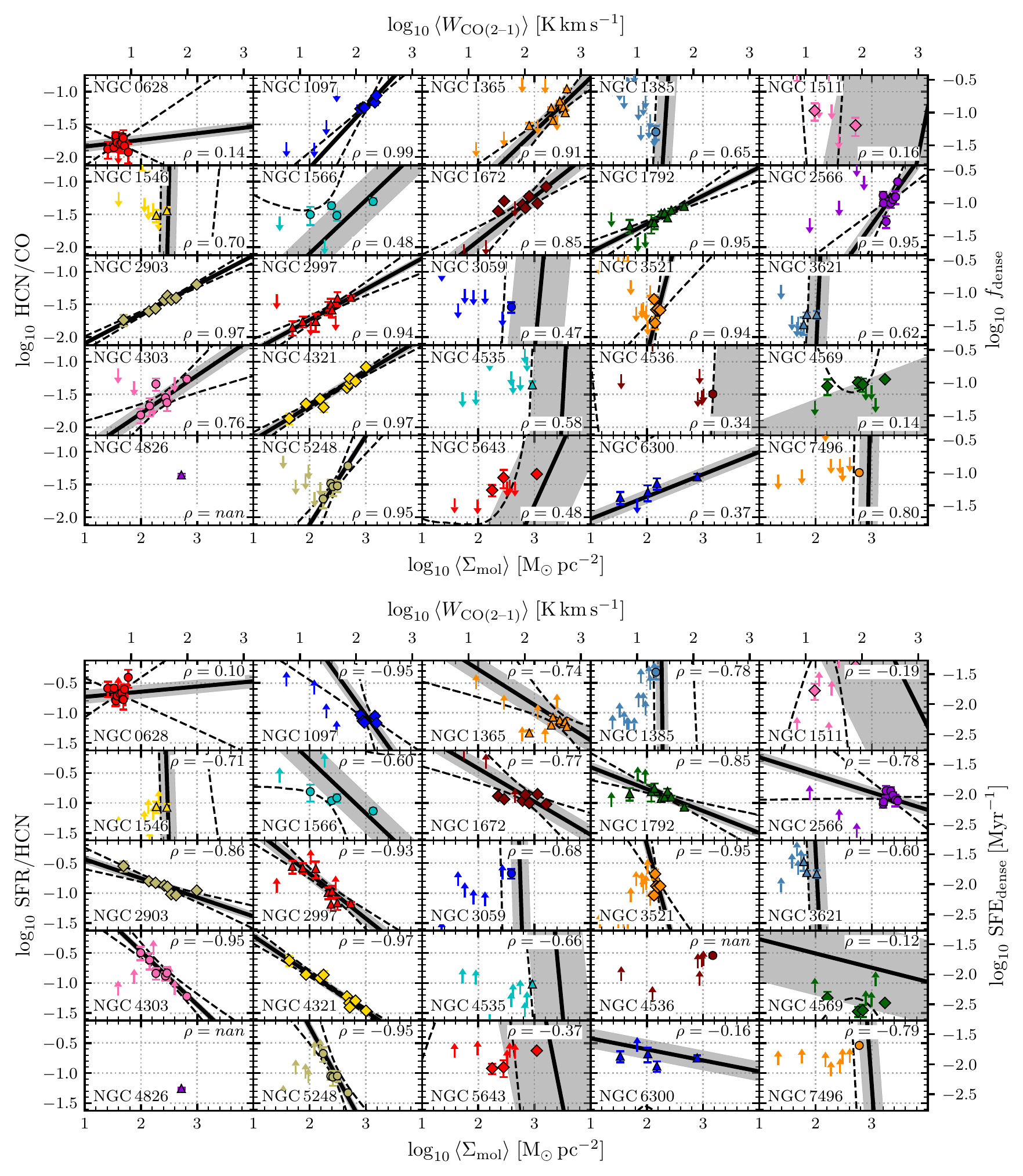}
    \caption{HCN/CO \textit{(top)} and SFR/HCN \textit{(bottom)} vs. \sigmolavg at \SI{2.1}{\kilo\parsec} and \SI{150}{\parsec} scales, plotted and fitted individually for each galaxy. The solid line shows the best fit line where the dotted line is the $1\sigma$ uncertainty. The grey shaded area indicates the scatter of the significant data about the fit line.}
    \label{FIG:HCN_corr_plots_galaxies}
\end{figure*}

%\begin{comment}

%%%%%%%%%%%%%%%%%%%%%%%%%%%%%%%%%%%%%%%
\section{Supplements: ALMOND atlas}
\label{SEC:appendix:supplements}
\begin{figure*}
    \includegraphics[scale=0.87]{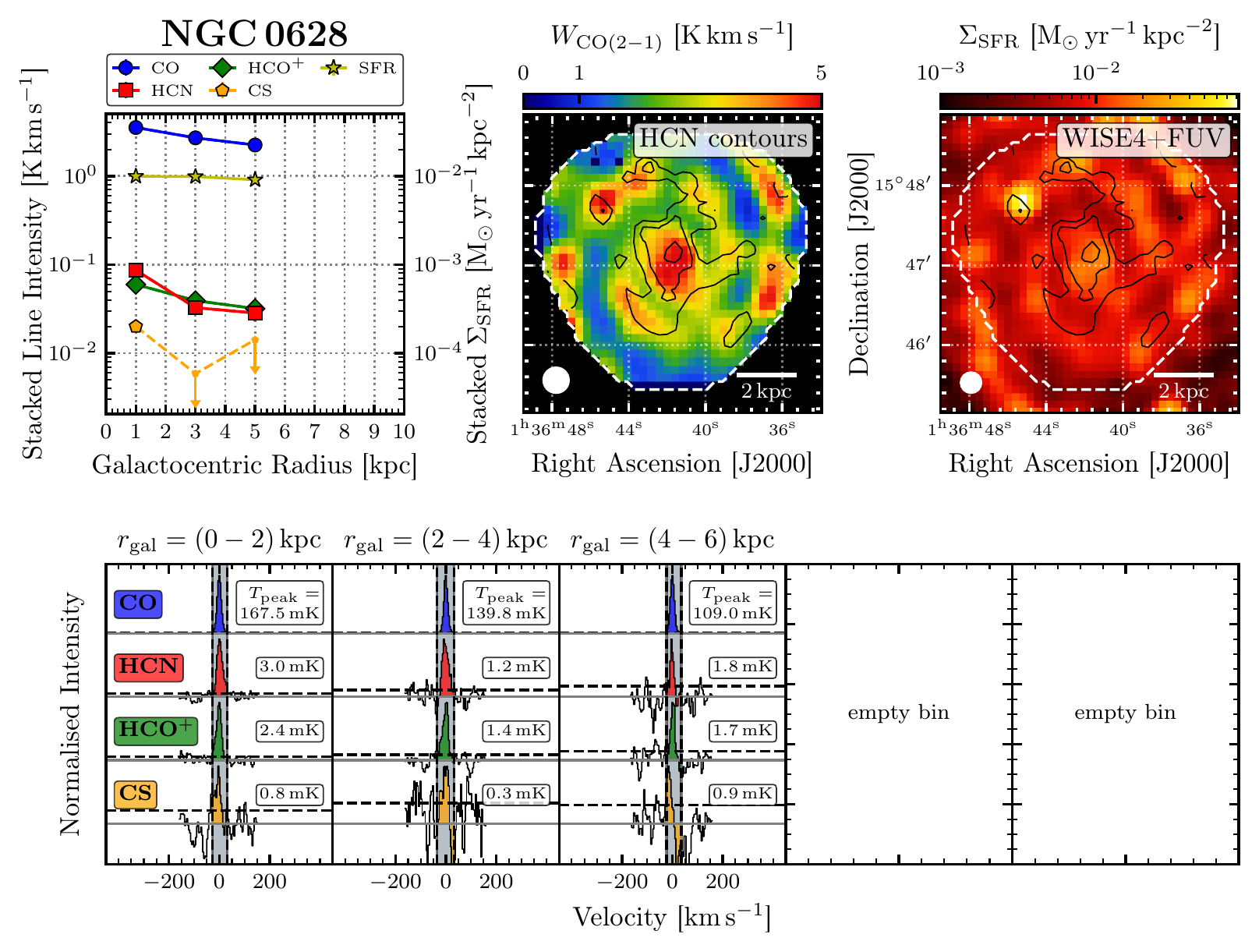}
    \caption{Analogues to Figure~\ref{FIG:ngc4321_stacking} for NGC\,628.}
    \label{FIG:almond_catalogue_first}
\end{figure*}

%%%%%%%%%%%%%%%%%%%%%%%%%%%%%%%%%%%%%%%
In Figures~\ref{FIG:almond_catalogue_first} to ~\ref{FIG:almond_catalogue_last}, we show supplemental plots analogues to Figure~\ref{FIG:ngc4321_stacking} for the remaining 24 galaxies of the ALMOND sample.

\begin{figure*}
    \includegraphics[scale=0.87]{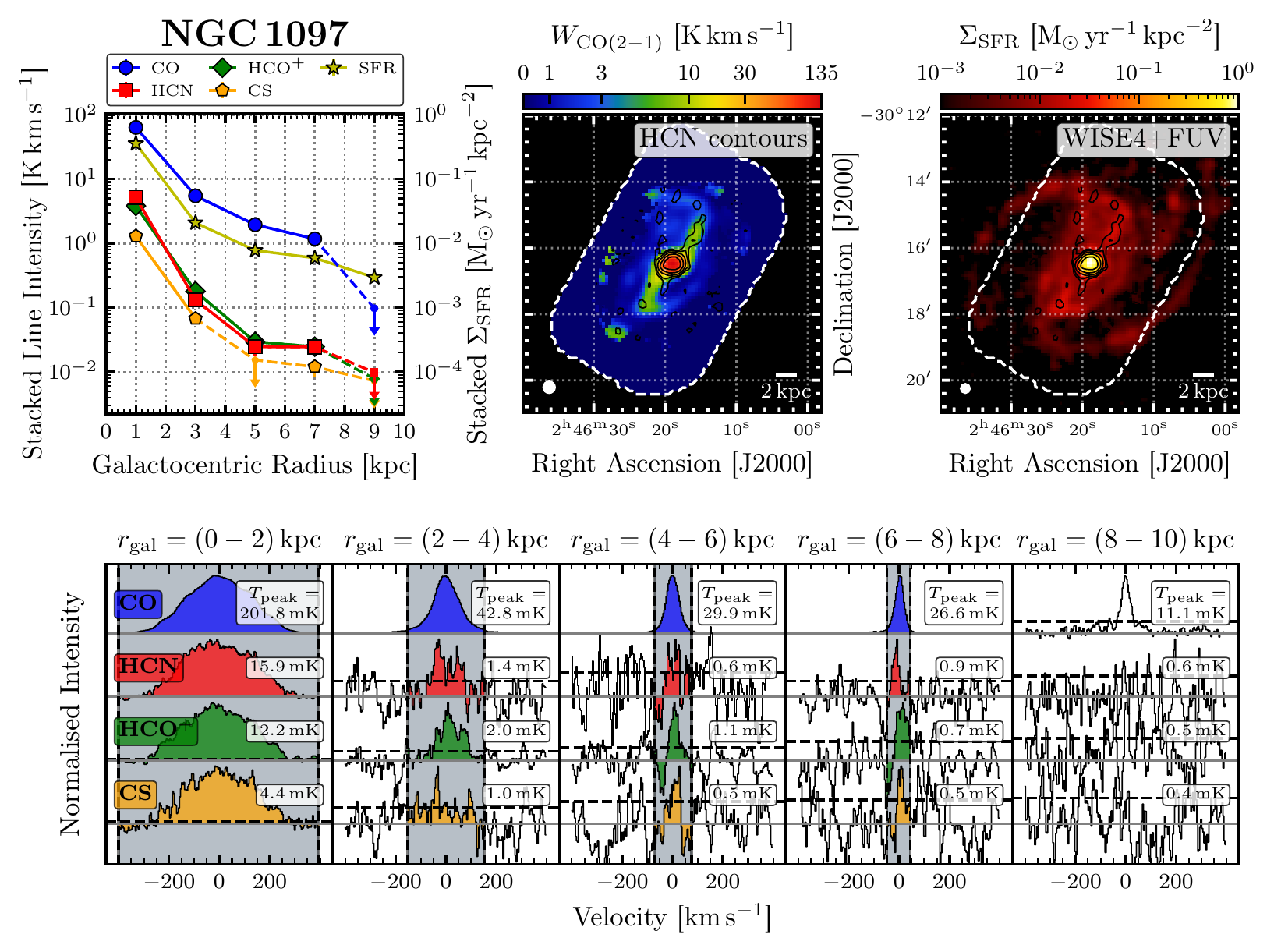}
    \caption{Analogues to Figure~\ref{FIG:ngc4321_stacking} for NGC\,1097.}
\end{figure*}

\begin{figure*}
    \includegraphics[scale=0.87]{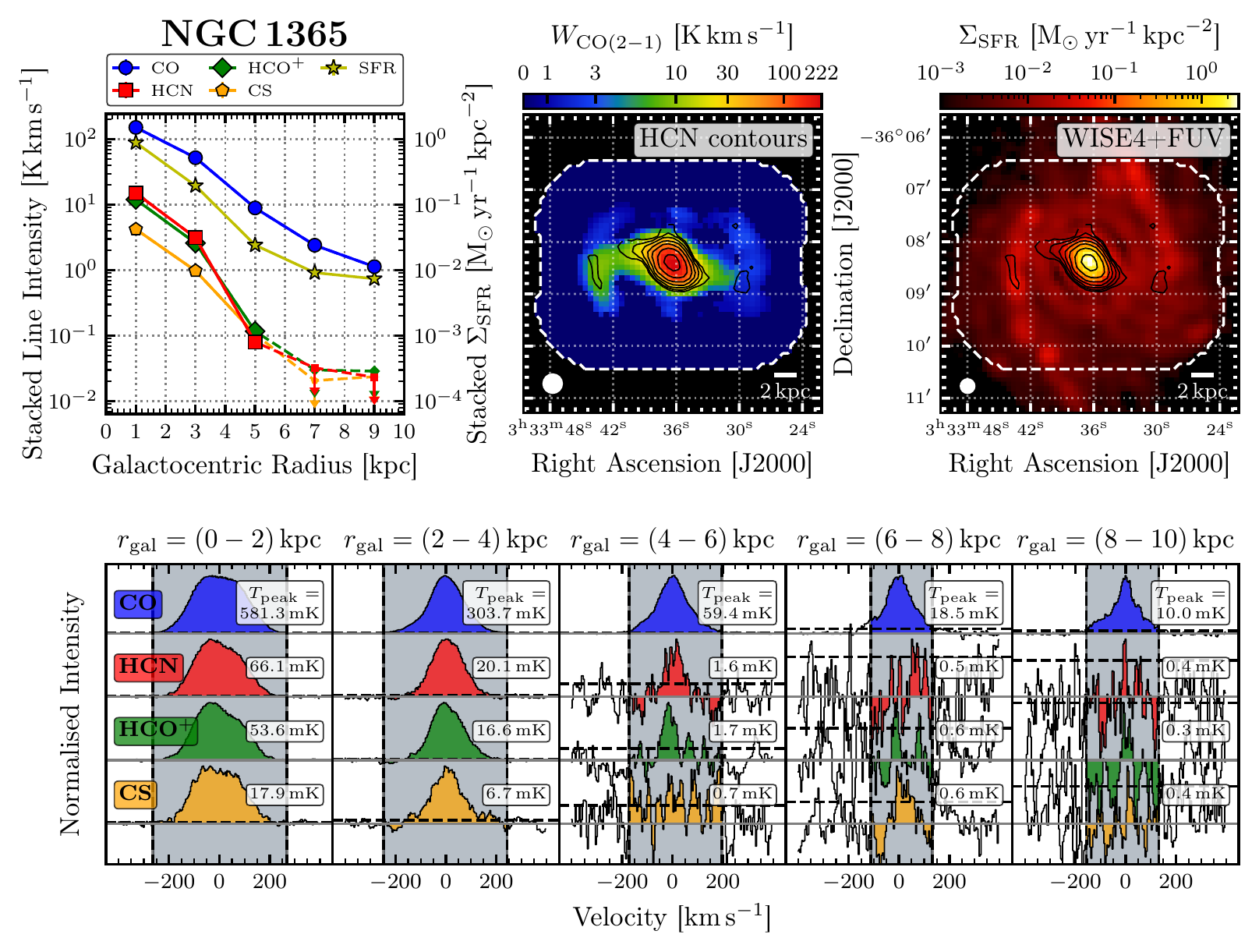}
    \caption{Analogues to Figure~\ref{FIG:ngc4321_stacking} for NGC\,1365.}
\end{figure*}

\begin{figure*}
    \includegraphics[scale=0.87]{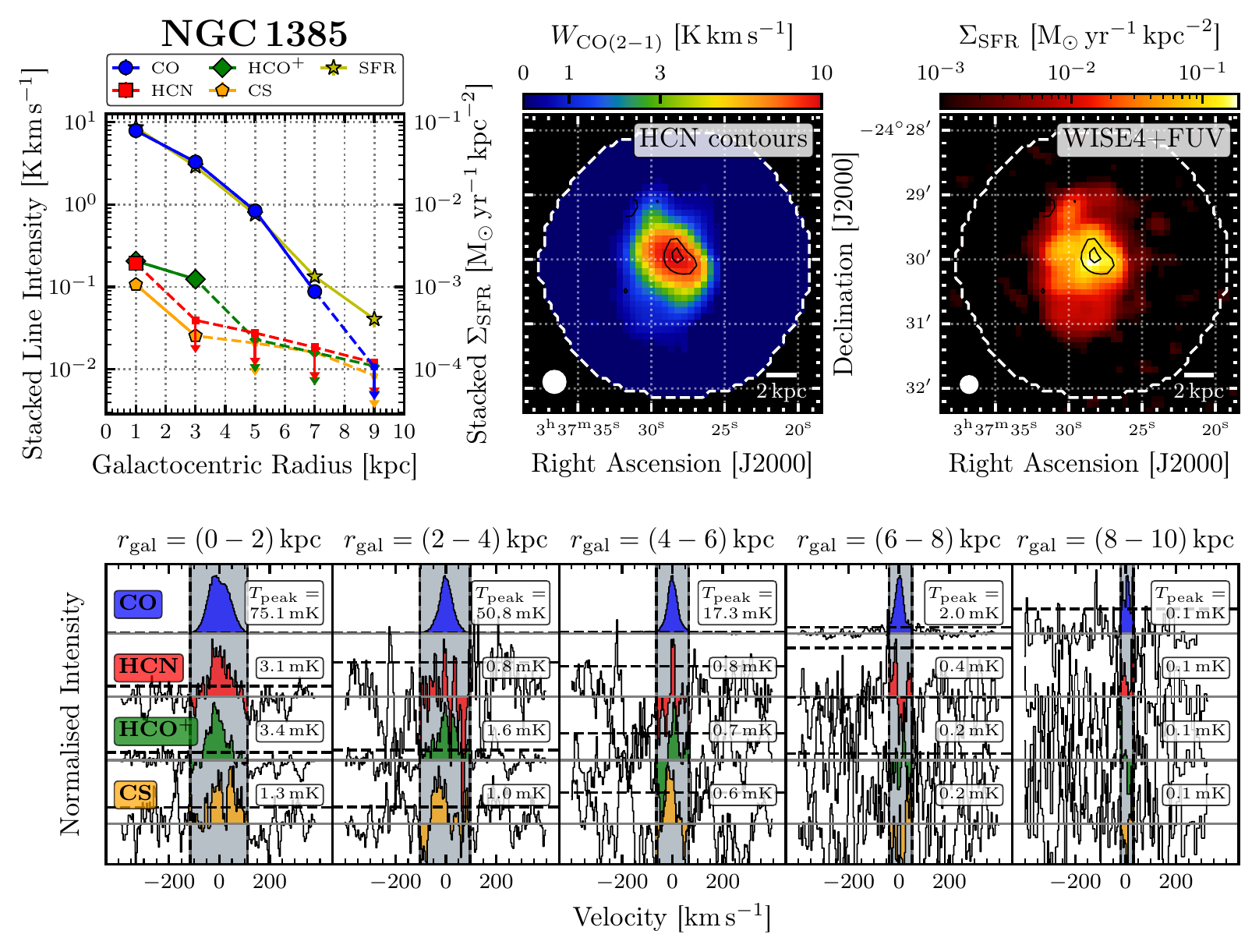}
    \caption{Analogues to Figure~\ref{FIG:ngc4321_stacking} for NGC\,1385.}
\end{figure*}

\begin{figure*}
    \includegraphics[scale=0.87]{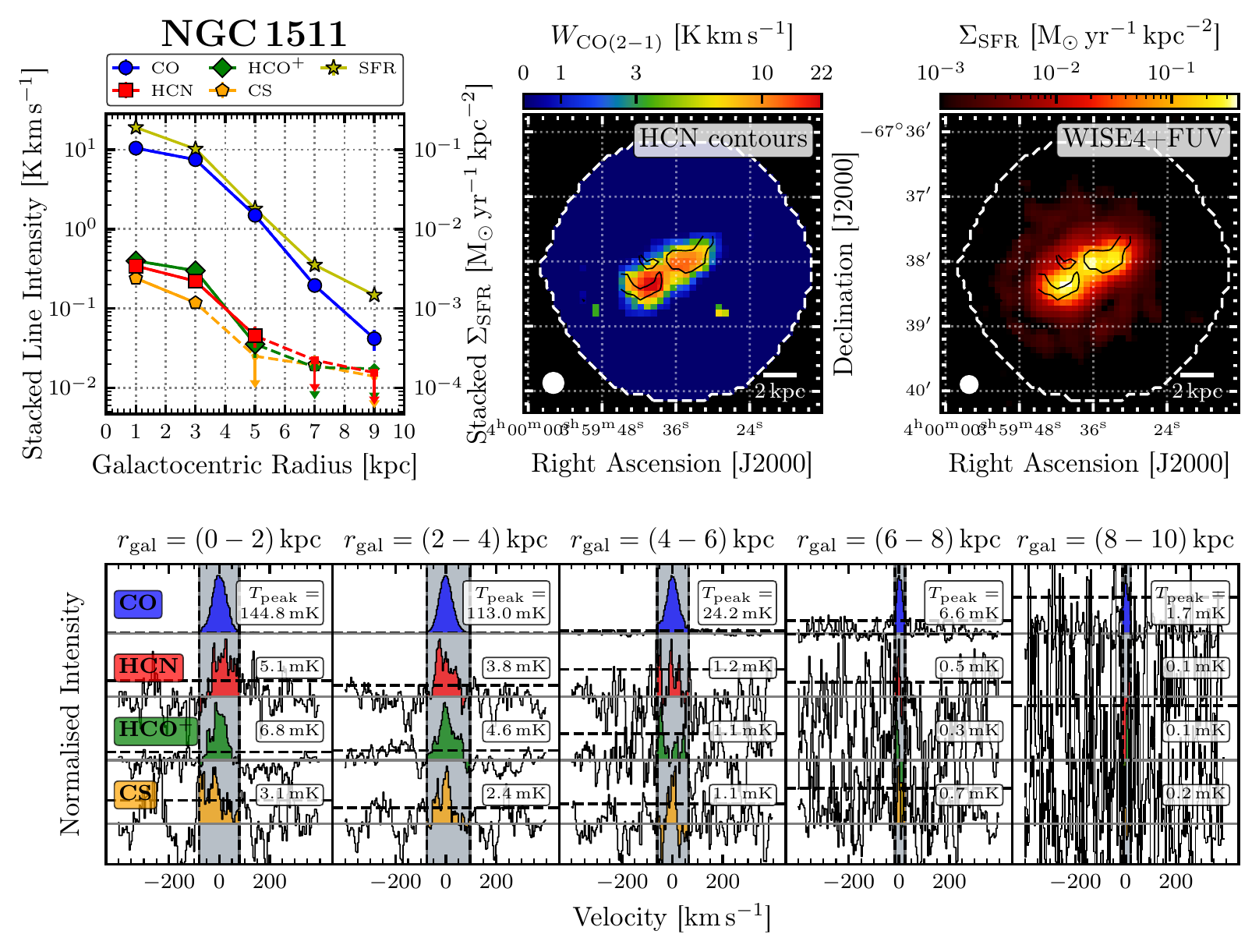}
    \caption{Analogues to Figure~\ref{FIG:ngc4321_stacking} for NGC\,1511.}
\end{figure*}

\begin{figure*}
    \includegraphics[scale=0.87]{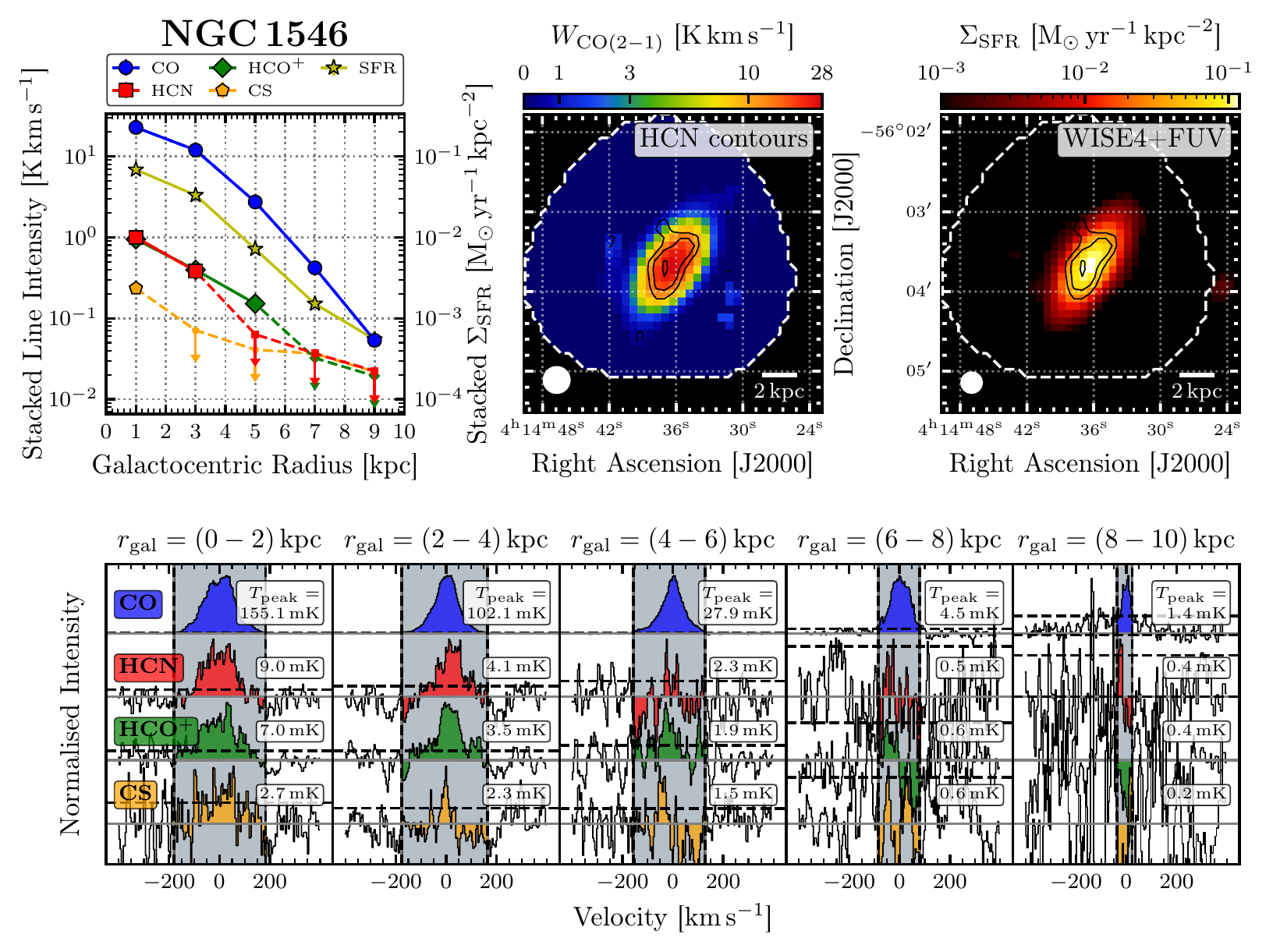}
    \caption{Analogues to Figure~\ref{FIG:ngc4321_stacking} for NGC\,1546.}
\end{figure*}

\begin{figure*}
    \includegraphics[scale=0.87]{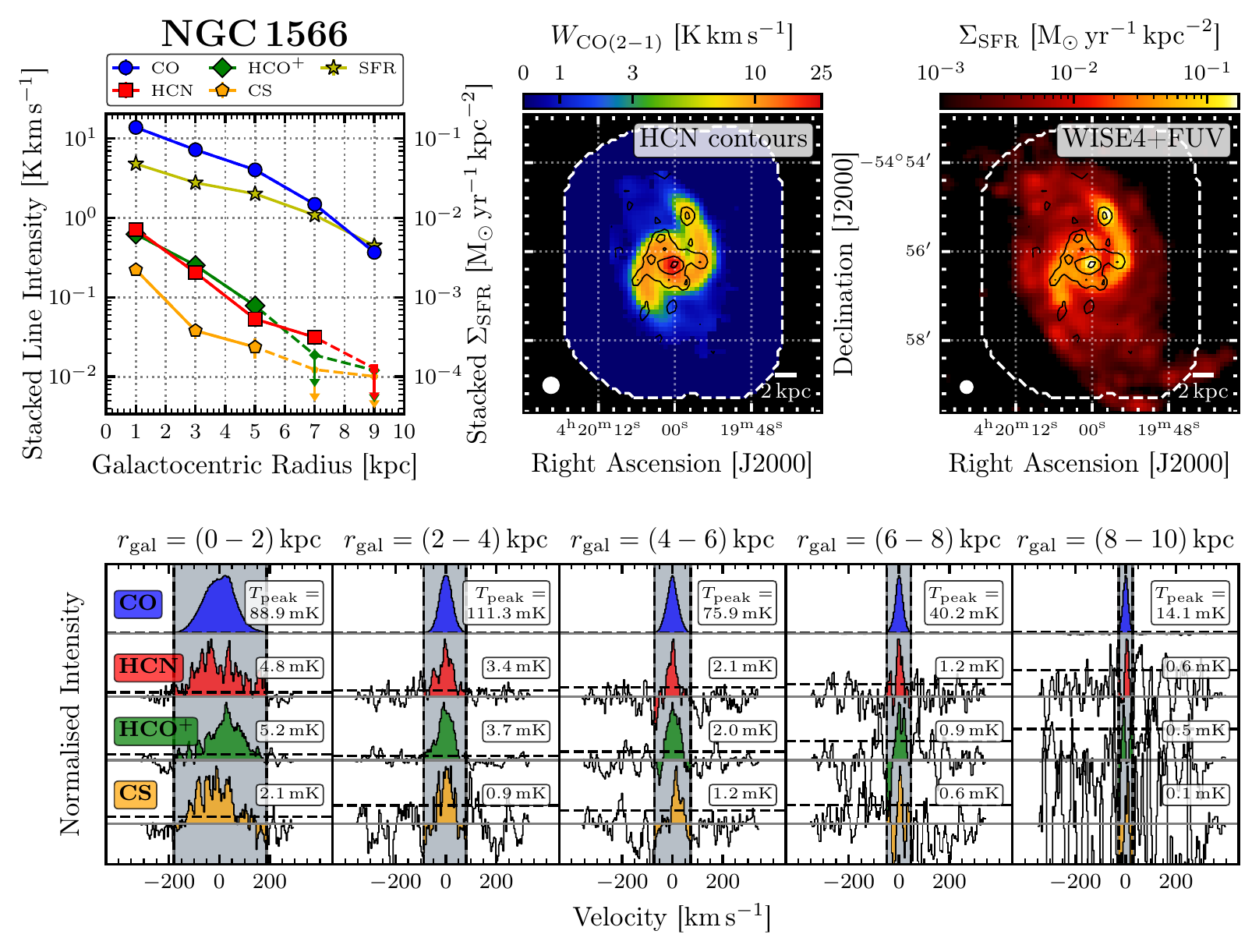}
    \caption{Analogues to Figure~\ref{FIG:ngc4321_stacking} for NGC\,1566.}
\end{figure*}

\begin{figure*}
    \includegraphics[scale=0.87]{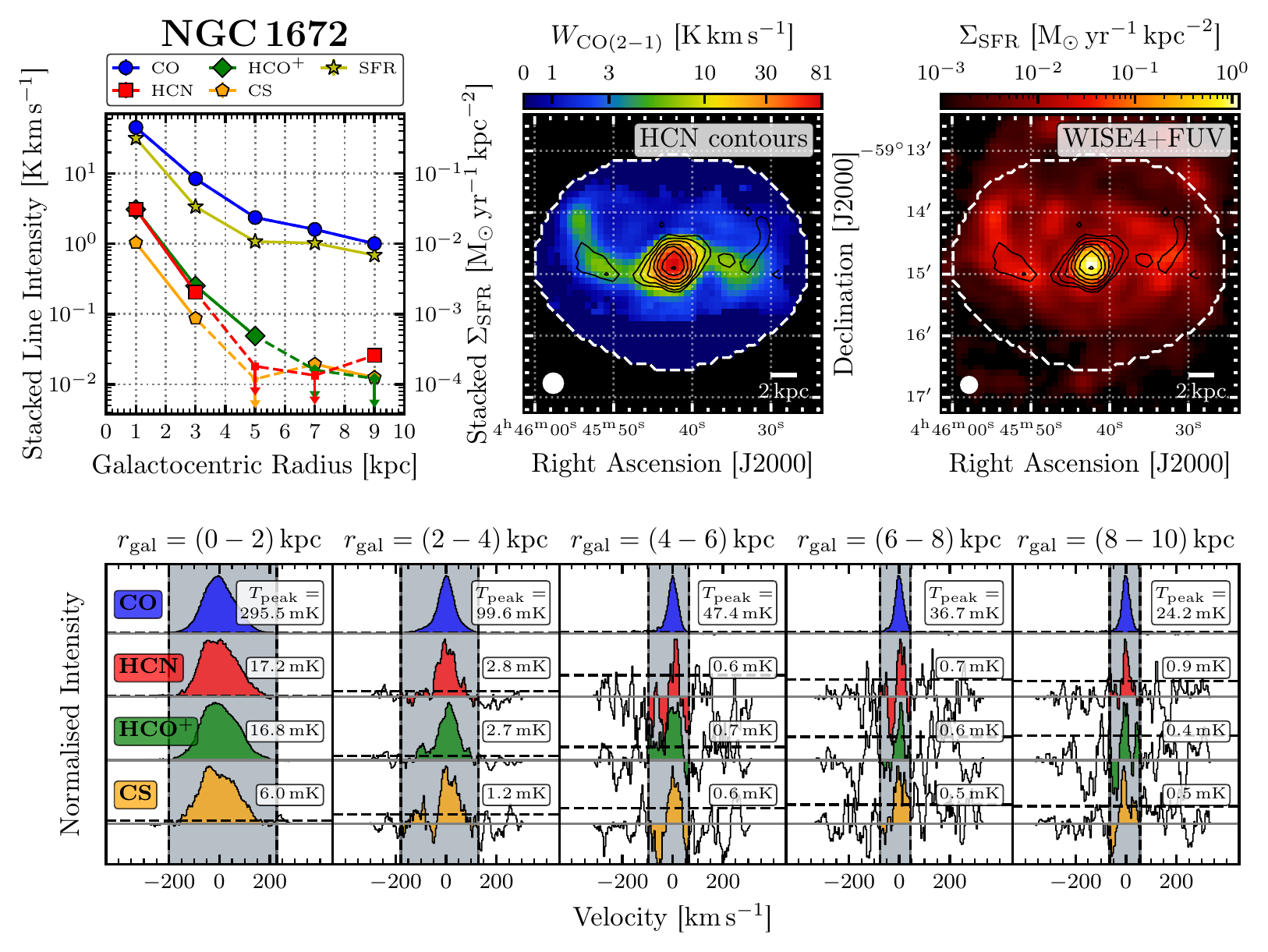}
    \caption{Analogues to Figure~\ref{FIG:ngc4321_stacking} for NGC\,1672.}
\end{figure*}

\begin{figure*}
    \includegraphics[scale=0.87]{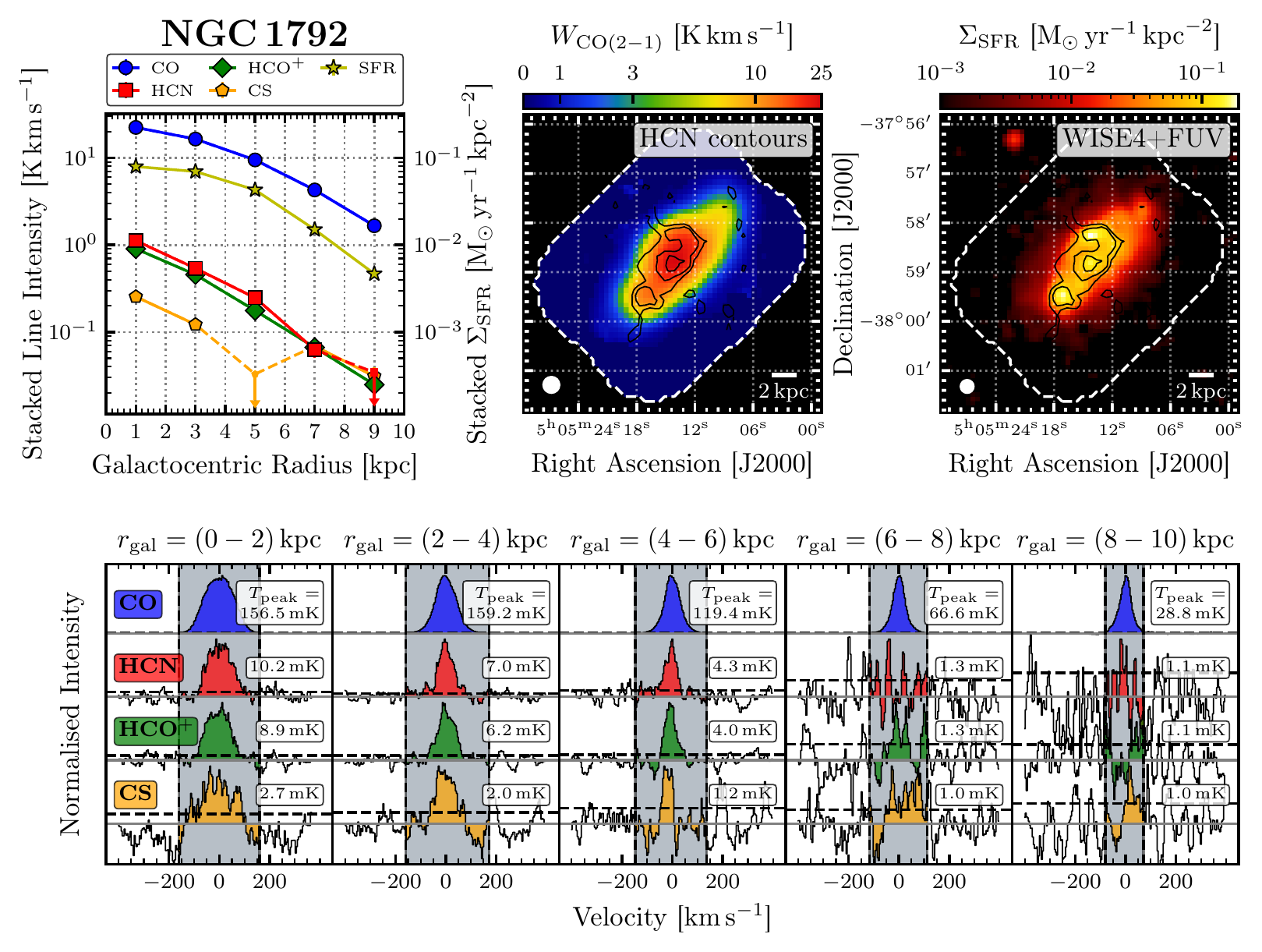}
    \caption{Analogues to Figure~\ref{FIG:ngc4321_stacking} for NGC\,1792.}
\end{figure*}

\begin{figure*}
    \includegraphics[scale=0.87]{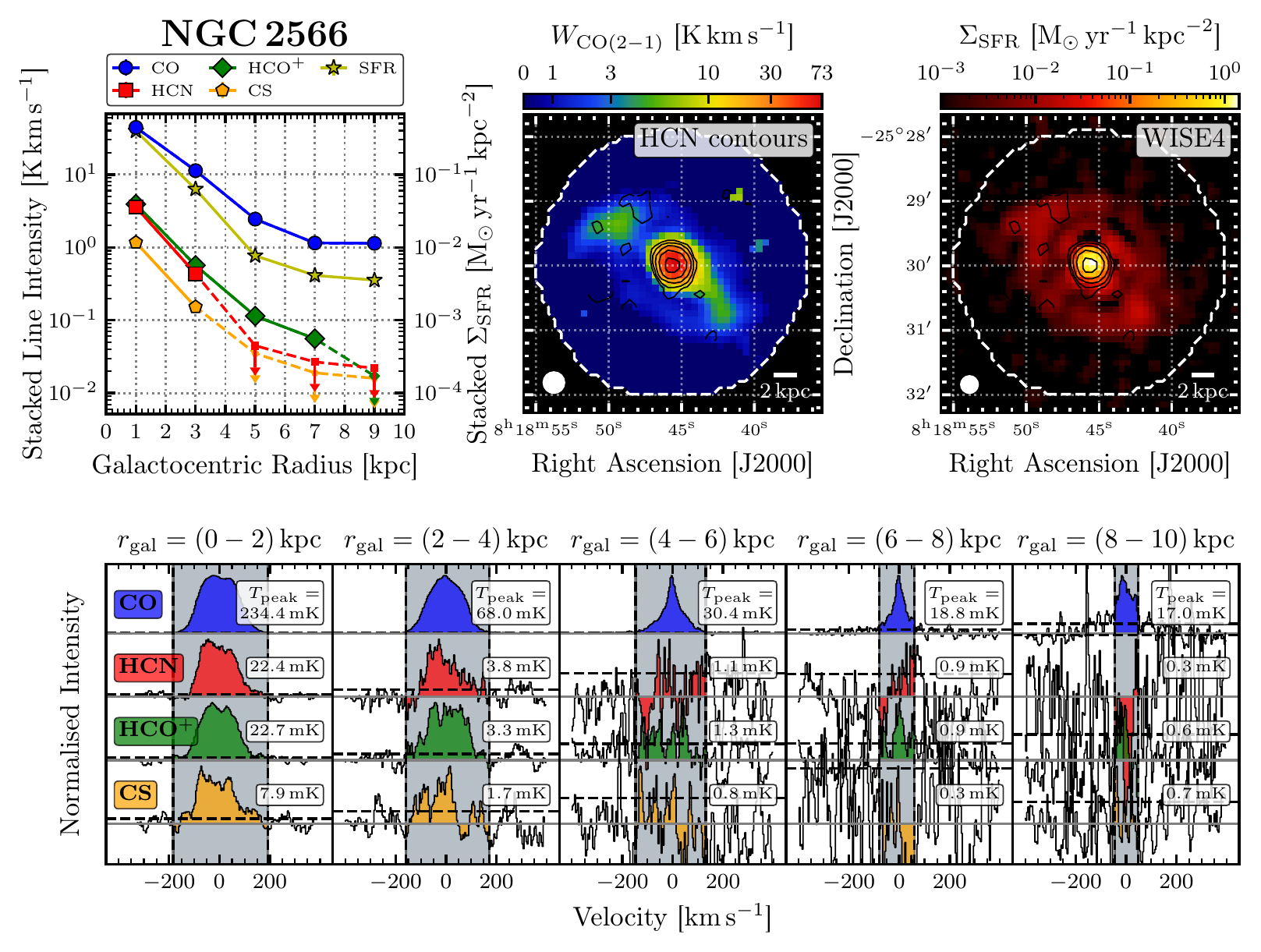}
    \caption{Analogues to Figure~\ref{FIG:ngc4321_stacking} for NGC\,2566.}
\end{figure*}

\begin{figure*}
    \includegraphics[scale=0.87]{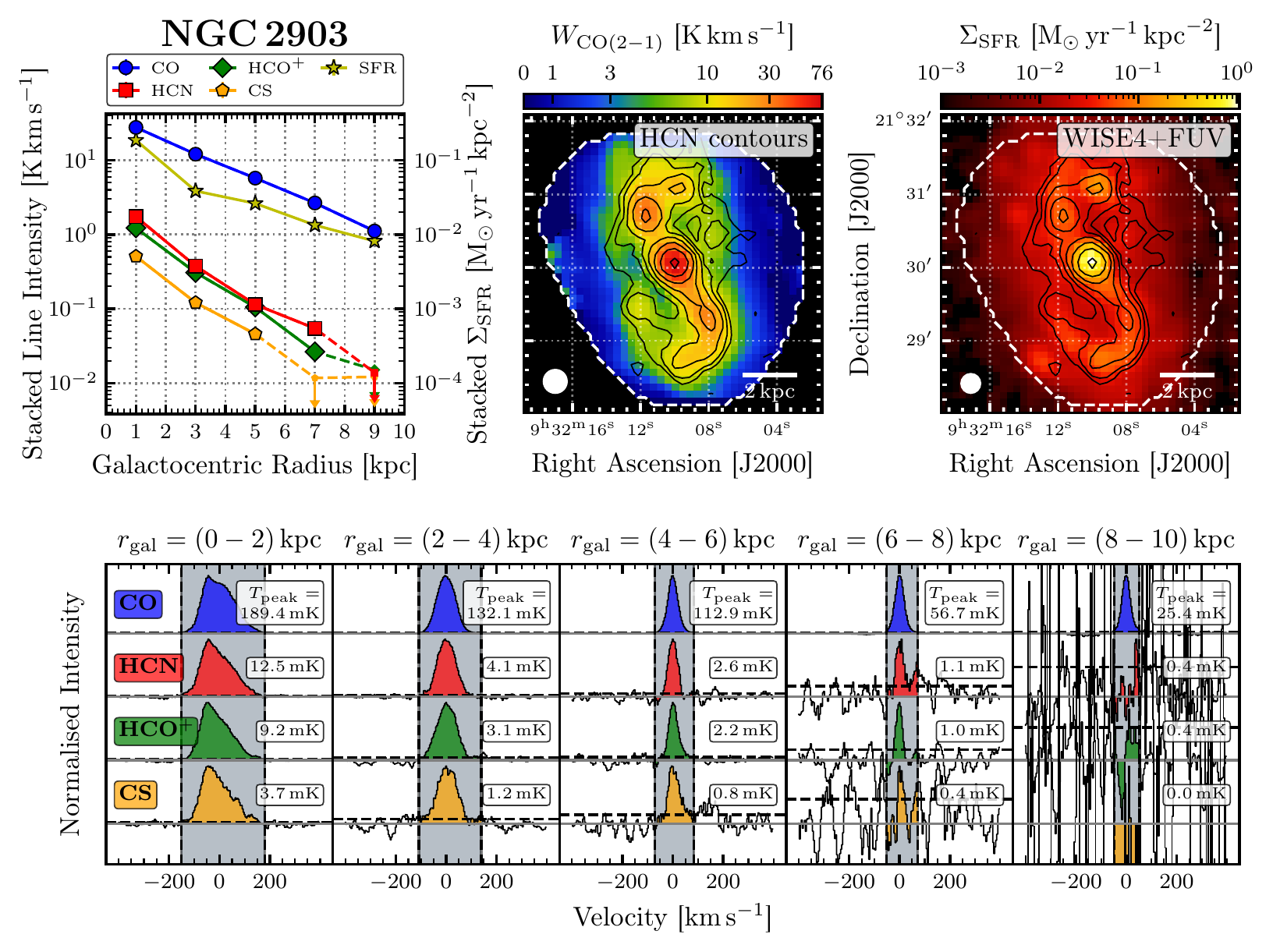}
    \caption{Analogues to Figure~\ref{FIG:ngc4321_stacking} for NGC\,2903.}
\end{figure*}

\begin{figure*}
    \includegraphics[scale=0.87]{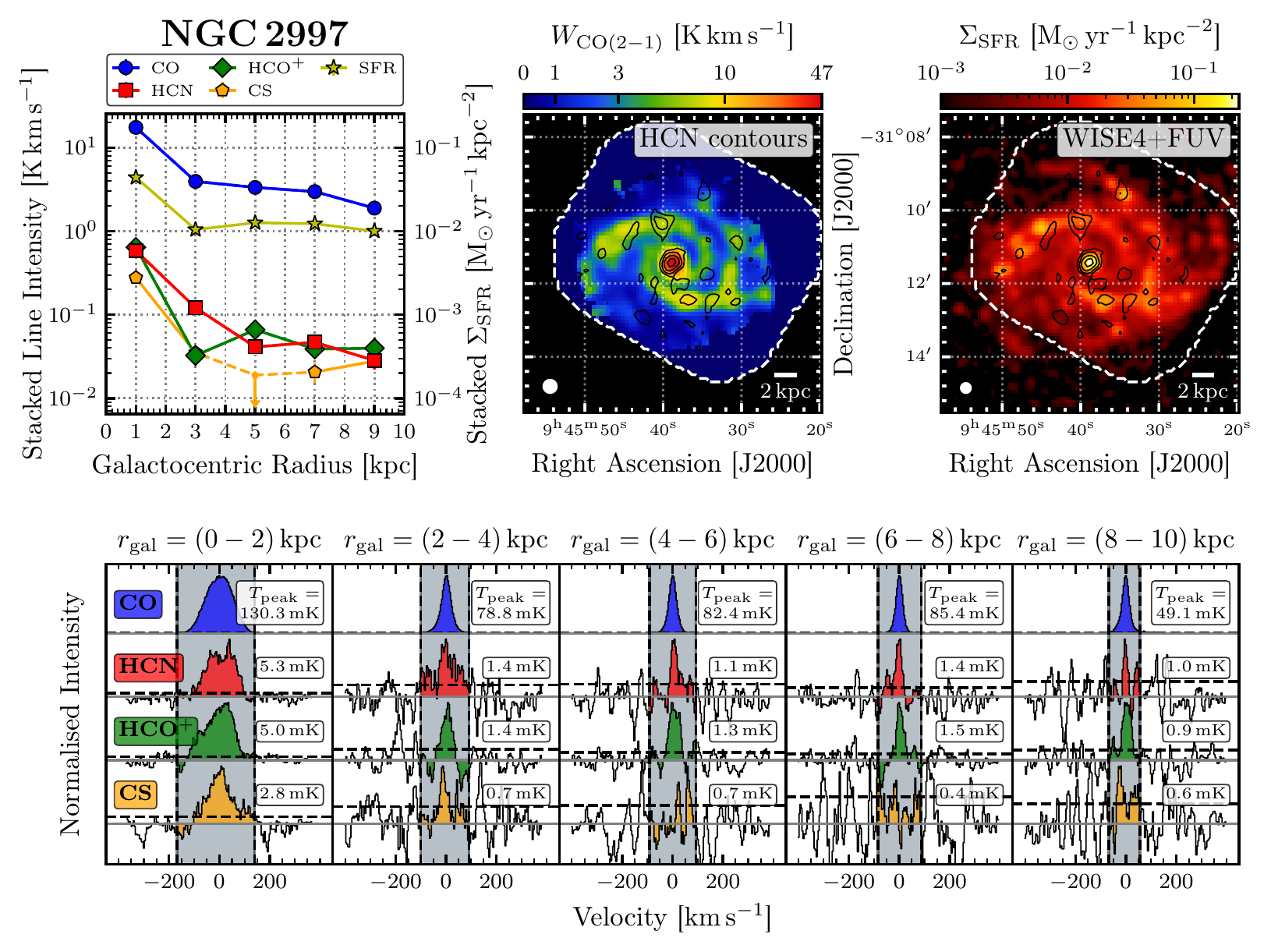}
    \caption{Analogues to Figure~\ref{FIG:ngc4321_stacking} for NGC\,2997.}
\end{figure*}
\begin{figure*}
    \includegraphics[scale=0.87]{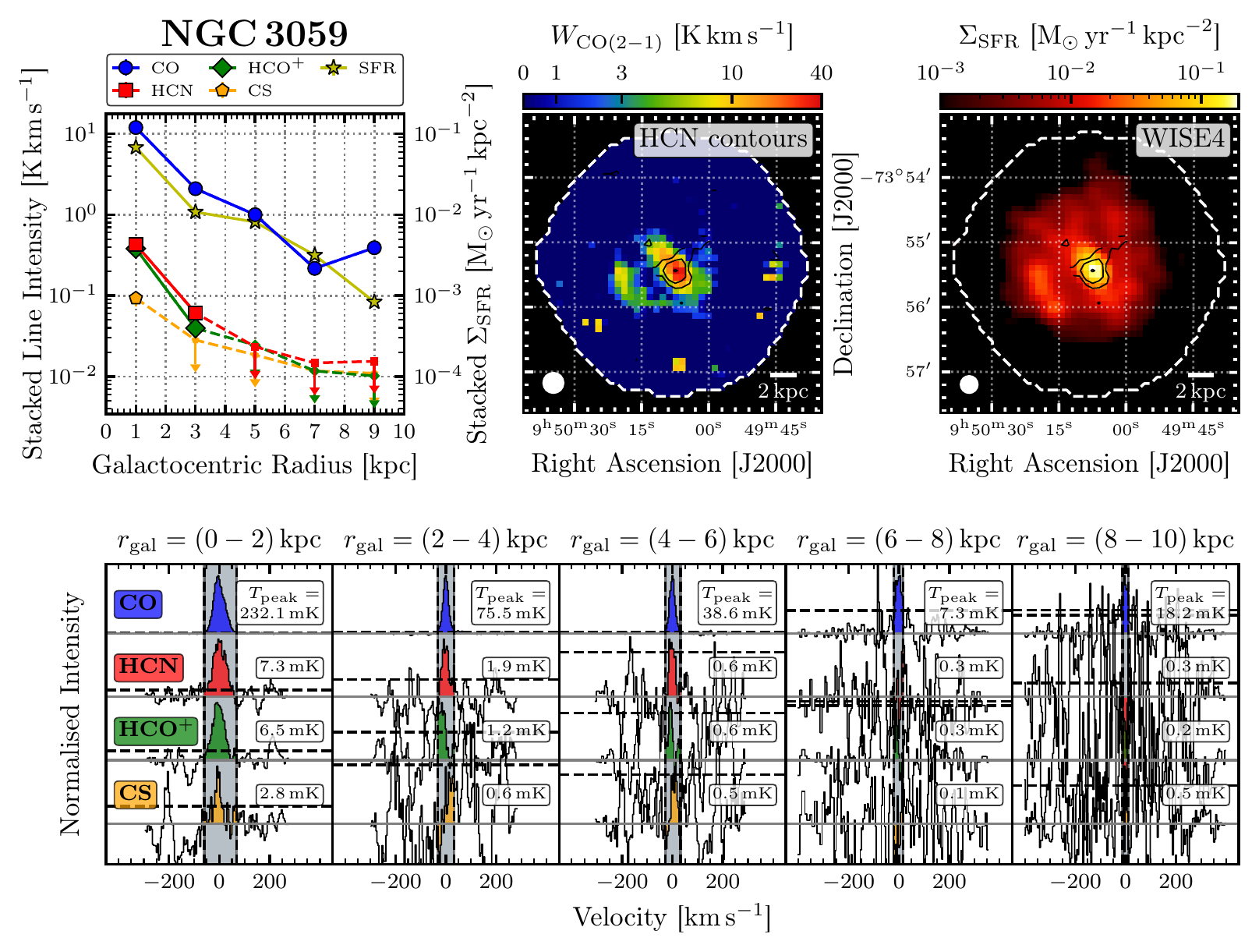}
    \caption{Analogues to Figure~\ref{FIG:ngc4321_stacking} for NGC\,3059.}
\end{figure*}

\begin{figure*}
    \includegraphics[scale=0.87]{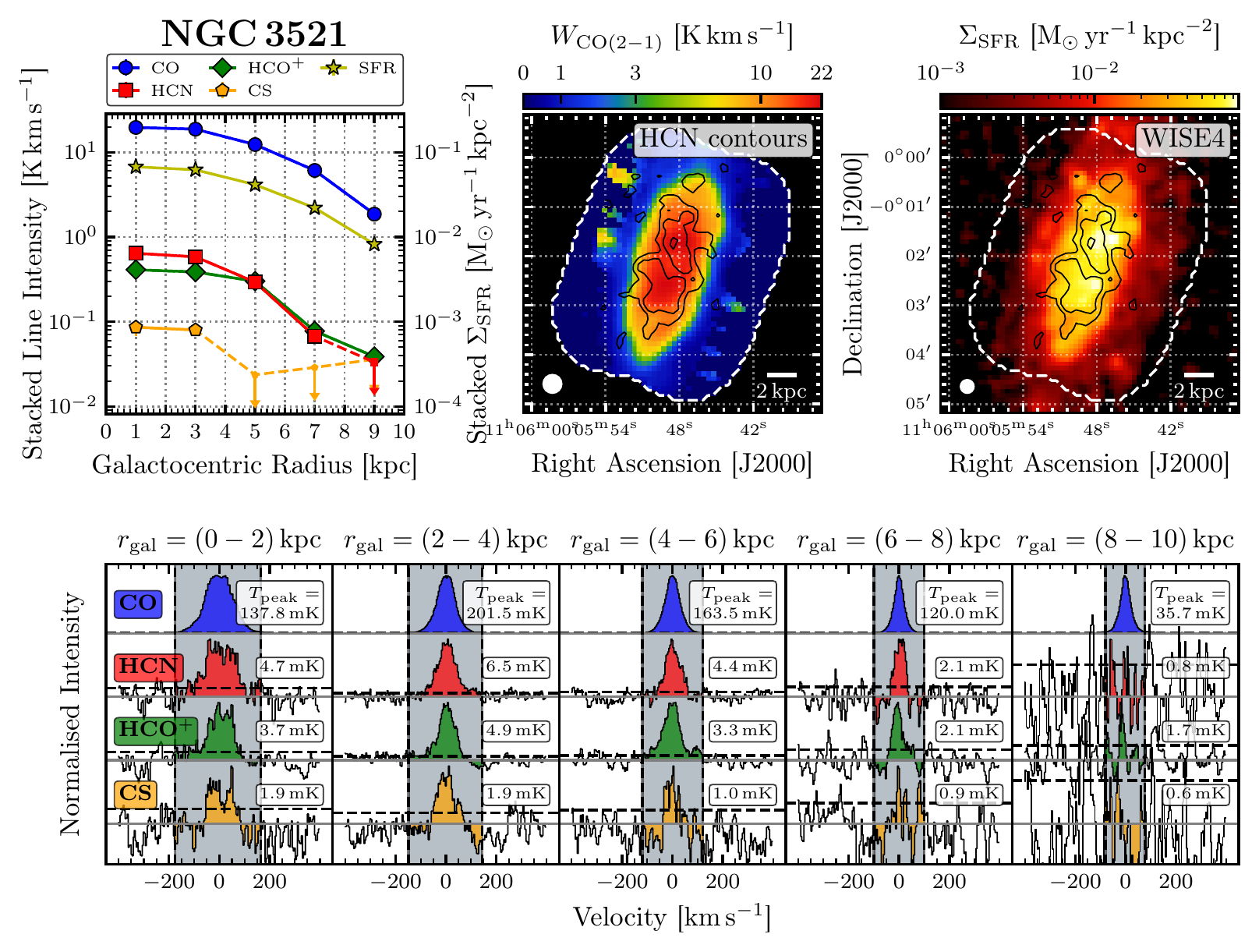}
    \caption{Analogues to Figure~\ref{FIG:ngc4321_stacking} for NGC\,3521.}
\end{figure*}

\begin{figure*}
    \includegraphics[scale=0.87]{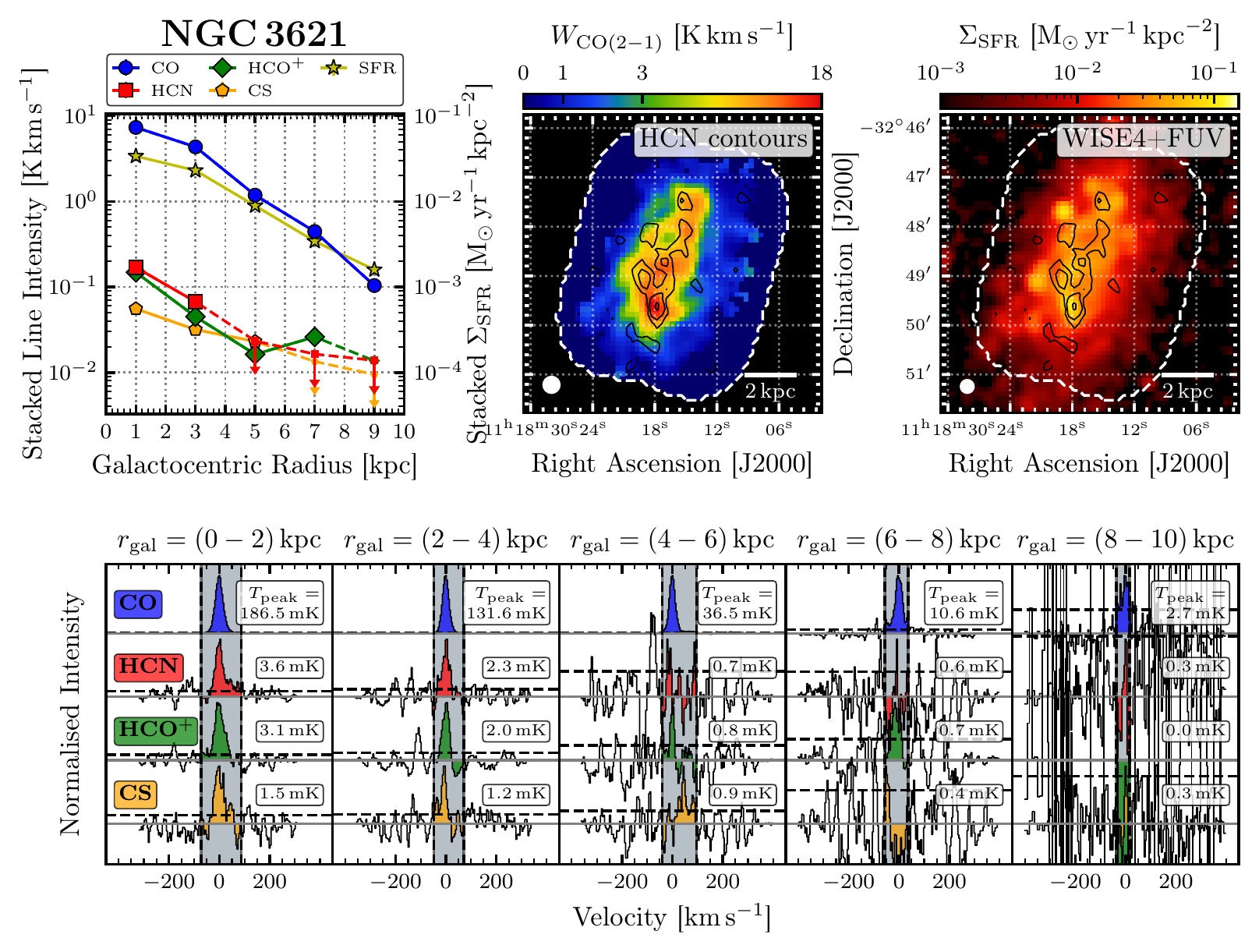}
    \caption{Analogues to Figure~\ref{FIG:ngc4321_stacking} for NGC\,3621.}
\end{figure*}

\begin{figure*}
    \includegraphics[scale=0.87]{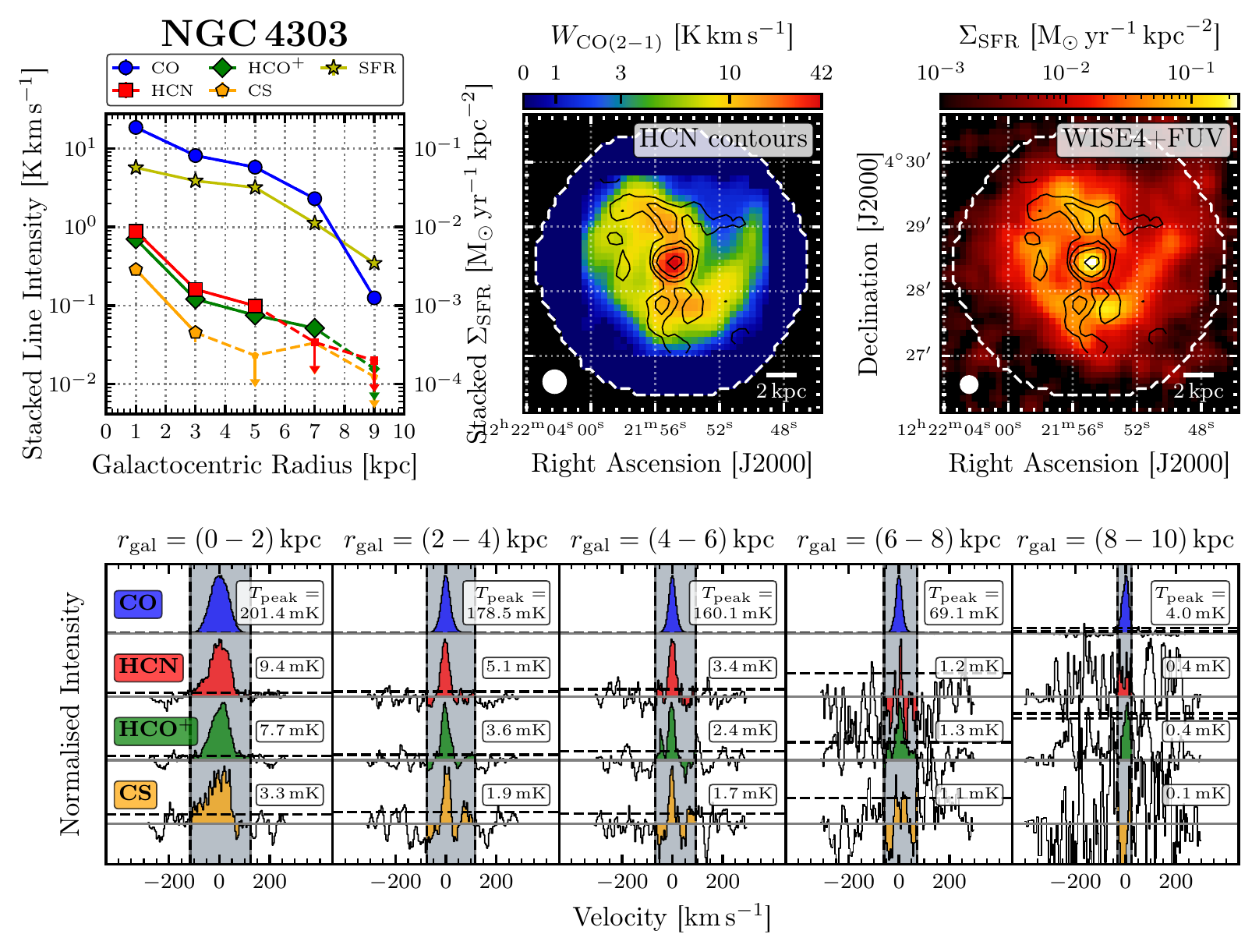}
    \caption{Analogues to Figure~\ref{FIG:ngc4321_stacking} for NGC\,4303.}
\end{figure*}

\begin{figure*}
    \includegraphics[scale=0.87]{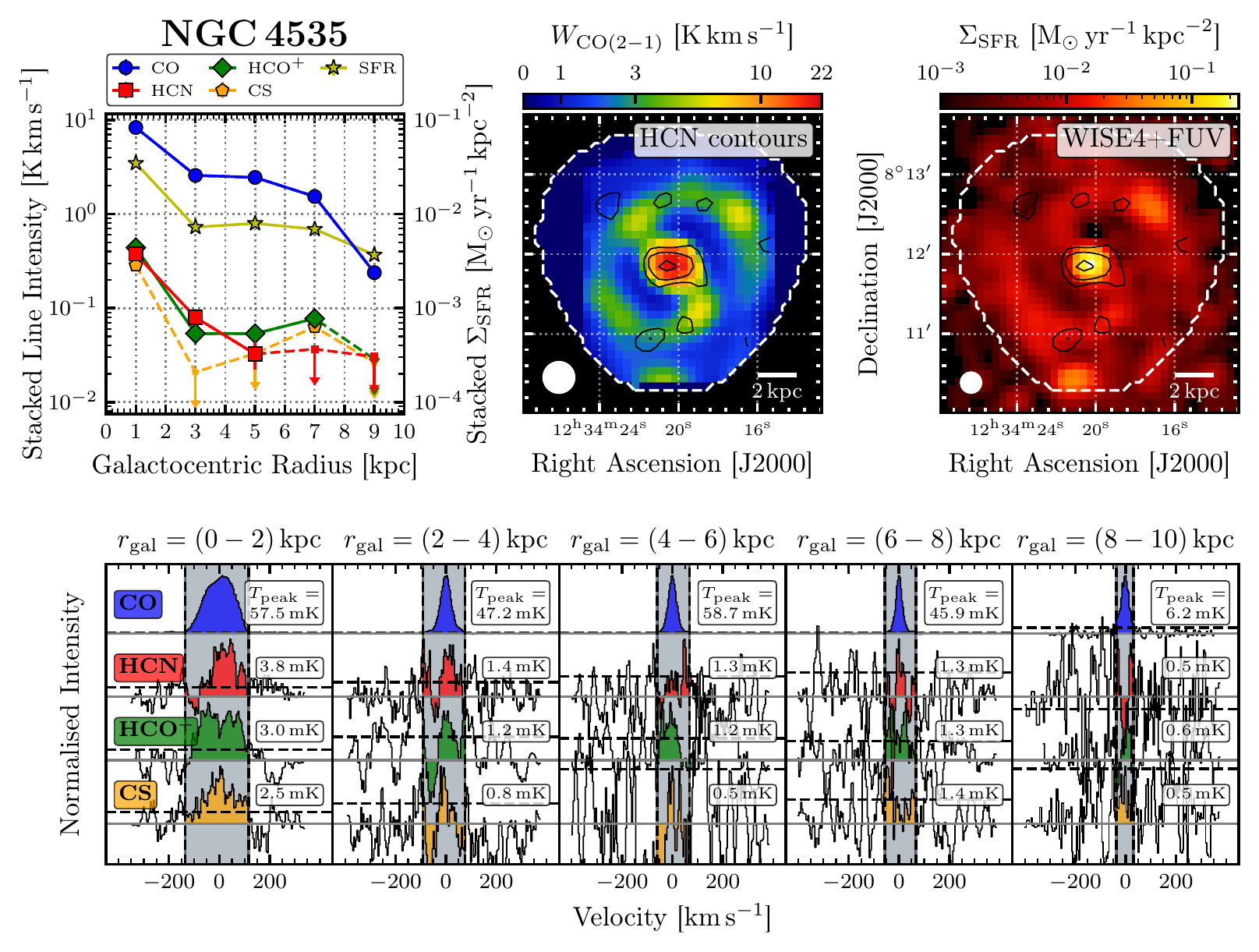}
    \caption{Analogues to Figure~\ref{FIG:ngc4321_stacking} for NGC\,4535.}
\end{figure*}

\begin{figure*}
    \includegraphics[scale=0.87]{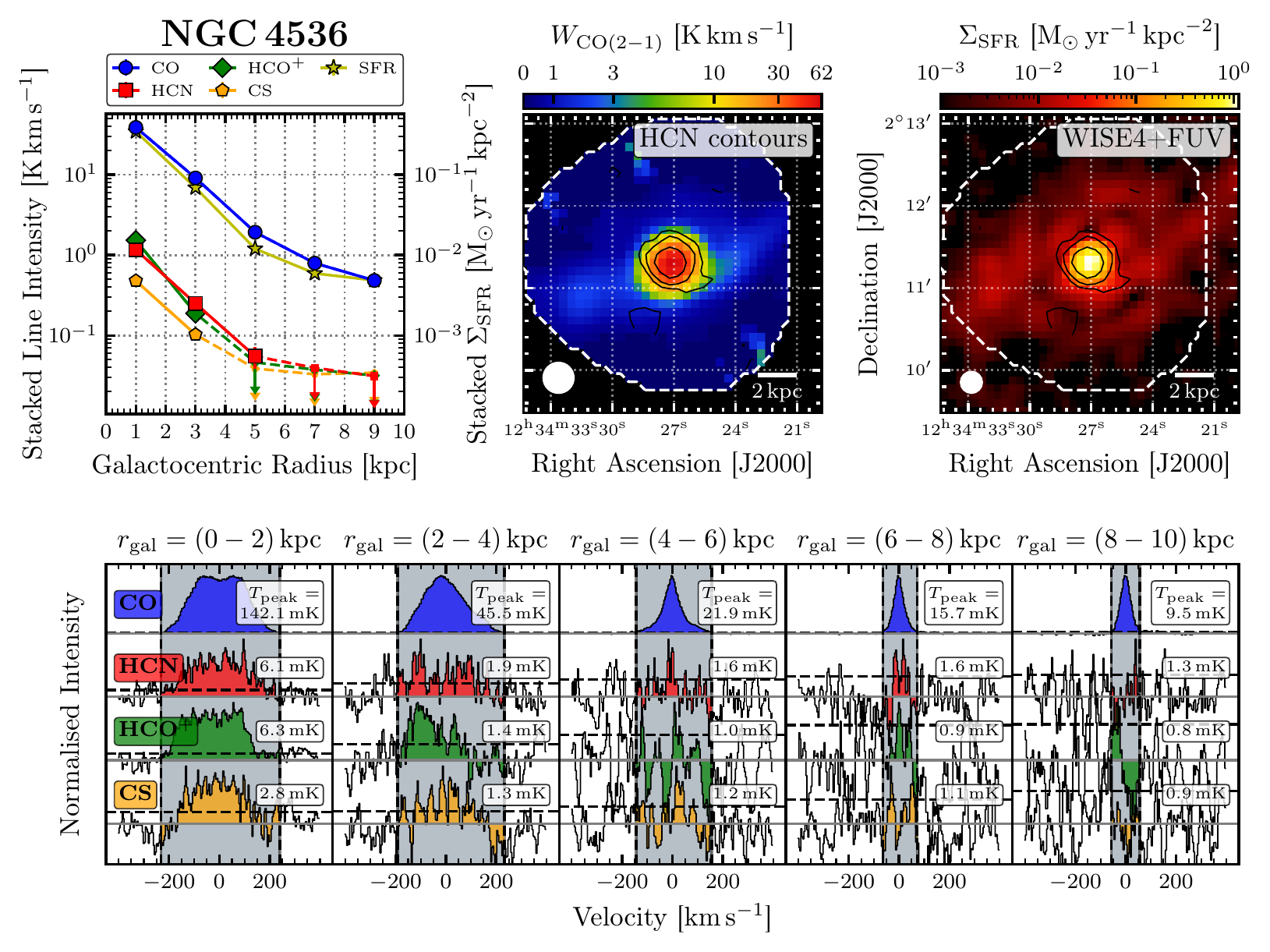}
    \caption{Analogues to Figure~\ref{FIG:ngc4321_stacking} for NGC\,4536.}
\end{figure*}

\begin{figure*}
    \includegraphics[scale=0.87]{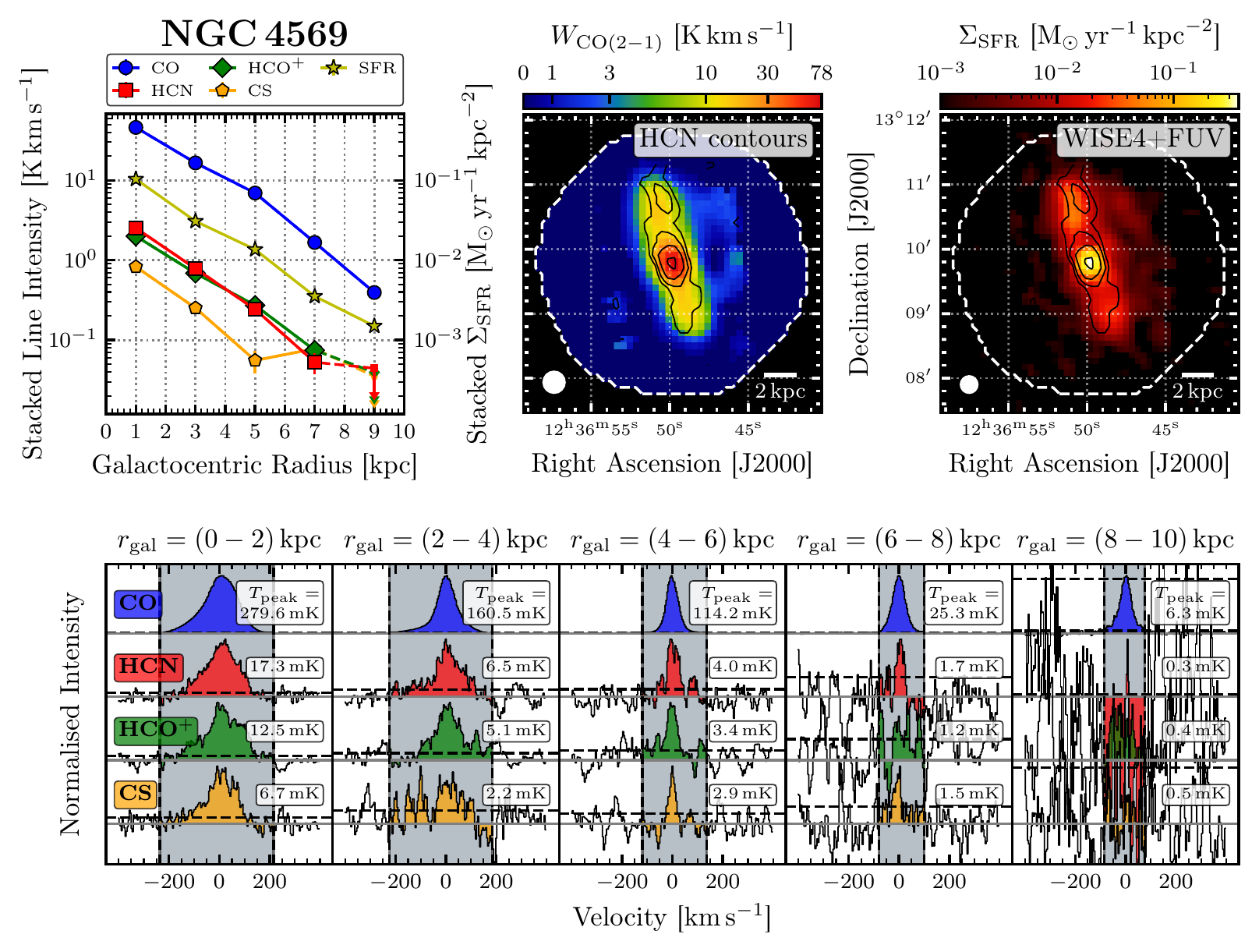}
    \caption{Analogues to Figure~\ref{FIG:ngc4321_stacking} for NGC\,4569.}
\end{figure*}

\begin{figure*}
    \includegraphics[scale=0.87]{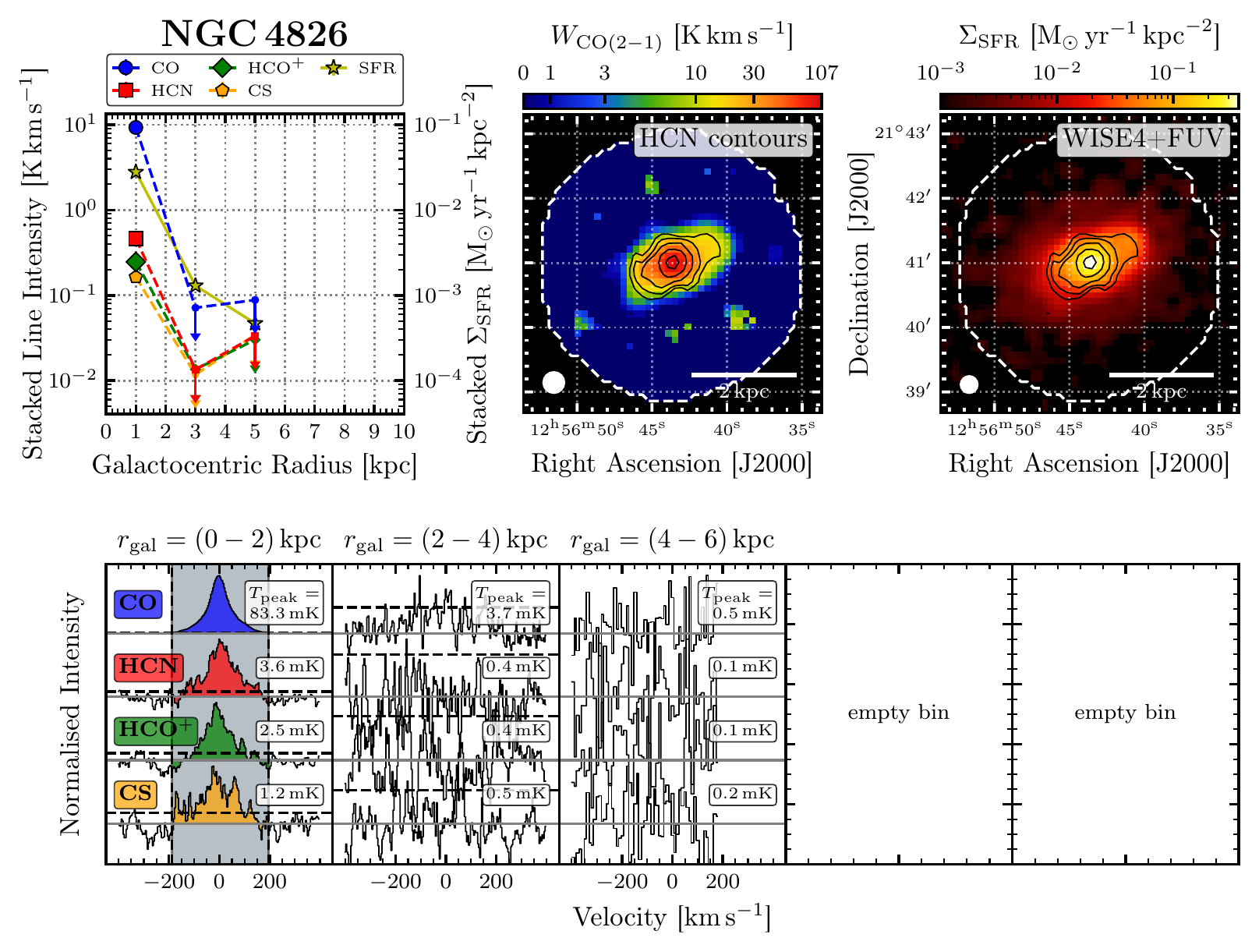}
    \caption{Analogues to Figure~\ref{FIG:ngc4321_stacking} for NGC\,4826.}
\end{figure*}

\begin{figure*}
    \includegraphics[scale=0.87]{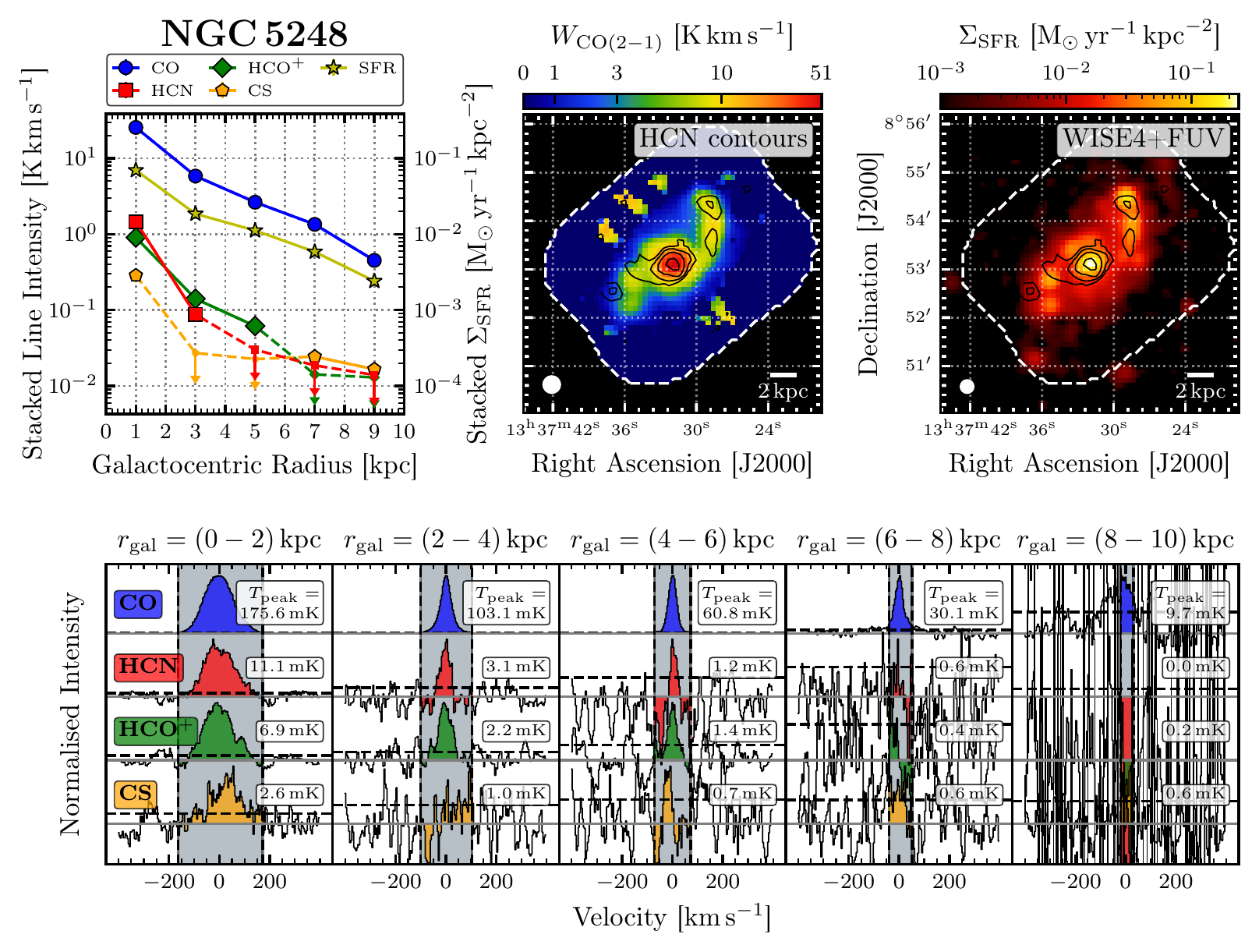}
    \caption{Analogues to Figure~\ref{FIG:ngc4321_stacking} for NGC\,5248.}
\end{figure*}

\begin{figure*}
    \includegraphics[scale=0.87]{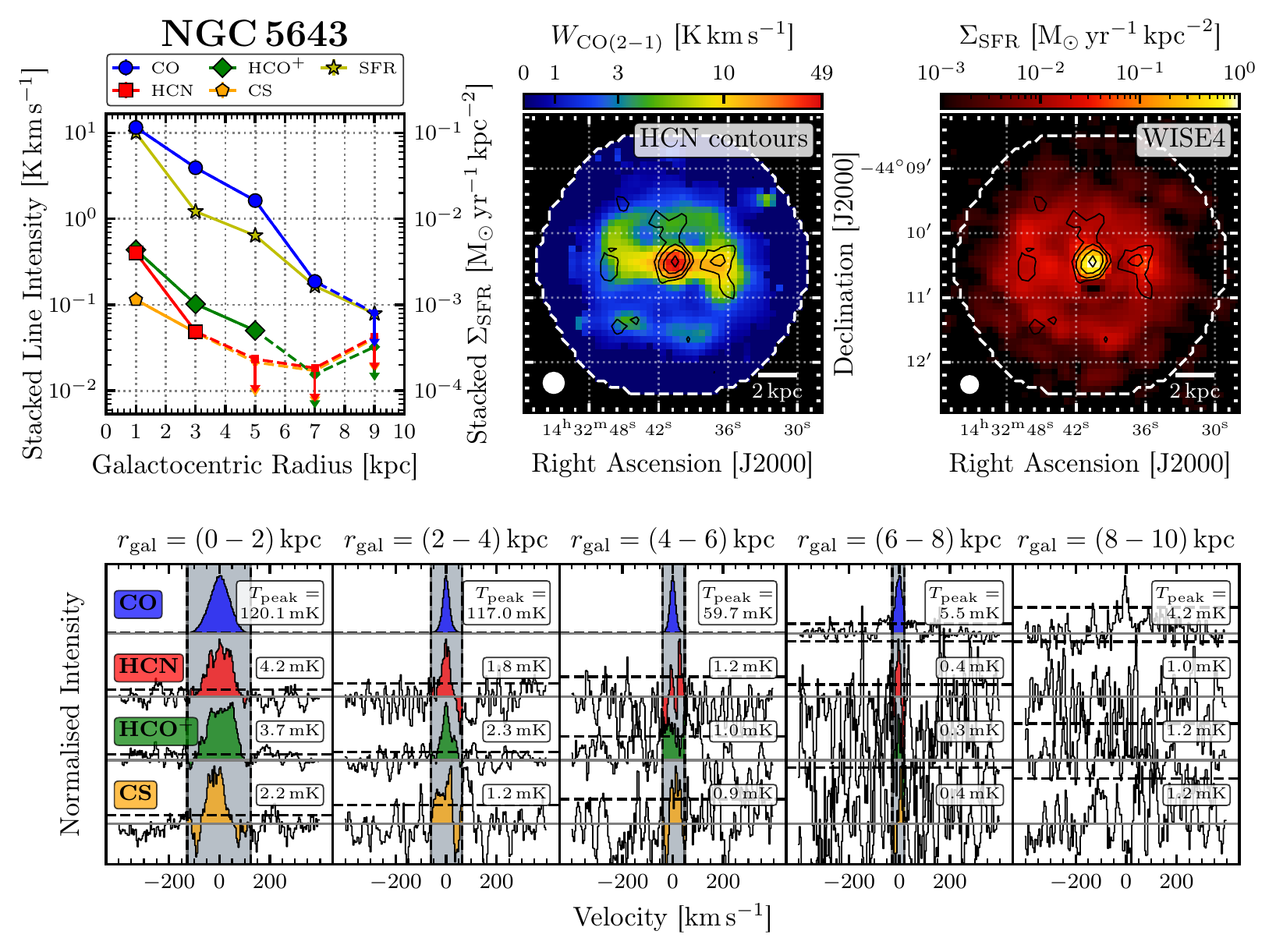}
    \caption{Analogues to Figure~\ref{FIG:ngc4321_stacking} for NGC\,5643.}
\end{figure*}

\begin{figure*}
    \includegraphics[scale=0.87]{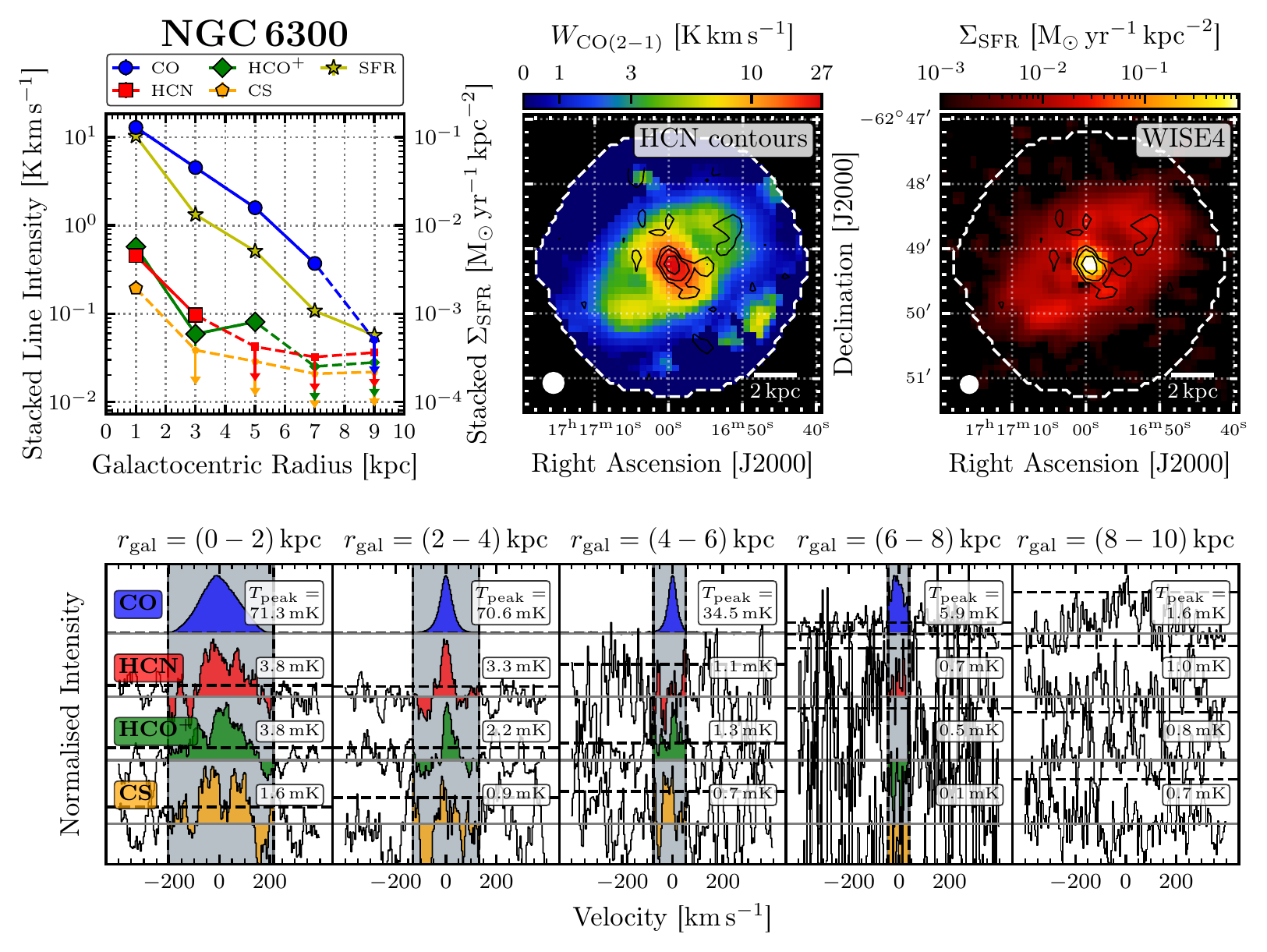}
    \caption{Analogues to Figure~\ref{FIG:ngc4321_stacking} for NGC\,6300.}
\end{figure*}

\begin{figure*}
    \includegraphics[scale=0.87]{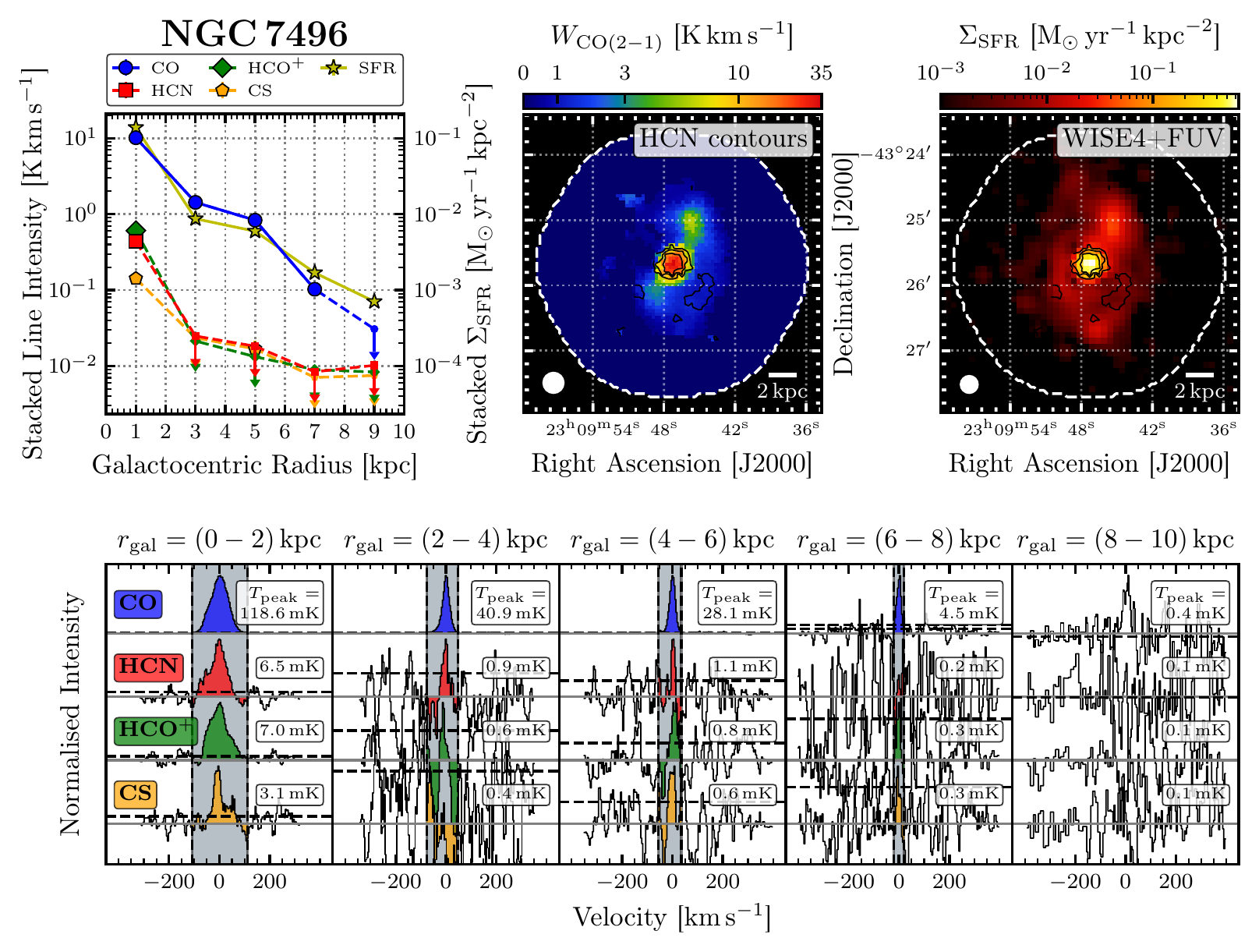}
    \caption{Analogues to Figure~\ref{FIG:ngc4321_stacking} for NGC\,7496.}
    \label{FIG:almond_catalogue_last}
\end{figure*}
    
%\end{comment}

%%%%%%%%%%%%%%%%%%%%%%%%%%%%%%%%%%%%%%%%%%%%%%%%%%

% Don't change these lines
\bsp	% typesetting comment
\label{lastpage}
\end{document}